\documentclass{ws-rv961x669}
\usepackage{ws-rv-van}     
\usepackage{ws-rv-thm}     
\usepackage{subfigure}     
\makeindex

\usepackage[T1]{fontenc}
\usepackage[utf8]{inputenc}

\graphicspath{{figs/}} 

\usepackage[export]{adjustbox}

\usepackage{relsize}
\makeatletter
\providecommand{\ion}[2]{#1$\;$\textsmaller{\@Roman{#2}}}
\makeatother

\newcommand{\msun}     {\ensuremath{{M}_{\scriptscriptstyle \odot}}}
\newcommand{\subsun}     {\ensuremath{_{\scriptscriptstyle \odot}}}
\newcommand{\kms}      {\ensuremath{\mathrm{km~s^{-1}}}}

\newcommand{\ergs}     {\ensuremath{\mathrm{erg\,s^{-1}}}}

\newcommand{\msigma}   {\ensuremath{M}{--}\ensuremath{\sigma}}

\newcommand{\mbh}      {\ensuremath{M_{\mathrm{BH}}}}

\newcommand{\mgii} {Mg\,{\sc ii}}
\newcommand{\feii} {Fe\,{\sc ii}}
\newcommand{\civ} {C\,{\sc iv}}
\newcommand{\ciii} {C\,{\sc iii]}}
\newcommand{\carboniii} {C\,{\sc iii}\,$\lambda$977}

\newcommand{\ha} {H\,$\alpha$}
\newcommand{\hb} {H\,$\beta$}
\newcommand{\oiii} {[O\,{\sc iii}]}
\newcommand{\oii} {[O\,{\sc ii}]}
\newcommand{\neiii} {[Ne\,{\sc iii}]}
\newcommand{\nev} {[Ne\,{\sc v}]}

\newcommand{\ovi} {O\,{\sc vi}\,$\lambda$1035}
\newcommand{\nv} {N\,{\sc v}}

\newcommand{\heii} {He\,{\sc ii}}
\newcommand{\neviii} {Ne\,{\sc viii}\,$\lambda$780}
\newcommand{\hii}{H\,{\sc ii}}

\newcommand{\arcsec}{\ensuremath{^{\prime\prime}}}

\newcommand{\units}[1]  {\ensuremath{\mathrm{{#1}}}}

\begin{document}
\chapter[Massive black holes in galactic nuclei: observations]{Massive black holes in galactic nuclei: Observations\label{chapt:mbh_obs}}

\author[M.\ Vestergaard and K.\ G{\"u}ltekin]{Marianne Vestergaard and Kayhan G{\"u}ltekin
}

\address{The Niels Bohr Institute, University of Copenhagenn, \\
Jagtvej 155, 2200 Copenhagen N, Denmark, \\
mvester@nbi.ku.dk
}
\address{Department of Astronomy, University of Michigan,\\
1085 S.\ University Ave., Ann Arbor, MI, USA\\
kayhan@umich.edu
}

\begin{abstract}
Since a black hole does not emit light from its interior, nor does it have a surface on which light from nearby sources can be reflected, observational study of black hole physics requires observing the gravitational impact of the black hole on its surroundings. A massive black hole  leaves a dynamical imprint on stars and gas close by. Gas in the immediate  vicinity  of an accreting massive black hole can, due to the presence of the black hole, shine so brightly that it outshines the light of the billions of stars in its host  galaxy and be detected  across the Universe. By observing the emission from stars and gas and determining their kinematics scientists can extract vital information not only on the fundamental properties of  the black  holes themselves but also the impact they have on their surroundings. As it  turns out, supermassive  black holes  appear to play a vital role in shaping  the Universe as we know it, as they can profoundly impact the star formation history in galaxies. As a consequence, these black holes indirectly impact the cosmic build up of chemical elements heavier than Helium and thus affect when and where life can form. For these reasons alone, observations of massive  black holes constitute a very active research area of modern astrophysics.

In this chapter we aim to provide a general overview---fit for a non-expert---of what scientists have learned,  and hope to learn, from analyzing observations of massive  black holes and the material around them. We deliberately do not provide a review of the vast literature on this topic but refer to relevant sample journal articles or reviews, when available, for readers interested in exploring the topics in greater  detail.

\end{abstract}

\medskip{} \noindent
This chapter is the pre-print of the version currently in production. Please cite this chapter as the following: M.  Vestergaard and K.  G\"ultekin. “Massive black holes in galactic nuclei: Observations,” in The Encyclopedia of Cosmology (Set 2): Black Holes, edited by Z. Haiman (World Scientific, New Jersey, 2023)

\medskip{} \noindent
The Chapter contents are as follows: After an introduction in \S~1, we outline the observational evidence for accretion onto massive black holes (\S~2). The next sections cover the measurements of the black hole fundamental properties of mass (\S~3) and spin (\S~4) using various techniques. 
Section 5 is devoted to a characterization of the black hole environment; here we discuss observations of the central black holes in the Milky Way, Sagitarius A$^{\ast}$, and in the nearby massive elliptical galaxy, Messier 87 (M87). 
The sections that follow address the co-evolution of massive black holes and their host galaxies (\S~6); black hole fueling at low redshift (\S~7); observational evidence for black hole feedback (\S~8); observations of massive black holes in the high redshift universe (\S~9) and black hole growth over cosmic time (\S~10).
Intermediate mass black holes (\S~11) and massive black hole pairs (\S~12) are also covered. Cosmic distance measurements using massive black holes are described in \S~13. The Chapter is concluded with a view to the immediate future (\S~14).


\body

\section{Where do we see massive black holes in the Universe? \label{sec:intro}}

Massive ($10^4 
\lesssim \mbh/\msun \lesssim 10^6$) and supermassive black holes ($\mbh \gtrsim 10^6$ \msun) reside in the centers of galaxies. The nearby dwarf galaxy RGG\,118 hosts the least massive black hole known  to us, with a secure measurement\citep{2015ApJ...809L..14Baldassare} of $\sim$50,000 \msun{}, while the most massive ones power the luminous quasars in distant young galaxies. 
Our own galaxy, the Milky Way, is host to a 4 million solar mass black hole (\S~\ref{sec:MWBH}). When the black hole accretes gas at a  sufficiently high rate, it is commonly known as an active galactic nucleus, AGN, because copious amounts of energy are produced from the tiny volume around the black hole, commonly known as the AGN (or black hole) central engine. Since an accreting black hole gains mass, it has long been a working hypothesis that the active phase of a  black hole represents an early evolutionary phase of the galaxy hosting the black hole\cite{1989ApJ...347...29S}\,---\,consistent with AGN and quasars preferentially residing at large cosmic distances where the Universe was much younger than it is today.  At later times the massive black hole stops growing and turns quiescent.

\begin{figure}
       \centering
       \includegraphics[width=0.95\textwidth]{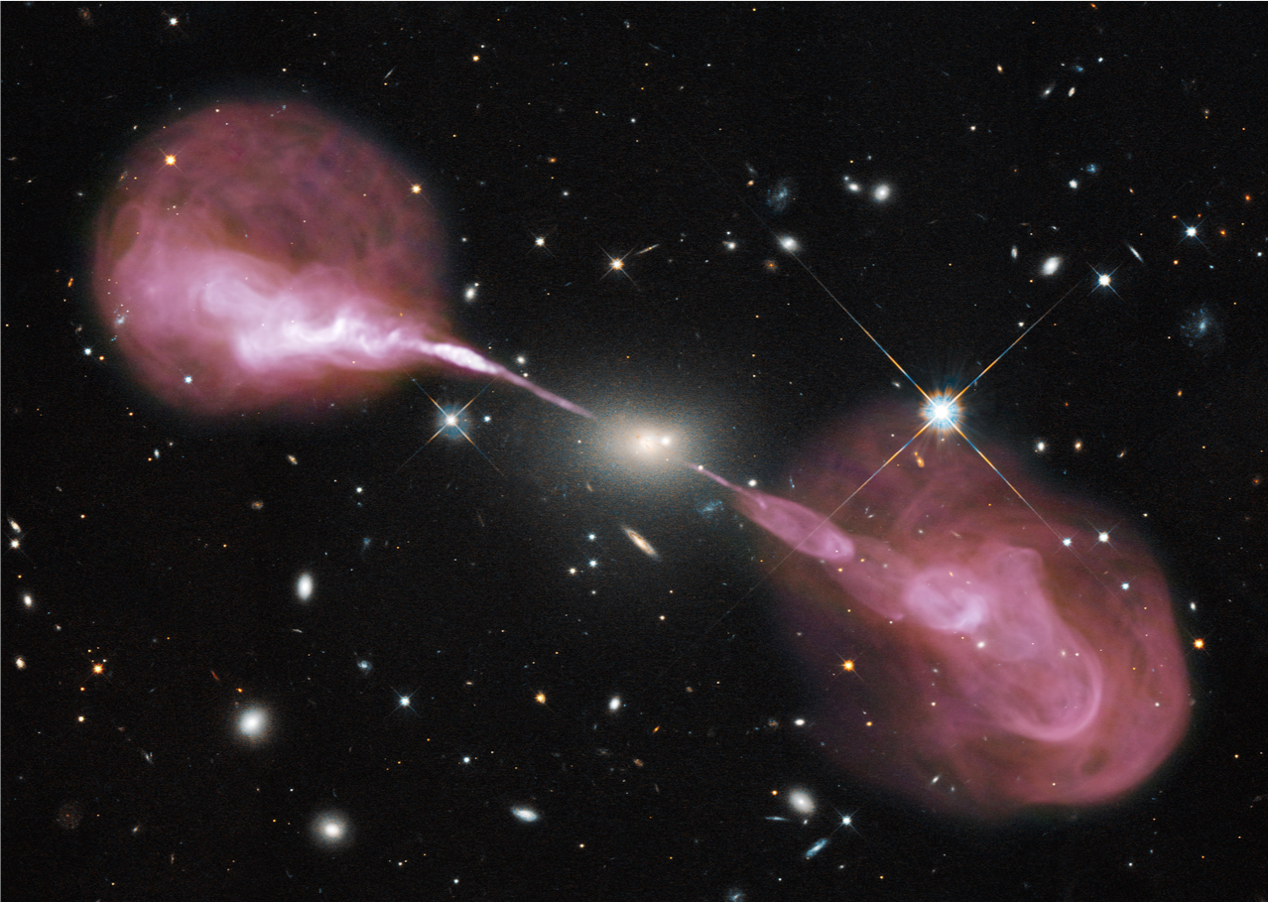}
       \caption{Black holes are nature's most powerful particle accelerators. This composite view of the Hercules A radio galaxy shows synchrotron radio emission (in red) from accelerated charged particles, spiralling along magnetic field lines emanating from regions near the black hole. The incredible power of the central black hole is capable of ejecting the particles to distances several times  the size of the host galaxy itself, shown in the center  (in white). The starlight (white/yellow) is observed with the {\it Hubble Space Telescope} and the radio emission  (red) is obtained with the Jansky Very Large Array in New Mexico. Credit: NASA, ESA, S. Baum and C. O'Dea (RIT), R. Perley and W. Cotton (NRAO/AUI/NSF), and the Hubble Heritage Team (STScI/AURA)}
       \label{fig:HerculesA}
   \end{figure}

While indisputable evidence for 
supermassive black holes has only been obtained within the past few years through the detection of the shadow of the central supermassive black hole of, first, the nearby elliptical galaxy Messier 87
(\S~\ref{sec:characterizing-m87eht}) and, then, the Milky Way (Sagitarius A$^{\ast}$), astronomers have long suspected their existence. The emergence of radio astronomy in the 1930s, paved the way for discovering, first, radio galaxies in the 1950s (e.g., Cygnus A\cite{1953Natur.172..996J}) and later quasars in the early 1960s (e.g., 3C\,273\cite{1963Natur.197.1037H, 1963Natur.197.1040S}). 
Quasars were first detected as powerful point-like or compact, single-jetted radio sources\cite{2019ARA&A..57..467B}. While optical spectroscopic data indicated the presence of a very compact, redshifted source\citep{1963Natur.197.1040S}
coinciding with the radio source, it took another $\sim$20 years to securely connect the quasar  phenomenon to the centers of galaxies\cite{1978ApJ...223..747Stockton,1982Natur.296..397B}. 
Similar to Cygnus A, many galaxies with powerful radio jets reaching far beyond the extent of the stellar body of the galaxy were revealed. An example is Hercules A, shown in Figure~\ref{fig:HerculesA}. The immense power required to eject the  relativistic, radio-emitting particles to kilo-parsec scales and the small volume from where the jet emanates from (confirmed by the short variability time scales at X-ray and Gamma-ray energies\cite{2019ARA&A..57..467B}%
) suggests a powerful, compact source. Soon after the discovery of radio-jetted  objects it became apparent that the most likely source is an accreting massive black hole\citep{1969Natur.223..690L}.%

Using techniques we detail in \S\ref{sec:mass}, nearby galaxies without quasars were examined for their so-called `relics': the massive, quiescent black holes at the galaxy centers.  The search was possible only for the nearest galaxies with the angular resolution available on ground-based telescopes.  Many galaxies were  found to host central black holes.  With the angular resolution of \emph{Hubble Space Telescope}, the volume of space that could be tested grew dramatically.   Surprisingly (at the time), it turned out that nearly every galaxy at or above the mass of the Milky Way galaxy has a central black hole\cite{1998Natur.395A..14R, 1995ARA&A..33..581K}.

\section{Observational evidence for black hole accretion \label{sec:BHevidence}}
Supermassive black holes in centers of galaxies not only grow in mass but their activity is also fuelled when gas in the galaxy is transported to the center and onto the black hole. In the following we provide a brief and general description of how this is believed to occur.
As material is driven from large galactic scales into the very center of the galaxy, it first settles into a dense, thick planar structure -- often referred to as the `(obscuring) torus' -- before it moves into a central thin accretion disk. Both structures orbit the black hole. The `torus' is believed to consist of dense molecular gas and dust that obscures our direct view of the nuclear emission at optical, UV and soft X-ray wavelengths\citet{1995PASP..107..803U}.
The infalling material loses a lot of its angular momentum and rotation energy, as required if it is to  fall onto the black hole, eventually. The original rotation energy can be removed in one of two ways. A large part is dissipated as heat, radiating into space. The heat is generated when the infalling gas collides with the disk material. In addition, the disk gas undergoes differential rotation and is thereby heated through viscous processes: as the radial distance from the black hole decreases, the disk annuli orbit increasingly faster.  In addition,  angular momentum and rotation energy are removed when material is ejected from the central region, either as winds of plasma or ionized gas\cite{1991ApJ...373...23W, 2008MNRAS.386.1426K, 2015Natur.519..436T, 2020ApJS..247...37A, 2020MNRAS.495..305X} 
or as outflows of radio and X-ray emitting plasma\cite{2019ARA&A..57..467B}. 
%
Jets of molecular gas have even been observed emanating from the center.\cite{2013ApJ...776...27V, 2020A&A...640A.104A} %
The powerful radio jets,  observed in a subset of AGN, extract rotation energy from the black hole itself via the Blandford-Znajek mechanism\cite{2019ARA&A..57..467B}. 

The accretion disk  does not extend all the way  to the black hole but has an inner radius, the innermost stable circular orbit. Inside this radius the material cannot maintain a stable orbit and it falls directly onto the black hole\cite{1985apa..book.....F}, being heated to a plasma of electrons and protons in the process.
The temperature in the disk increases toward the center and the inner parts of the accretion  disk are so hot that they emit far-ultraviolet (FUV) photons. 
Some of these photons are detected directly in the observed spectra of AGN as FUV continuum emission and some travel to ionize gas at large radial distances, known as the broad-line region -- as this region emits the  broad emission lines, characteristic of AGN, with velocity widths of several thousand \kms{}. Some of the FUV disk photons scatter off fast-moving electrons, located in the central region, that transfer much of their energy to the photons. In this process, called `inverse Compton scattering', the photons are `kicked' to much higher (keV) energies. We observe these photons in the X-ray wavelength regime\cite{1993ApJ...413..507H}.








So in short, our observational evidence for black hole accretion is (a) the  detection of FUV, UV and optical continuum emission from the accretion disk (direct evidence); and (b) indirect evidence in the form of (i) X-ray emission from the central corona; and (ii) line emission at UV, optical and infrared wavelengths from the broad emission line region. The X-ray emission is often considered direct evidence because it is directly powered by the ionizing photons from the inner accretion disk and the emission signature (the power and the hard spectrum) of the X-ray component is unique. While the broad emission lines are also powered by the FUV ionizing photons, their emission requires the additional presence of broad-line gas.

In  the following sections we describe in more detail some of the key observational properties  of galaxies with accreting supermassive black holes.

\subsection{Emission is linked to the mass accretion rate \label{sec:L-mdot}
}

The luminosity $L$ of an AGN is directly linked to the mass accretion rate $\dot{M}$ of the central black hole: $L = \eta \dot{M} c^2$, where $\eta$ is the energy conversion efficiency and $c$ is the light speed. Black holes are eminent at converting mass to energy with $\eta$ in the range of 10\% -- 40\%, and even higher if the black hole is spinning\cite{1974ApJ...191..507T}. The integrated light over cosmic history (known as  the So\l tan argument; \S~\ref{sec:cosmic-growth}) implies an average value of $\eta \approx$10\% 
with values up to $\eta \approx$ 15\% reported\cite{1982MNRAS.200..115S, 2002MNRAS.335..965Y, 2002ApJ...565L..75E}. 

\begin{figure}[h]
    \centering
    \includegraphics[width=0.45\textwidth]{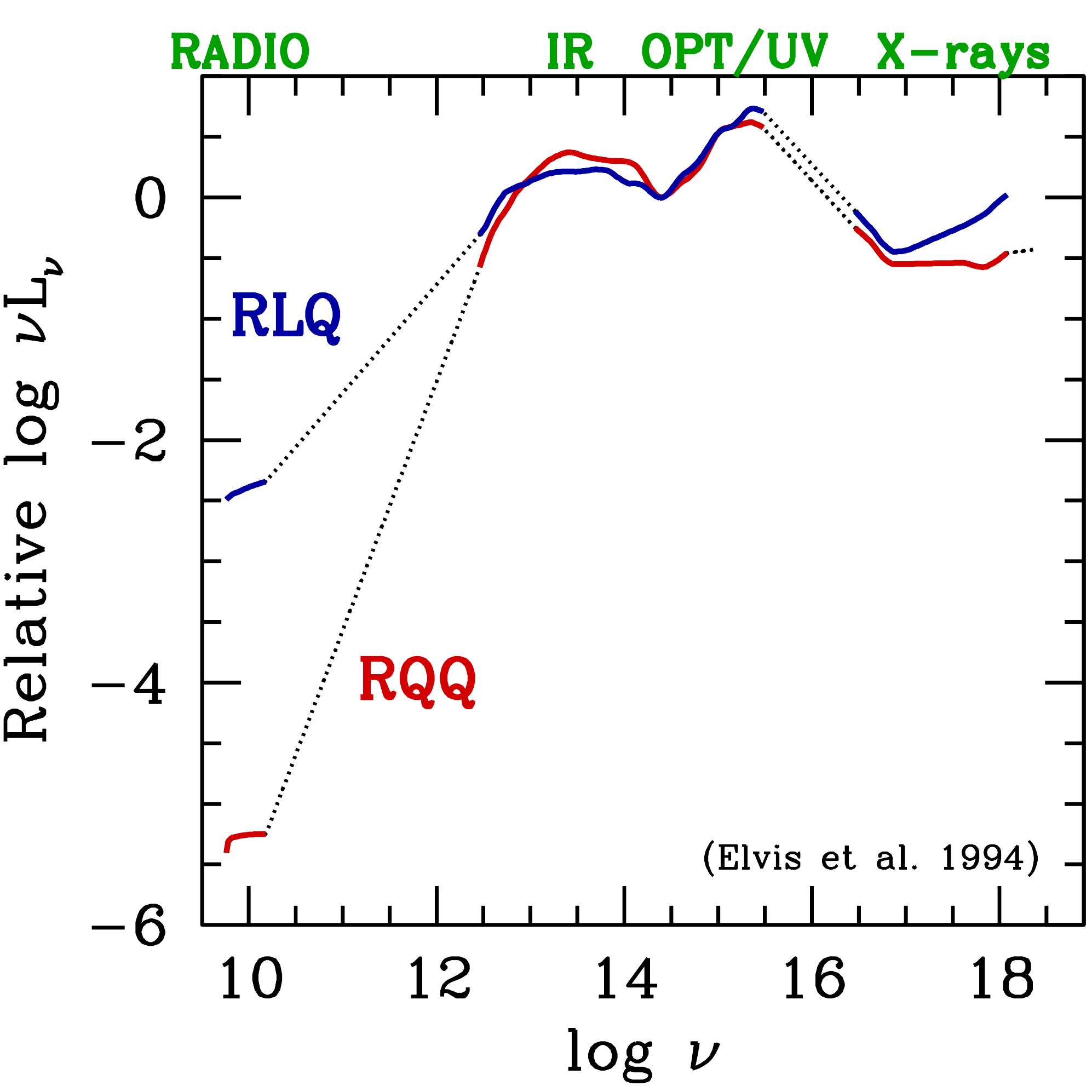}
    \vskip -0.2cm
    \caption{Mean radio to X-ray spectral energy distribution of powerful AGN. Radio-quiet AGN (marked RQQ, red curves) are remarkably similar to radio-loud AGN (marked RLQ),  except at radio wavelengths. The minimum observed  between the bumps at infrared and optical wavelengths is the 1\,$\mu$m inflection point; here, the dust sublimates due to the high temperatures. Dotted curves indicate missing data due to Earth's atmosphere and atomic Hydrogen in the Milky Way. Figure by authors based on the catalog of Elvis et al.\citep{1994ApJS...95....1Elvis}}
    \label{fig:elvisSED}
\end{figure}

The emission from AGN spans the entire electromagnetic spectrum from radio to gamma ray emission.
The typical observed spectral energy distribution (SED) of AGN is shown in Figure~\ref{fig:elvisSED}. While these observations were made of X-ray bright AGNs (so to be well detected in X-rays) they hold true to first order for all relatively bright AGN\citep{1994ApJS...95....1Elvis,2006ApJS..166..470Richards}. The SED shows four main components in the continuum emission: (a) an X-ray power-law; (b) an optical-UV bump dominated by emission from the central accretion disk that feeds the black hole; (c) an infrared  bump of emission from dust at a range of temperatures (here dust from the host galaxy may also contribute); (d) radio emission. Some ($\sim$\, 5\% -- 20\%,  depending on source redshift\citep{2007ApJ...656..680Jiang}) display  powerful non-thermal emission from radio jets while most emit only weakly at radio wavelengths. The nature (thermal versus non-thermal, stellar versus non-stellar  origin) of this radio emission is still debated.\citep{2019NatAs...3..387Panessa}. 

AGNs that deviate from this general picture are LINERs (Low-Ionization Narrow Emission-line Regions) and blazars (\S~\ref{sec:AGNtypes}). LINERs display an extremely weak UV continuum relative to  the amplitude of the non-thermal X-ray power-law component, indicating that the mass accretion rate is extremely low\citep{1996ApJ...462..183Ho}. The observed emission of blazars, highly polarized  and highly variable, is dominated by relativistically Doppler-boosted  and -beamed emission as we view  these objects  within a few degrees of the jet  axis. An overview of the physics of blazars and the vast literature on this subject can be obtained via a recent review of Madejski \& Sikora\citep{2016ARA&A..54..725Madejski}.

\begin{figure}[ht]
    \centering
    \includegraphics[width=0.95\textwidth]{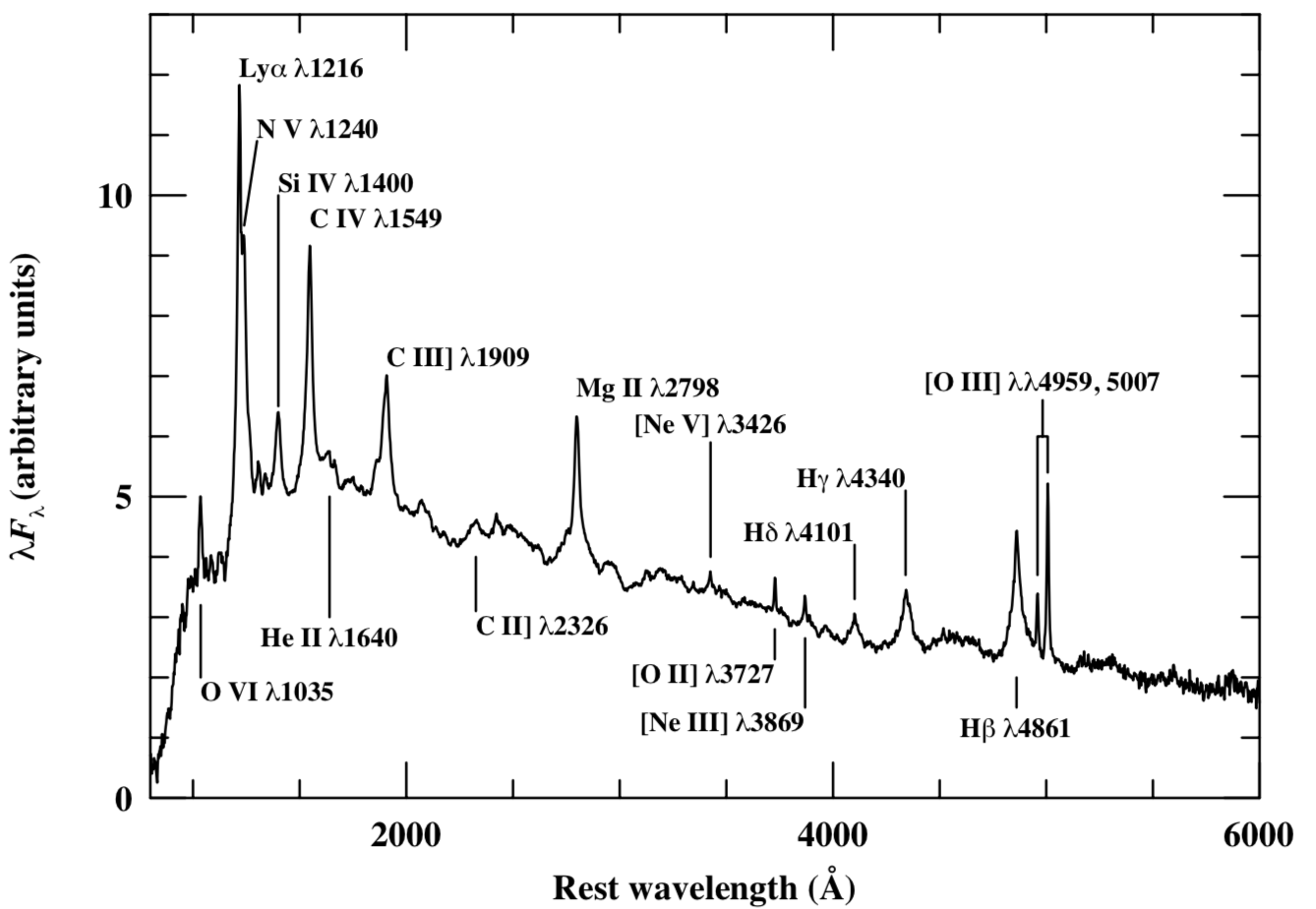}
    \vskip -0.2cm
    \caption{Composite spectrum of high luminosity quasars, covering the ultraviolet and optical region. The major broad and narrow emission lines are marked. Examples of narrow emission lines (widths less than 1000\,km\,s$^{-1}$) are \oiii{}, \oii{}, \neiii{}, and \nev . The spectrum is created from 718 individual spectra obtained as part of the Large Bright Quasar Survey\citep{1991ApJ...373..465Francis}. Figure  from Peterson\citep{1997iagn.book.....Peterson}, based on data provided by P.J. Francis and C.B. Foltz.
    }
    \label{fig:francis}
\end{figure}

In addition to the power-law continuum emission from the central accretion disk, the UV-optical spectra of AGN and quasars are littered with broad and narrow emission lines of varying strength (Figure~\ref{fig:francis}). The broad emission lines with widths of several thousand km s$^{-1}$ originate in the broad-line region, a dense gas region located somewhere inside the obscuring torus.  Narrow lines with widths less than 1000 km s$^{-1}$ are emitted from a diffuse and extended gas region located beyond the torus on kilo parsec scales. 
To first order, the UV-optical spectra appear somewhat similar, even when compared across redshifts of $z =$ 0.5 to $z = $ 4.5 for a given luminosity\citet{2002ApJ...581..912Dietrich}. 
However, there are also object-to-object variations. A few dominant trends and correlations have been identified. The decreasing equivalent width of the broad \civ{} $\lambda$1549 emission line with increasing AGN luminosity is known as the {\it Baldwin Effect}\citet{1977ApJ...214..679B,2002ApJ...581..912Dietrich}. 
In early 1990s, Boroson \& Green\citet{1992ApJS...80..109BorosonGreen} identified the strongest spectral line correlation known (dubbed `Eigenvector 1') that links the strength of the \oiii{} $\lambda$5007 emission line, the strength of the optical \feii{} emission complex and the broad \hb{} emission line asymmetry. The second strongest correlation connects the  strength of the optical luminosity and the \heii{} $\lambda$ 4686 line emission. These two correlations are now understood to be driven by the Eddington luminosity ratio $L_{\rm Bol}/L_{\rm Edd}$ of the central source emission and the mass accretion rate\citet{2002ApJ...565...78Boroson}.
Subsequent studies (see e.g., ref.~\citet{2000ApJ...536L...5S, 2015FrASS...2....6Sulentic} and references therein) have added observed properties to the original findings. These correlations are sometimes referred to as the `Fundamental Plane of the Broad-line Region'. 

%


\subsection{Observed X-ray properties}
Since the launch in 1970 of $Uhuru$, the first Earth-orbiting X-ray telescope\footnote{https://heasarc.gsfc.nasa.gov/docs/uhuru/uhuru.html}, scientists have worked hard to characterize the observed X-ray spectrum of AGN and  to understand the underlying physics. The spectrum is a superposition of the emission and absorption of several physical components along our line of sight,  each of which vary on their own time scales.  Owing to this complexity we still do not have a detailed understanding  of the observed  X-ray properties, yet a general picture has emerged. The  0.3 $-$ 80\,keV emission spectrum consists of (i) a power-law continuum with a slope $\Gamma \sim 2$ from the X-ray corona; (ii) soft X-ray emission at $E \lesssim 1$\,keV, likely thermal in nature, rising steeply above an extrapolation of the power-law continuum at higher energies; 
(c) a reflection component rising  above the power-law continuum at $E > 10$\,keV; (d) X-ray emission lines of which iron lines -- in particular Fe K$\alpha$ (at $E  \sim$ 5--7 keV) -- are prominent. 
As noted earlier, it is commonly thought that extreme-UV photons from the accretion disk scatter off (possibly relativistic) electrons in a corona  surrounding the disk, although a thermal origin of the X-ray emission is also a possibility\citep{1997iagn.book.....Peterson}. 
The X-ray corona is believed to be the  origin of the primary X-ray power-law continuum that  will irradiate optically thick material in its vicinity,  thereby creating spectral signatures of X-ray reflection. Reflection from distant, cold and low-velocity material, such as the purported dense, obscuring, rotating torus will mainly reveal itself as a narrow iron fluorescence line. Reflection from the ionized inner accretion disk is expected to produce a strongly broadened and redshifted Fe K$\alpha$ emission line and additional ('excess') emission at soft ($E < $1 keV) and hard ($E > $10 keV) energies.  
The X-ray continuum is expected to have a high-energy roll-over at a few hundred keV.
If the emitted spectrum is viewed through disk winds launched off  the central accretion disk, the observed X-ray spectrum will also have an absorption spectrum superimposed (with numerous absorption lines and bound-free edges). Absorbing material with intermediate temperatures, also known as {\it warm absorbers}, may also be connected to such disk winds.  

The literature is vast on this topic and many models of the X-ray emission exist. A good overview can be obtained by reviews\citep{2016AN....337..404Reynolds, 2007ARA&A..45..441Miller, 1993ARA&A..31..717Mushotzky} and textbooks on the topic\citep{1997iagn.book.....Peterson,1999qagn.book.....Kembhavi}.


\subsection{Variability of black hole activity}
Gas accretion onto black holes does not occur through a constant, smooth mass flow. As a result, the flux of ionizing photons generated by the central disk changes over time. It can vary from a slow trickle to a gushing flow where even the photons are 'flushed' into the black hole (`adiabatic accretion'\citep{2014ARA&A..52..529Y}%
).  
Although random and unpredictable, the flux does largely tend to vary by a certain margin around a slowly varying mean luminosity. This is often described as a `Damped Random Walk'\cite{2011ApJ...735...80Z, 2015ApJ...811...42S}. The timescale over which the ionizing continuum varies depends on the photon energy---the X-ray flux varies on very short timescales compared to the far-UV flux---and the target source. It appears that supermassive black holes of relatively lower mass, $\mbh \lesssim 10^8$ \msun, that power Seyfert galaxies, vary on shorter timescales and with larger amplitudes than the more massive black holes\cite{2001MNRAS.324..653L, 2004ApJ...601..692V, 2007ApJ...659..997K}. 
The (intrinsically) longer time scales and lower variability amplitudes of quasars is considered a manifestation of the larger physical scales in these systems connected with the larger black hole masses. 



The variable nature of the accretion process has been utilized over the past few decades to map the physical gas structure in the vicinity of the black hole  for a sample of nearby AGN. Starting in the mid to late 1980s scientists formed large scale collaborations to execute monitoring campaigns simultaneously on ground-based telescopes and the space-borne  telescope,  the {\it International Ultraviolet Explorer}  ({\it IUE}\cite{1991ApJ...366...64C, 1994ApJ...425..582R, 1994ApJ...425..609S, 1991ApJ...368..119P}) for their studies. Because the different gas components, located physically away from the inner disk, reprocess the ionizing continuum into emission at other wavelengths, scientists are able to study how each of the X-ray corona and the broad-line region respond to the continuum variations and on what timescale. This form of light echo mapping (a.k.a reverberation mapping; \S~\ref{sec:RM}) can be used to establish the structure of the gas components in the central engine.  The  recent review by Cackett, Bentz \&  Kara\cite{2021iSci...24j2557C} provide a good overview of what is currently possible with this technique.

\subsection{Emission-line diagnostics reveal the growing black hole \label{sec:bpt}}
Emission line diagnostic diagrams are often used to distinguish between galaxies with star forming regions from galaxies with central  AGN activity. Gas will emit emission lines if energy -- in the form of heat (e.g., increased pressure due to shocks propagating into the gas) or  ionizing radiation -- is injected into the gas. 
The emergent line emission will depend on the gas composition, density and temperature and the shape of the spectral energy distribution of the radiation incident on the gas.
High energy photons from high mass stars can photoionize nearby gas, known as ionization nebulae or \hii{} regions, that in turn emit emission lines. The ionizing radiation from the central engine of an AGN is much more energetic than that produced by stars and is therefore able to ionize and excite gas to much higher energy levels. As a result, AGNs can produce emission from multiply ionized species such as \neviii{}, \carboniii{}, \ovi{}, 
and \nv{}$\lambda 1240$, and the relative strength of the emission lines are also different than those emitted from gas excited by stars. This result of photoionization physics\cite{1989agna.book.....O, 2006agna.book.....O} is used to distinguish star-forming galaxies from AGN or to identify the ionizing source of line emitting gas within individual galaxies that may have both star formation and AGN activity\citep{2014MNRAS.444.3961D, 2019MNRAS.486..123R}.





\begin{figure}
    \centering
    \includegraphics[width=0.62\textwidth]{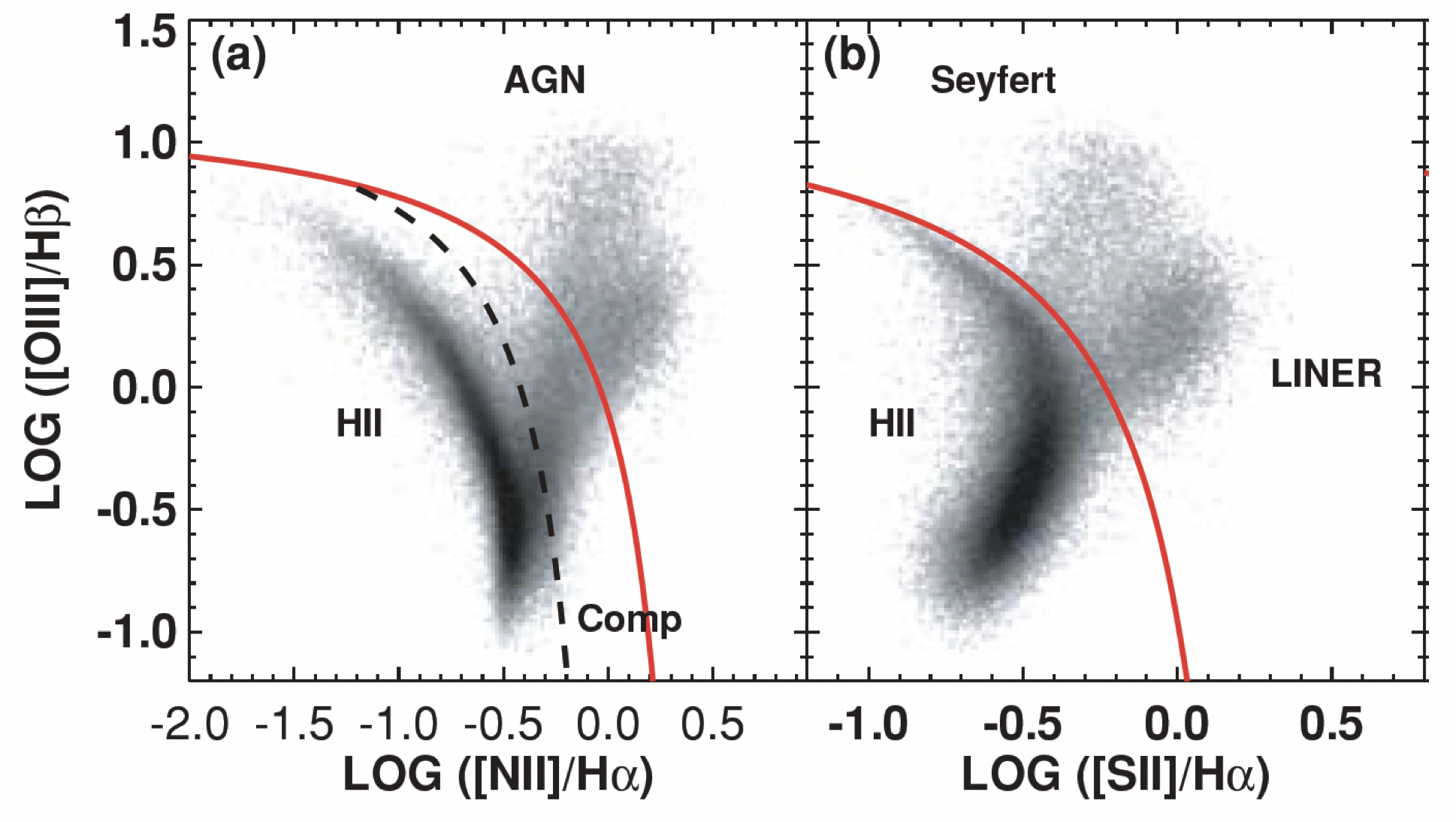}
    \includegraphics[width=0.35\textwidth]{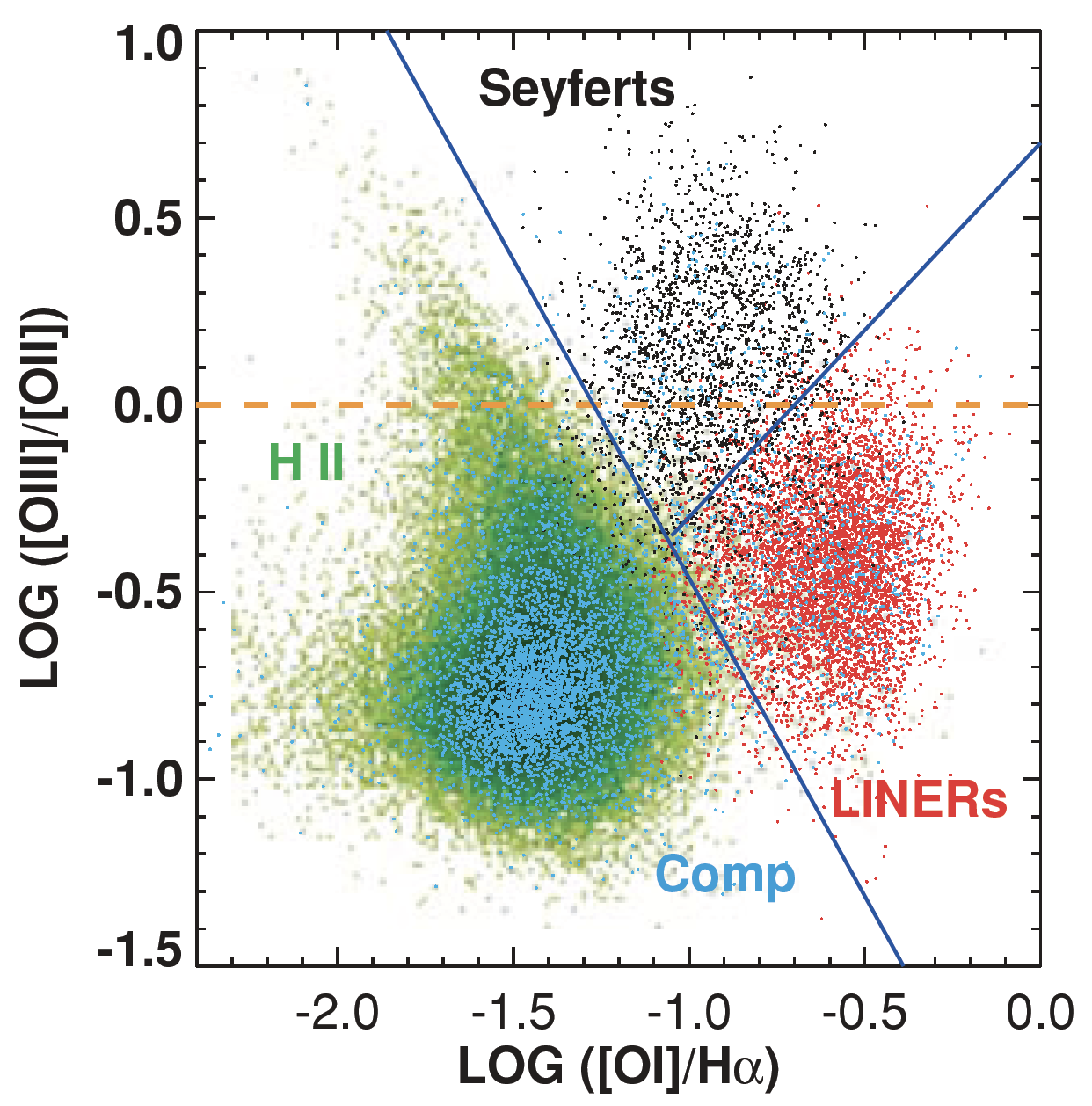}
    \caption{Emission line diagnostic diagrams for Sloan Digital Sky Survey galaxies\cite{2006MNRAS.372..961K}. 
    The solid and dashed curves indicate the division between source  classes  (labelled). Adapted from Figure 1 (two left-most panels) and Figure 5 (right panel) of Kewley et al.\cite{2006MNRAS.372..961K}.}
    \label{fig:bpt}
\end{figure}

A key diagnostic is the line ratio  pair of \oiii{} $\lambda$5007/\hb{} and [N\,{\sc ii}] $\lambda$ 6584/\ha{} (Figure~\ref{fig:bpt}a,b)
where the two different ionizing sources (stars and AGN)  are well separated\cite{1981PASP...93....5B, 1987ApJS...63..295V}. However, other line ratios (e.g. [S\,{\sc ii}] $\lambda\lambda$ 6717,6731/\ha{}, [O{\sc i}] $\lambda$6300/\ha{}, \oiii/[O{\sc ii}] $\lambda$3727, are also useful for separating galaxy types\cite{2006MNRAS.372..961K}  such as Seyferts and  LINERs.  Sample diagrams are shown in Figures~\ref{fig:bpt}b,c. 
Because some galaxies contain both a central, weak AGN and have significant star formation, there is not a sharp division between galaxies classified as AGN and those classified as star forming ({\it H\,{\small II} regions}) in this diagram. 
It is not known if such composite galaxies are in a transition phase from an active black hole phase (AGN) to a quiescent one (or visa-versa), or if they are galaxies that spend their entire lives with a mixture of star formation and nuclear activity. Since a black hole will switch to an active phase whenever it has gas within its gravitational reach, it is quite likely that the class of composite galaxies (labelled `comp' in Figure~\ref{fig:bpt}) is a mixture of all three categories (Seyfert, LINER, H\,{\sc ii} region).


\subsection{AGN types  \label{sec:AGNtypes}}

The first AGN were discovered before anyone knew what they were. In 1909 Fath \citep{1909LicOB...5...71Fath} detected strong emission lines in spectroscopic observations of the nuclei of two nearby galaxies, NGC\,1068 and Messier\,81.  This was at a time when scientists had yet to discover the expansion of the Universe.  Over time, scientists discovered a range of new objects with unusual properties -- including spiral and elliptical galaxies with nuclear line and continuum emission that was not understood.

Carl Seyfert\cite{1943ApJ....97...28S}  presented observations of nearby spiral galaxies with a bright central source emitting strong and extremely broad emission lines of widths $\sim$8000\,\kms.  Such galaxies are now known as Seyfert galaxies. With the emergence of radio astronomy,  many new objects were discovered, including radio galaxies: nearby elliptical galaxies with powerful radio emission from their centers and  large-scale radio jets emanating from the galaxy center (Figure~\ref{fig:HerculesA}).  The radio galaxies that also emit narrow emission lines were called narrow-line radio galaxies, NLRG, and those detected in X-rays were classified as narrow-line X-ray galaxies, NLXG.  Large radio surveys, such as the  canonical 3C radio survey\,---\,the Third Cambridge Catalogue of Radio Sources\cite{1962MmRAS..68..163B, 1983MNRAS.204..151L}---\,turned up a number of radio sources with point-like optical counterparts, that were then classified as  quasi-stellar radio sources — QSRS, now known as quasars\cite{1964PhT....17e..21C}.   Optically luminous, stellar-like sources with very blue spectral slopes and broad emission lines were in subsequent years found in large numbers in optical surveys (e.g., the Palomar-Green Bright Quasar Survey\citep{1986ApJS...61..305G}; Durham-AAT\cite{1990MNRAS.243....1B}; The Large Bright Quasar Survey\citep{1995AJ....109.1498H}; the 2dF survey\cite{1990MNRAS.243....1B}). While these objects appeared to have similar properties to the QSRS,  they had no radio counterpart, and were then classified ‘quasi-stellar objects’ (QSOs).  In recent years, the term quasar is used for both the radio-loud and the radio-quiet variety.
Other types of peculiar sources were since discovered, such as BL Lac objects and optically violent variables, OVVs\,---\,collectively known as blazars. These are objects with high-amplitude variability on very short time scales (of order a day) and the emission is highly polarized (up to a few percent). Furthermore, BL Lacs do not have emission lines in their optical-UV spectra. 
The AGN taxonomy can be confusing, but the many AGN types are the result of scientists not knowing the underlying physics.  Yet, it was quite clear that the bright nuclear emission with a rather hard (i.e., blue) spectrum is not of stellar origin (\S~\ref{sec:bpt}).









   
It appears that all massive galaxies  contain a central massive black hole. When gas is transported to the center and onto the black hole, the nucleus turn into an AGN. So all the different AGN types are to first order the same type of object, namely a black hole powered source. The object properties that we see depend primarily on our viewing angle, the mass of the central black hole and its available fuel. The angular dependence is described by the Unified Model\cite{1993ARA&A..31..473A, 1995PASP..107..803U}: Depending on our viewing angle, we may see the central engine directly (a type 1 AGN with broad and narrow emission lines) or our line of sight to the ionizing source goes through the obscuring torus, hiding the broad lines (a type 2 AGN with only narrow emission lines) and absorbing the central optical-UV continuum and the soft X-ray emission.  If we view a radio source directly down the jet (to within a few degrees) we will see a blazar\cite{1997iagn.book.....Peterson}.  
The black hole mass and its fuel supply determine the luminosity of the AGN. The more massive black hole can accrete more gas and thus sustain a higher luminosity, which can eject or ionize gas in a larger central volume of gas — the receding torus scenario — providing a clear view of the central engine  over a wider range of viewing angles\cite{2000ApJ...545...63E}.  The power of the AGN is what sets Seyfert galaxies apart from quasars, the most luminous subset of AGN. While all AGN have relatively similar spectral energy distributions (barring blazars; \S~\ref{sec:L-mdot}), they display different luminosities, owing predominantly to the mass accretion rate.

The recent discovery of especially optical ‘changing-look AGN’\cite{2014ApJ...788...48S, 2014ApJ...796..134D, 2015ApJ...800..144L} has emphasized the additional need for a time or evolutionary phase dependence in the Unification model\cite{2014MNRAS.438.3340E}. For these sources, changes in the accretion state or in the properties of the ionizing source occur on timescales between a few years to a few decades\cite{2014ApJ...796..134D, 2015ApJ...800..144L, 2016MNRAS.455.1691R, 2016A&A...593L...8M}, presumably due to changes in fuel supply or disk instabilities\cite{2014MNRAS.438.3340E, 2018MNRAS.480.3898N, 2020ApJ...900...25J}.  In some cases within a matter of a few years, the AGN turns on (NGC\,2617\cite{2014ApJ...788...48S}) or off (Mrk\,590\cite{2014ApJ...796..134D, 2016MNRAS.455.2745K}), or even off and on again (NGC 4151\cite{2014ApJ...796..134D}; Mrk\,590\cite{2019MNRAS.486..123R}; Mrk\,1018\cite{2016MNRAS.457..389M}). Although changing-look AGN are now found in large numbers\cite{2016MNRAS.457..389M} we still do not understand the phenomenon. For example,  accretion state changes due to a dwindling fuel supply\cite{2014MNRAS.438.3340E} is expected to take place on timescales of millions of years or more\cite{2016MNRAS.455.2745K}, not a few years.

%










\subsection{Fundamental Plane of Black Hole Accretion}
\label{sec:fpbha}
The fundamental plane\footnote{The fundamental plane described here is different from the 'Fundamental Plane of the Broad-line Region' mentioned in \S~\ref{sec:L-mdot}. } of black hole accretion is a less-well understood relation that ties together bulk energetic emission in radio and X-ray, normalized by the mass of the accreting black hole.
The term `fundamental plane' is used in astronomy to indicate a two-dimensional manifold, embedded in a three-dimensional space.  If any two of the properties are known, then the third can be predicted with the fundamental plane.  

\begin{figure}
    \centering
    \includegraphics[width=0.49\textwidth]{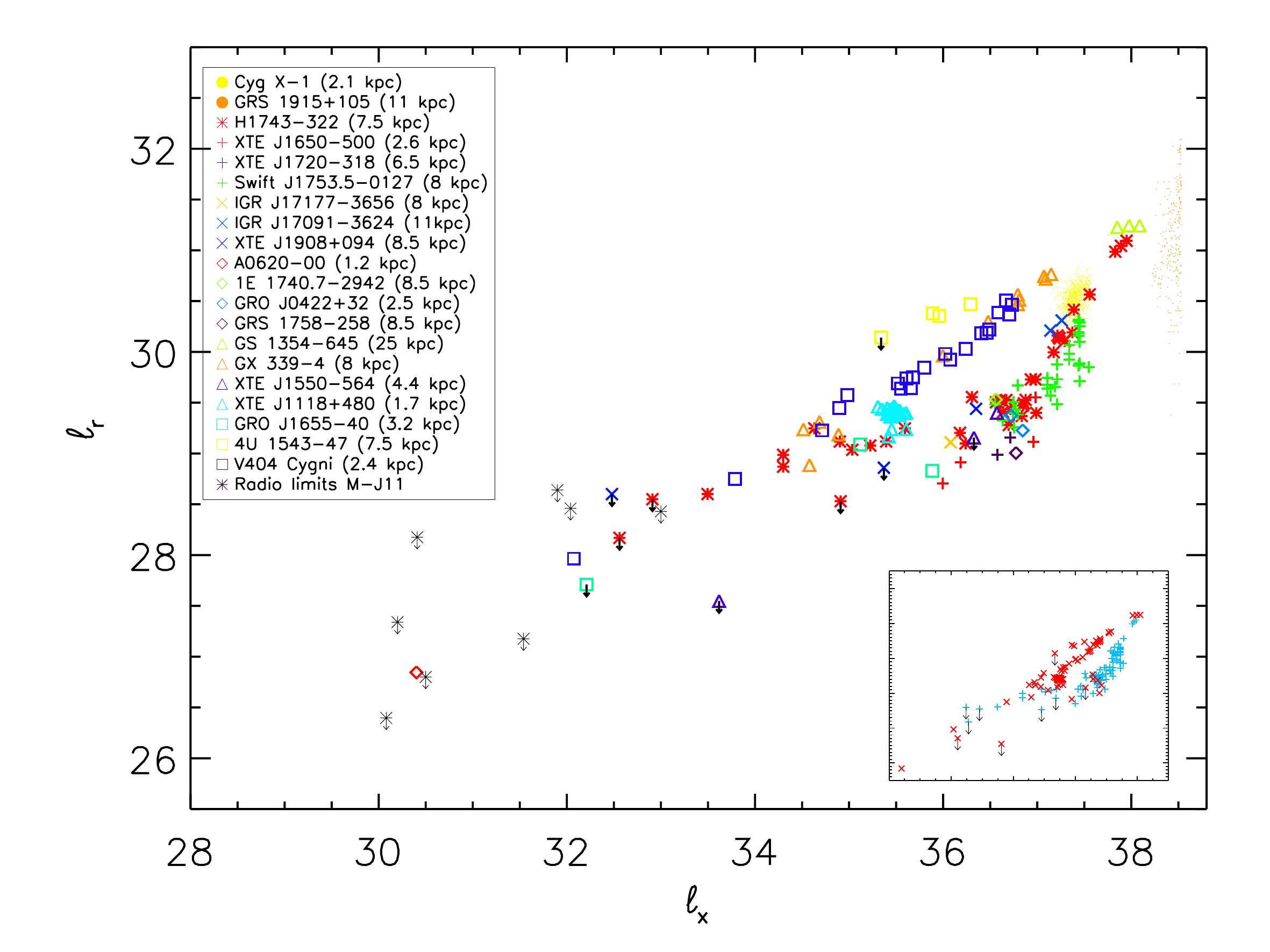}
    \includegraphics[width=0.49\textwidth]{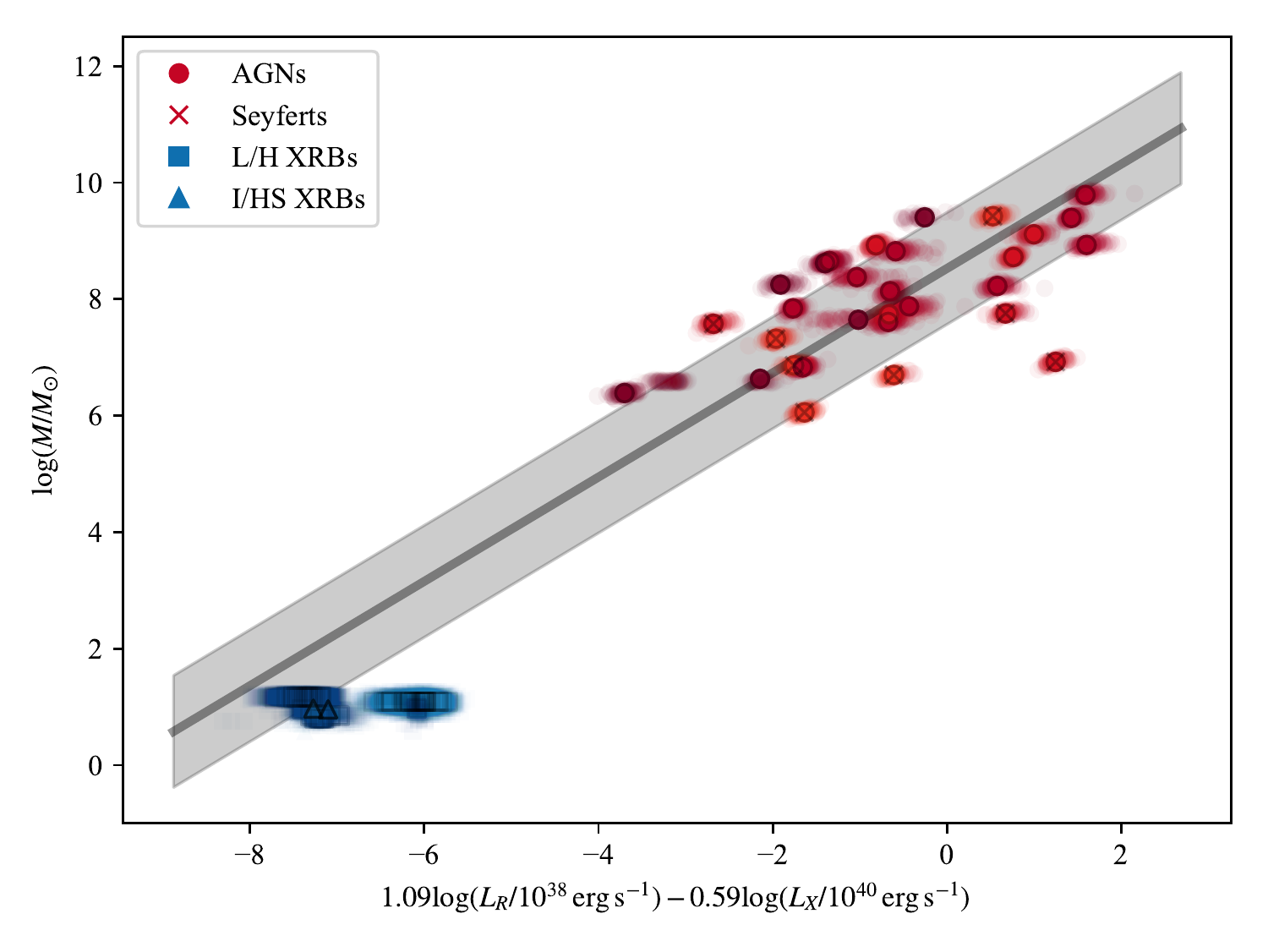}
    \caption{\textit{Left panel}: Radio luminosity as a function of X-ray luminosity for a sample of X-ray binaries.  What was originally thought to be a single, universal relation is showing signs of more complexity with two tracks (indicated by the colors of the points in the inset).  From Gallo et al.\cite{2012MNRAS.423..590G}. \textit{Right panel}: The fundamental plane of black hole accretion, oriented with mass as the dependent variable and viewed edge-on. Each source is represented by its measurement uncertainty distribution (shown as a partially transparent color relevant for its object classification) and a black symbol superimposed at the nominal value of the mentioned probability distribution. The figure shows that a single relation can describe the general properties of black holes spanning many orders of magnitude in mass.  From G{\"u}ltekin et al.\cite{2019ApJ...871...80G}}
    \label{fig:fundplane}
\end{figure}

The three dimensions of the fundamental plane of black hole accretion are logarithmic mass ($\mu \equiv \log{(M/10^{8}\ \msun)}$), logarithmic radio luminosity ($\log{(L_R / 10^{38}\ \units{erg\ s^{-1}})}$), and logarithmic X-ray luminosity ($\log{(L_X / 10^{40}\ \units{erg\ s^{-1}})}$).  The normalizations are chosen to be reasonable for massive black holes, but the relation applies to black holes of all masses.  Typically, the core radio luminosity is calculated as $L_R \equiv \nu L_{\nu}$ at 5 GHz; and the X-ray luminosity is a bandpass luminosity calculated in a hard X-ray band, usually 2--10 keV.  The origins of the fundamental plane of black hole accretion come from studying the relation between $L_R$ and $L_X$ for X-ray binaries\cite{2003MNRAS.344...60G}.  The range of masses in X-ray binaries is small enough to ignore, and for these it was found that $L_R \propto L_X^{0.7}$.  Follow-up work has shown that what appeared as a universal relation may actually be at least two different tracks (Figure \ref{fig:fundplane})\cite{2012MNRAS.423..590G}.  Thus what originally seemed to be a simple, universal relation concerning the regulation of energy emitted in two different bands may be far more complex. 

In order to isolate flux coming from an accreting black hole, relatively high spatial resolution is needed, especially at low accretion rates, 
when low-luminosity AGN (LLAGN) have X-ray luminosities comparable to the brightest X-ray binaries, and in galaxies with high star formation rates, which can produce significant radio emission.

The pioneering studies to examine the fundamental plane of black hole accretion across the mass scale did so with aims to explain the underlying physics.  One explanation is that in sub-Eddington jet-dominated systems, both the radio and X-ray emission originate in the jet\citep{2004A&A...414..895F}.  In this model, black holes that accrete at higher Eddington rates are not expected to follow the same relation and yet there is evidence that Seyferts do follow it\citep{2019ApJ...871...80G}.  Another idea is that X-ray emission depends only on the mass accretion rate and jet properties (and, hence, radio emission) depend on the mass of the black hole, the accretion rate, and the distribution of energy of electrons in the jet \citep{2003MNRAS.345.1057M}.  The underlying physics of the fundamental plane of black hole accretion is still not understood.  Given the large scatter and the wide variety of black holes and accretion modes that can be empirically explained by a three-dimensional fit in logarithmic space, the plane may be a blunt tool for understanding the details of accretion and outflow physics.  

The fundamental plane of black hole accretion can be written as
\begin{equation}
    \log{\left(\frac{M_{\rm BH}}{10^{8}\ \msun}\right)} = \mu_0 + \xi_{M,R} \log{\left(\frac{L_R}{10^{38}\ \units{erg\ s^{-1}}}\right)} + \xi_{M,X} \log{\left(\frac{L_X}{10^{40}\ \units{erg\ s^{-1}}}\right)}.
\label{eq:fp}
\end{equation}
G{\"u}ltekin et al.\citet{2019ApJ...871...80G} used a sample of LLAGN whose black holes had direct, dynamical mass measurements and best-fit parameters of $\mu = 0.55 \pm 0.22$, $\xi_{M,R} = 1.09 \pm 0.10$, $\xi_{M,X} = -0.59^{+0.16}_{-0.15}$.  The measured intrinsic scatter of the relation in the direction of $\log{(M_{\rm BH} / 10^{8}\ \msun)}$ mass is Gaussian in shape with the natural logarithm of the dispersion estimated to be $\ln{\epsilon_M} = -0.04^{+0.14}_{-0.13}$.  The form of the fundamental plane with mass as the dependent variable, eqn.~(\ref{eq:fp}), was used so that measurements of radio and X-ray luminosities could be used to estimate black hole mass, a notoriously difficult quantity to measure (see next section), but with the substantial intrinsic scatter it is only a rough estimate.  It is, however, useful for distinguishing between X-ray binaries and massive black holes.

\section{Measurements of Fundamental Properties of Black Holes: Mass}
\label{sec:mass}
In general relativity, the spacetime around a black hole can be described using only three parameters: mass ($M_{\rm BH}$), spin ($a$), and electrical charge ($q$).  In nature, there will always be enough free electrical charges to neutralize a charged black hole.  We will consider only how we measure $M_{\rm BH}$ in this section and $a$, in the following section.  Mass is always measured, directly or indirectly, by measuring the speeds of material surrounding the black hole.  Dimensionally, we have
\begin{equation}
    M_{\rm BH} = \frac{V_{\rm vir}^2 R}{G},
\end{equation}
and we see that we only need to measure $V_{\rm vir}$, the virial velocity of the material (e.g., the Keplerian velocity), and $R$, the separation of the material from the black hole.  For stellar and gas dynamical mass measurements, $R$ is measured with superb angular resolution, and for reverberation mapping measurements it is measured by timing the light-travel time from the inner parts of the accretion disk to the gas emitting the broad emission lines, namely the broad-line region, as described in Section~\ref{sec:RM}.  In all cases, velocities are measured as radial velocity line widths --- the projection of the virial velocity onto our line  of sight.
 \subsection{Stellar dynamics}
 Stellar dynamical mass measurement techniques are any that use the motions of stars as the tracer of the black hole's gravitational potential.  The advantage of using stars is that their motions are  influenced only by gravity, not by winds or radiation  pressure.  A complication is that, with exception at the Galactic center, we cannot see individual stars and thus must do something to infer the ensemble kinematics of possibly many different  stellar orbits.  In general, one wants to be able to have angular resolution capable of isolating stars whose orbits are in a gravitational potential dominated by the black hole and not by the rest of the galaxy.  This region is called the black hole's `sphere of influence.'  There are different ways of measuring the size of the black hole's sphere of influence, but the best is to use the radius at which the enclosed mass of stars is equal to the mass of the black hole: $M_{*}(r_{\mathrm{infl}}) = M_{\rm BH}$.  A commonly used alternative is to define $r_{\mathrm{infl}} \equiv G M \sigma^{-2}$, where $\sigma$ is the central stellar velocity dispersion.
 The equal mass prescription may be much larger than the velocity dispersion prescription in galaxies with shallow central density profiles.  Even then, with sufficiently good description of the stellar density, one does not strictly need to resolve the sphere of influence; it is only an order-of-magnitude rule of thumb for resolution requirements. 

   \subsubsection{Schwarzschild Method  \label{sec:mass-Schw}}
   The Schwarzschild orbit library method\cite{1979ApJ...232..236S} is the most general stellar dynamical equilibrium model of a self-gravitating system that can reasonably be used to measure a black hole mass.  
   First, imaging data of the galaxy is used to measure the stellar surface brightness.  Imaging in multiple bands is preferred so as to mitigate extinction by dust.  The surface brightness distribution is converted to a stellar luminosity density distribution with assumptions about the geometry and viewing angle. The geometry assumed is usually either axisymmetric or triaxial, though very recently stellar bars have been included.  The viewing angle can be parameterized and iterated over for comparison.  Then the luminosity density is converted to a mass density with an assumed mass-to-light ratio ($\Upsilon$).  The mass-to-light ratio is typically assumed to be constant and parameterized and iterated over for comparison, but stellar evolution models are also sometimes used to calculate the expected value of $\Upsilon$.  Once the stellar density distribution is calculated, it is converted to a gravitational stellar potential by solving Poisson's equation.  To the stellar gravitational potential, one typically adds a dark matter halo potential and a central point mass.  The dark matter halo can be parameterized in multiple different ways, but two common parameterizations are as a Navarro-Frenk-White\cite{1997ApJ...490..493N} (NFW) profile or a 'cored logarithmic profile'. The central point mass is the black hole.
   
   Thus, by using imaging data and some assumptions about the geometry and mass-to-light ratio along with a handful of parameters (viewing angles, $\Upsilon$, dark matter halo parameters, and $M_{\rm BH}$) one has a complete mass model of the galaxy.  All that remains to be done is to compare the expected velocity profiles of the model to the data.  The velocity data are high-spatial-resolution spectra of stellar absorption lines.  Typical lines used are the \ion{Ca}{2} triplet ($\lambda\lambda$8498, 8542, and 8662 \AA), the CO `bandhead' (i.e., a series of broad stellar absorption lines at 2.3--2.5 \units{\mu{m}}), and the 
   \ion{Ca}{2} H\&K lines ($\lambda\lambda$3968.5 and 3933.7 \AA).  These lines are used as they are strong features in the K- and M-giants that dominate the stellar spectrum.  The spectra also need to be taken at large radial distance from the center of the galaxy because the model covers the entire galaxy.  When long-slit spectra are used, this is usually done with major and minor axis slit placements; when integral field units (IFUs) are used, spectroscopic data are obtained at each position in the image covering 
   the entire galaxy.  Once the spectra are taken, the absorption features are deconvolved to yield the line-of-sight velocity distributions (LOSVDs) in each spatial bin in the spectra.  
   
   To create the LOSVDs of the model to compare to the data, a library of orbits is calculated.  For each potential considered, one calculates the orbits of representative stars.  One must be sure to sufficiently cover orbital parameter space in the geometry.  While the orbits are calculated, time and line-of-sight velocity spent in each spatial bin is recorded.  The next step consists of weighting each of the calculated orbits such that the sum of the light added to each bin (proportional to the time spent in that bin and the weight assigned) reproduces the observed light distribution of the galaxy.  Because typically $10^3$--$10^4$ orbits are calculated, this is an underconstrained problem: there are many combinations of weights that can reproduce the imaging data.  Finally, using the combination of weights that can reproduce the imaging data, the LOSVD of each bin of the model (proportional to the time spent in that bin and the weight assigned to each orbit) is compared to the observed LOSVD.  By comparing the goodness-of-fit across parameter space, one may infer the value of black hole mass (and other parameters) that best explains the data, given the model.  Note that this Schwarzschild method is self-consistent in that it uses the light of the galaxy to calculate a potential in which orbits are calculated that must, through some weighting, combine 
   to reproduce the potential in which they are calculated.  
   
   \begin{figure}
       \centering
       \includegraphics[height=0.49\textwidth]{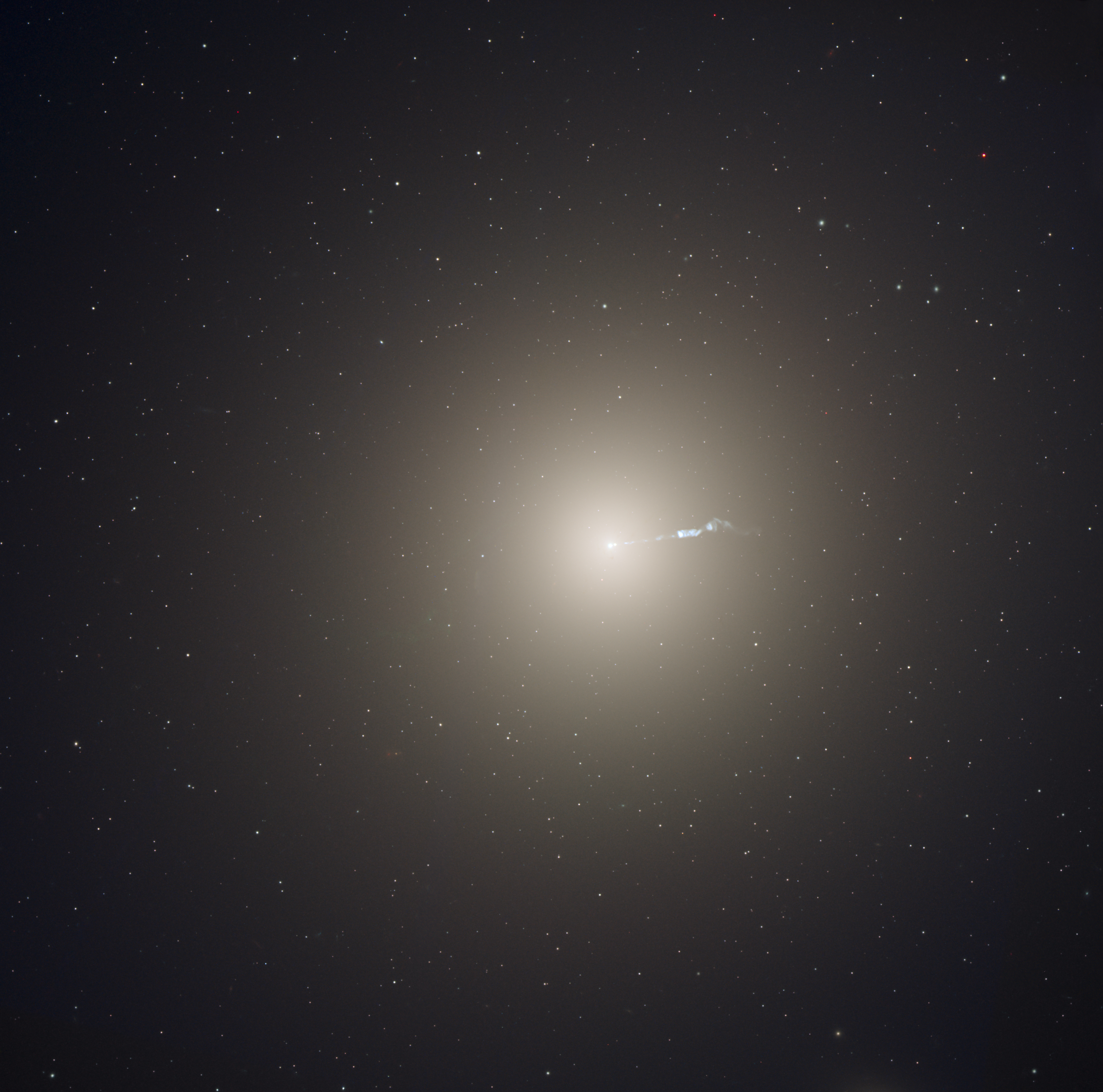}
       \includegraphics[height=0.49\textwidth]{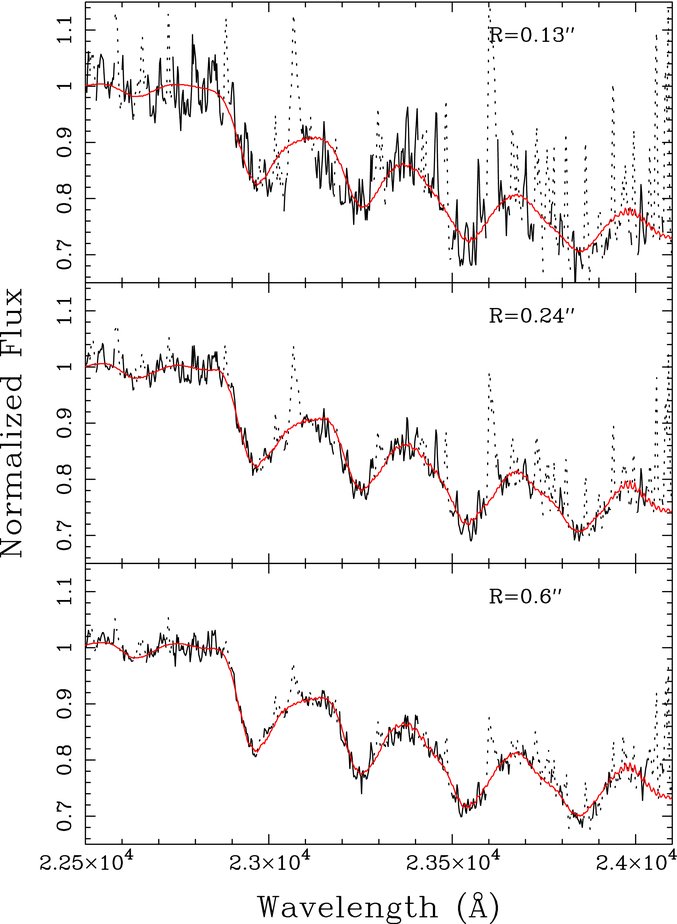}
       \caption{\textit{Left panel}: M87, the giant elliptical galaxy in the Virgo cluster.  The image is approximately $10$ kpc on a side at the distance of M87.  Clearly seen is the prominent jet coming from the galaxy's center.  The jet is powered by the massive black hole at the center.
       \textit{Right panel}: Spectra of M87 taken with Gemini NIFS behind adaptive optics.  The specrtra shown come from three different distances from the center of the galaxy as indicated in each panel.  The CO bandhead absorption lines are apparent in each panel and are broadened by the line-of-sight velocity distribution (LOSVD), which is deconvolved from the data.  The LOSVD is then modeled with the Scwarzschild method to get a best-fit mass of  $M_{\rm BH} = (6.6 \pm 0.4) \times 10^{9}\ \msun$.  From Gebhardt et al.\cite{2011ApJ...729..119G}}
       \label{fig:m87image}
       \label{fig:m87spectra}
   \end{figure}
   
   One example of the Schwarzschild method is the measurement of the black hole in M87\cite{2011ApJ...729..119G}, the giant elliptical in Virgo seen in Figure \ref{fig:m87image}.  Measurement of M87's black hole mass required ground-based spectroscopy with adaptive optics and a large aperture telescope because of the relatively low central surface brightness.  Several spectra at different distances from the center of M87 are shown in Figure \ref{fig:m87spectra}, revealing that the weak AGN contaminated the central spatial bin, reducing the effective angular resolution.  The other spatial bins, however, show obvious CO bandheads in absorption that were used to extract the LOSVDs.  The complete modeling of the galaxy resulted in a marginalized $\chi^2$ curve  that estimates the black hole mass\cite{2011ApJ...729..119G} to be $M_{\rm BH} = (6.6 \pm 0.4) \times 10^{9}\ \msun$, in agreement with the later mass estimate from the Event Horizon Telescope (\S\ref{sec:characterizing-m87eht}).

   \subsubsection{Jeans Anisotropic Modelling}
The Jeans anisotropic modelling (JAM) method is an additional stellar dynamical technique for modeling any gravitationally bound system of particles.  The modeling can recover a variety of kinematic properties of the systems under investigation, including dynamical mass-to-light ratio.  Inference of mass-to-light ratios leads to black hole mass estimates by finding evidence for a central point-source increase in the mass-to-light ratio.  This point-source increase in the mass-to-light ratio is then attributed to a black hole, even if a variety of now less plausible alternatives could also potentially explain it.  The attribution of a point-source increase in mass-to-light ratio to a black hole is a general feature of all stellar and gas dynamical mass measurements.  JAM works by using the Jeans equation solutions to the collisionless Boltzmann equation, but with an extra anisotropy term that parameterizes the kinematic deviations from a perfectly isotropic distribution function.  Combined with assumptions about the geometry and the variations in anisotropy, JAM can model high-spatial-resolution spectroscopy to estimate black hole masses that are in strong agreement with the more general Schwarzschild modeling estimates.

 \subsection{Gas dynamics}
 Gas dynamical mass measurement techniques refer to any technique that uses the motions of gas as the tracer of the black hole's potential.  An advantage of using gas is that, since it tends to form a disk of material, the geometry is easy to model.  A limitation of using gas is that non-gravitational forces,  non-circular motion, and three-dimensional effects  can be very difficult to take into account.  Another limitation is that while there are almost always stars near a central massive black hole, not all galaxies have nuclear gas disks.  This may explain why the best data and analysis of the gas disk around the central black hole in M87 finds a mass estimate a factor of $\sim2$ below the stellar dynamical and Event Horizon Telescope estimates \citep{2013ApJ...770...86W, 2011ApJ...729..119G, 2019ApJ...875L...6E}.  High spatial resolution is still required, and this plays into the strengths of radio interferometers sensitive to molecular gas and megamasers.
   \subsubsection{Ionized gas}
   The most widely used gas dynamical technique is ionized gas since the gas may emit brightly.  Typical emission lines used are H$\alpha$ (6563 \AA) and [\ion{N}{2}] (6548 and 6583 \AA).  To estimate the mass of the black hole, one usually assumes a razor-thin circular disk with an emissivity determined from narrow-band imaging.  To account for inclination effects, either IFU or multiple long-slit placements are needed.  From the spectra, one obtains line-of-sight velocity and velocity dispersion (and sometimes higher moments  of the line profile) at many locations in the ionized gas disk.  A model is then calculated for the circular velocity of the gas about the galaxy's systemic velocity.  The circular velocity is a function of enclosed mass, i.e., the sum of the black hole and the stellar mass, which is determined by imaging and a parameterized mass-to-light ratio.  Other parameters that need to be included in the model are inclination and position angle of the gas disk as well as an offset to account for sub-pixel localizing of the black hole.
   
\begin{figure}
       \centering
       \includegraphics[width=0.8\textwidth]{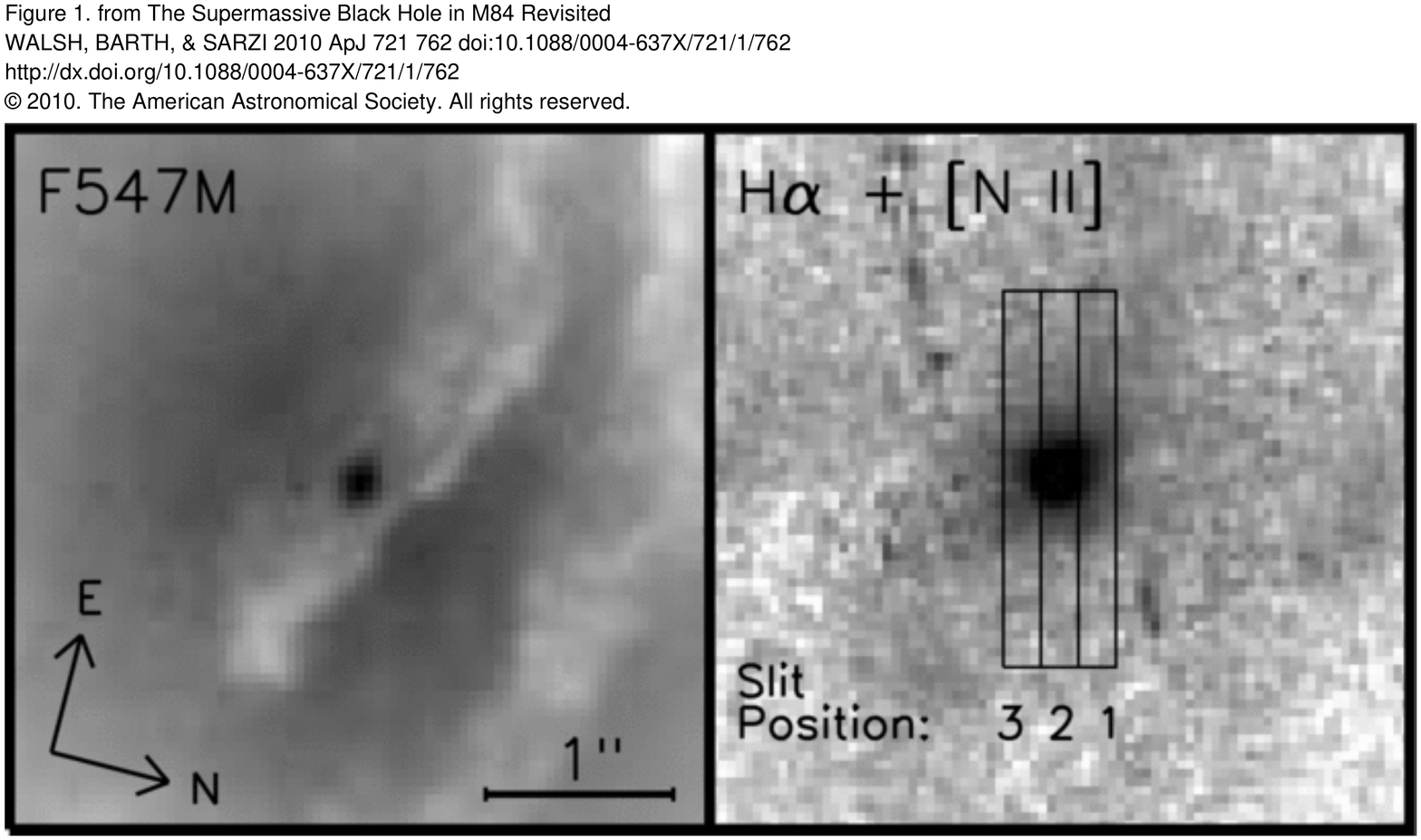}
       \includegraphics[width=\textwidth]{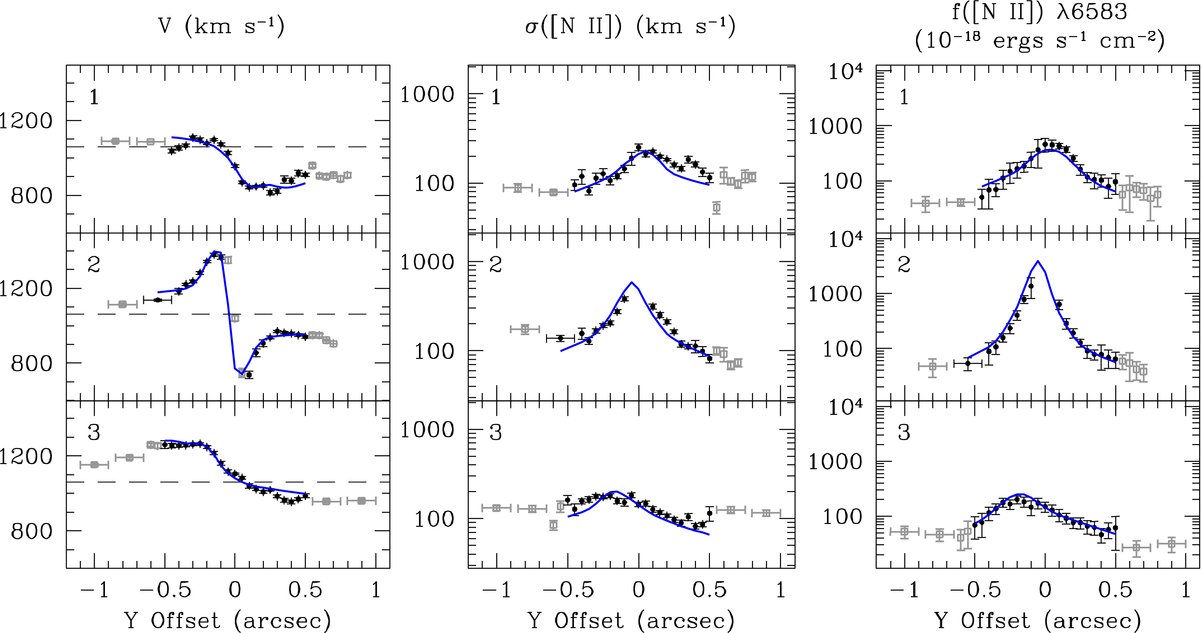}
       \caption{Top panels show Hubble Space Telescope imaging of the central part of M84 in F547M filter (left) and in ionized gas (right).  The bottom panels show the velocity, velocity dispersion, and line profiles for each of the indicated slit positions shown in the top-right panel.  The velocity profile for slit position 2 shows a clear indication of rotation about a central compact, massive object. (Imaging and velocity profiles from Walsh et al.\citep{2010ApJ...721..762W})}
       \label{fig:gasmass}
   \end{figure}

   We show an example of ionized gas modeling of M84 using a combination of H$\alpha$ and [\ion{N}{2}] emission in Figure \ref{fig:gasmass}\citep{2010ApJ...721..762W}. The best-fit model is an excellent description of the data and yields a mass estimate of $M_{\rm BH} = 8.5^{+0.9}_{-0.8} \times 10^{8}\ \msun$.  In order to model the data well, an intrinsic velocity dispersion was included in the gas motions.  This intrinsic velocity dispersion would not manifest itself if that gas were (i) only responding to an axisymmetric gravitational force while (ii) in a thin disk with (iii) purely circular motion. At least one of these assumptions is violated, but all have been suggested: (i) nongravitational pressure forces, (ii) flared disks, and (iii) net radial motion from spiral instabilities or significant inflow/outflow.
   
   \subsubsection{Molecular gas}
   The sensitivity and angular resolution of the Atacama Large Millimeter Array (ALMA) has made molecular gas measurements of black hole masses viable for a large number of galaxies.  The principle of this measurement and the modeling technique is very similar to the ionized gas case (Fig.~\ref{fig:gasmass}).  With cold molecular gas, however, it is generally assumed that pressure forces are completely subdominant relative to the gravitational forces.  The effects of three-dimensional geometry and radial motion, however,  still need to be taken into account.
Since ALMA came online, this technique has been applied predominantly to cold gas observations of quiescent black holes in centers of elliptical galaxies\cite{2019ApJ...881...10B, 2019ApJ...883..193N, 2021MNRAS.503.5984S} 
but also a few dwarf galaxies\cite{2020MNRAS.496.4061D, 2020ApJ...892...68N} and a low-luminosity AGN\cite{2016ApJ...823...51B}.




   \subsubsection{Megamaser \label{sec:megamaser}}
\begin{figure}
    \centering
    \includegraphics[width=0.95\textwidth]{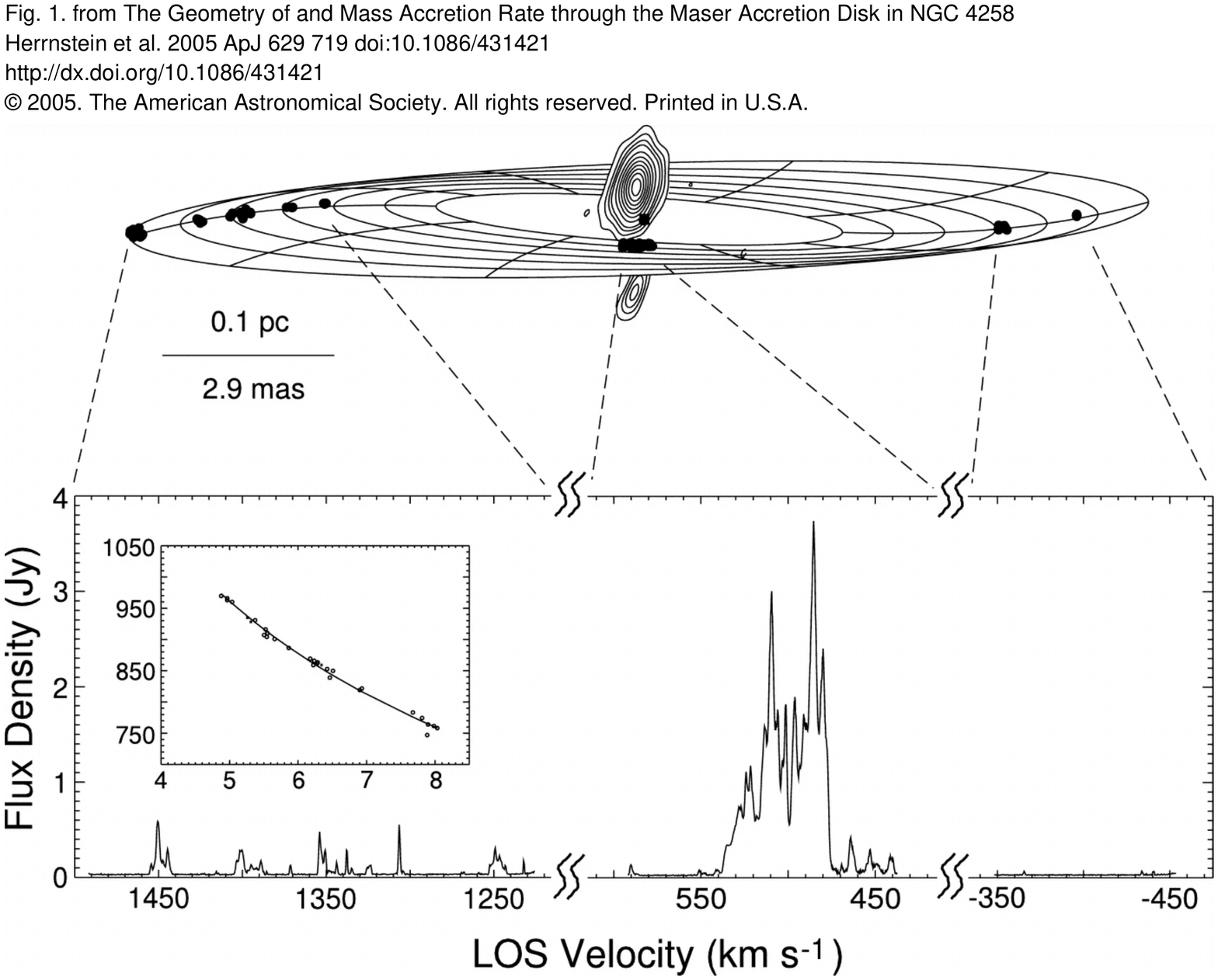}
    \caption{Diagram showing warped disk and water maser emission spots from NGC 4258.  The bottom panel indicates the spectrum, and the inset shows velocity of emission spots against projected distance as measured in milliarcseconds with a Keplerian rotation curve. (From Herrnstein et al.\ 2005\cite{2005ApJ...629..719H}).}
    \label{fig:masermass}
\end{figure}

   If a gas disk around a black hole is inclined to within a few degrees of edge-on towards our line of sight, it may be visible to us as a water megamaser, which allows precise black hole mass determinations.  Megamasers are identified by the 22 GHz water emission line, which is only possible when a high column of  water is visible along our line of sight.  The megamaser is useful for measuring black hole masses if it can be determined to be in a Keplerian disk.  The hallmark signs of a disk (as opposed to, e.g., gas inflow or outflow) in the spectrum are triple sets of emission lines within several hundred \units{km\,s^{-1}} of the galaxy's systemic velocity\,---\,one line arising from the approaching side of the disk, one from the receding side, and one dominated by the portion moving tangential to our line of sight.  These triple sets can be discovered with wide surveys using low-angular-resolution radio telescopes with good sensitivity.  After discovery, observations with high angular resolution (i.e., VLBA observations) can measure precise positions and velocities of individual masing spots.  The position--velocity diagram is then modelled as an  inclined, warped rotating disk whose motion is dominated by a central gravitating mass, that is, the black hole.  The position--velocity diagram fitting does not suffer from degeneracy with inclination angle because the latter is well constrained to be close to edge-on by the mere fact that it is a megamaser.  In some cases, multi-epoch observations can measure accelerations of individual masing spots to get a geometrical distance to the source.  This later fact is being exploited to measure the Hubble constant. 
   The hallmark target for megamaser observations is NGC\,4258  shown in Figure~\ref{fig:masermass}\cite{2005ApJ...629..719H}.
   
   
 \subsection{Gas Dynamics near Accreting Massive Black Holes \label{sec:Mass-AGNs}}
 When the supermassive black hole is accreting material, emission is released from the central regions of the galaxy. At a moderate to high accretion rate this emission is so strong that it washes out the faint spectral absorption features (Fig.~\ref{fig:m87spectra}) that are commonly used to measure the kinematics of stars orbiting the black hole. The central glare thus prohibits the use of stellar and ionized gas kinematics (as outlined above) to measure the mass of the black hole for AGN, unless the AGN is particularly weak, and alternative methods are used. 

 \subsubsection{Reverberation mapping \label{sec:RM}}
The reverberation mapping (hereafter, RM) technique is currently the most robust black hole mass determination method for powerful AGN.  RM does not require high spatial resolution, as do the stellar and gas kinematical methods, but relies on temporal resolution. Therefore, the RM technique has the added advantage of being applicable to black holes residing in galaxies at distances greater than $\sim$100 Mpc, that is, beyond our local Universe. The RM method assumes that the motion of the broad emission line gas is virial 
and is dominated by the black hole gravity:

\begin{equation}
    M_{\rm BH} = \frac{V_{\rm vir}^2 R}{G} = f \frac{W^2 R}{G},
\end{equation}

where $V_{\rm vir}$ is the virial (e.g., Keplerian) velocity of the broad line gas residing at the distance $R$ from the black hole and $G$ is the gravitational constant.  We are unable to measure $V_{\rm vir}$ directly, but we can determine the line-of-sight projection of this velocity in the emission line spectrum as the line width, $W$. Therefore, the approximation $V_{\rm vir} = \sqrt{f} W $  is used where $f$ is the scaling factor that accounts for our ignorance of the structure and kinematics of the broad line gas. This $f$-factor can be estimated in a couple of ways, addressed in \S~\ref{sec:f-factor}.

The broad line emission is powered by the ionizing radiation from the central continuum source of the inner accretion disk (\S~\ref{sec:BHevidence}). The accretion flow is not smooth (in mass or time) so the emitted continuum flux changes with time. Because the broad-line region is both extended and not co-spatial with the continuum source the light curve of the broad line fluxes is a slightly smeared copy of the continuum light curve, shifted in time. This time delay, $\tau$, is taken to represent the light travel time of the disk photons, thereby allowing an estimate of the characteristic size $R$ of the broad-line region, $R = c \tau $. Determining $\tau$ is in practise a non-trivial inversion problem involving a so-called transfer function that encompasses the physical processes of the light and matter interaction\cite{1993PASP..105..247P,  2004PASP..116..465H}. 
The transfer function connects the emission line light curve with the ionizing continuum light curve. The technical details of the RM technique, as described by Blandford \& McKee\cite{1982MNRAS.200..115S}, are summarized by Peterson\cite{1993PASP..105..247P}.


Since RM measures the distance to the gas that produces the variable emission lines, to measure the black hole mass requires a measure of the velocity of that same gas. From the spectroscopic monitoring data $W$ is measured as the width of the broad H$\beta$ emission line in the RMS spectrum (i.e., the one standard deviation relative to the mean spectrum based on all observed spectra during the observing campaign). Early work confirmed that the broad emission line gas kinematics is dominated by the black hole gravity, as several emission lines obey the virial constraint\cite{1999ApJ...521L..95P, 2000ApJ...540L..13P} $V_{\rm vir} \propto R^{1/2}$.
While there are multiple ways to measure the line width (e.g., the full width at half maximum, the line dispersion or the mean average deviation), the line dispersion is used since it is less susceptible to spectral noise and it is well defined for RMS profiles\cite{2004ApJ...613..682P, 2020ApJ...903..112D}. 
Moreover, it appears to be less biased\cite{2011MNRAS.415.2932R}.


Over the past $\sim$30 years numerous RM monitoring campaigns have reverberation mapped more than 60 AGN in the nearby Universe\citep{2004ApJ...613..682P, 2015PASP..127...67B} showing a range of black hole masses between 300,000 \msun{} and 1 billion \msun{} and an average of 30 million \msun{} (almost 8 times more massive than our Milky Way black hole). The smallest black hole in this sample resides in a low luminosity dwarf galaxy NGC 4395 where it accretes at a very low level\cite{2005ApJ...632..799P}. The most massive of these black holes is in the center of a local quasar, PG1426+015, discovered in the Palomar-Green survey\cite{1986ApJS...61..305G}. For comparison, distant quasars have black hole masses\cite{2004ApJ...601..676V, 2009ApJ...699..800V, 2010ApJ...719.1315K} between a billion and ten billion \msun{}. The exact value of the most massive black hole in the Universe is difficult to assess because  for black holes beyond our local Universe  we have so far mostly been able to {\it estimate} their mass (\S~\ref{sec:SEmass}) and these estimates have uncertainties of factors of a few\cite{2004ApJ...601..676V}. Very massive black holes of this caliber are also known to exist in local giant elliptical galaxies where they have ceased to accrete material. The black hole in the giant elliptical galaxy Messier 87 in the nearby Virgo Cluster (\S~\ref{sec:characterizing-m87eht}) is among the most massive known in the Universe at a whopping 6.5 billion solar masses\cite{2019ApJ...875L...6E}.


Until a few years ago, most existing RM studies targeted the H$\beta$ emission line in the restframe optical region because the forbidden [OIII] 5007\AA{} narrow emission line just redward of H$\beta$ offers an excellent way to flux calibrate the observed spectra to high precision. The reason is that the narrow line region emitting the [OIII] line resides at a sufficiently large distance that the [OIII] emission will not vary on the timescales of the duration of the monitoring campaigns\cite{2004ApJ...613..682P, 2013ApJ...779..109P}. Moreover, because of the low particle density  of the narrow-line region the  recombination timescale is  long, thus suppressing rapid variability.  Also, the [OIII] line profile provides a template to correct for the narrow line contribution to the observed H$\beta$ profile. While the H$\alpha$ emission line is stronger, observing H$\beta$ is more efficient as  in many cases it requires just a single spectral setup.


In addition to H$\beta$, other strong broad emission lines, such as \civ{} 1549\AA{}\cite{1995ApJS...97..285K, 2007ApJ...659..997K} 
and \mgii{} 2800\AA{}\cite{2006ApJ...647..901M},
have been reverberation mapped. For local AGN these lines require observations with space-based telescopes.  Because reverberation mapping campaigns are rather resource intensive they are not trivial to organize with space-based telescopes and as a result, far fewer AGN have been studied at ultraviolet wavelengths.  Nonetheless, eight nearby AGN have so far been monitored in the ultraviolet with the {\it International Ultraviolet Explorer} (1978 -- 1996) and/or the {\it Hubble Space Telescope} (launched in 1990); an overview is provided by De Rosa et al.\cite{2015ApJ...806..128D}  The most extensive and intensive space-based monitoring so far is undertaken in the recent campaigns of the AGN Space Telescope and Optical Reverberation Mapping (AGN STORM; PI: B.M. Peterson) projects targeting NGC\,5548 in 2014\cite{2015ApJ...806..128D, 2020ApJ...898..141D} and Mrk 817 in 2020--2022\cite{2020hst..prop16196P, 2021ApJ...922..151K}. 





The past 10 years have seen a strong increase in RM studies. Monitoring programs of large samples of distant AGN with the aim to determine broad-line region lags (time delays, $\tau$) and black hole masses have been launched as part of large surveys, such as the Dark Energy Survey (OzDES\cite{2015MNRAS.452.3047Y})
and the Sloan Digital Sky Survey (SDSS-RM\cite{2015ApJS..216....4S}).
These efforts are now increasing the number of AGN for which the size of the \civ{}\cite{2018ApJ...865...56L, 2019ApJ...887...38G, 2019MNRAS.487.3650H, 2022MNRAS.509.4008P}  and \mgii{}\cite{2016ApJ...818...30S, 2019ApJ...880...46C, 2021arXiv210502884H, 2021MNRAS.507.3771Y} line regions are estimated. While such surveys can monitor a much larger sample of objects at a time, they are often limited to a fixed exposure time for all targets in each survey field, regardless of the luminosity or variability properties of each target. Therefore, robust RM results are not guaranteed for all targets. While these surveys,  like  the original RM campaigns of  local AGN, will preferentially provide results for the brighter and most variable sources, the hope is that these surveys can target a broader and less biased subset of the AGN population\cite{2011AJ....141..167R, 2020ApJ...899...73F} 
than the original sample of RM AGN\cite{2004ApJ...613..682P}. 
Also, if the early predictions\cite{2015MNRAS.453.1701K} are correct, then RM analysis of large samples of distant AGN promise to measure cosmic distances and hence constrain cosmological parameters (\S~\ref{sec:cosdist}).





Because traditional spectroscopic RM studies are resource intensive a new technique is being developed that is showing quite some promise. This 'photometric reverberation mapping' method predominantly relies on monitoring observations through broad-band and narrow-band imaging\cite{2011A&A...535A..73H,  2016ApJ...818..137J}, 
which may prove useful with future telescopes such as the Large Synoptic Survey Telescope coming online.  


\subsubsection{Determination of the f-factor \label{sec:f-factor}}
If the intrinsic value of the black hole mass is precisely known, the $f$-factor can be determined. At present, the black holes with very precise and independent determinations of the mass --  e.g., via the megamaser method (\S~\ref{sec:megamaser}) or by measuring the extent of the  black hole shadow (\S~\ref{sec:characterizing-m87eht}) --  cannot be weighed with reverberation mapping.  For other objects, it is thus only possible to obtain an {\it estimate} of $f$. Efforts to use forward modeling of the broad-line region emission\cite{2017ApJ...849..146Grier,2019ApJ...871..108P}
have reported estimates for individual AGN,  assuming the model is representative of the  broad-line region. 
Spectropolarimetric studies\cite{2015MNRAS.454.1157P, 2016MNRAS.458L..69B, 2018MNRAS.473L...1S} have also contributed to estimates of $f$.  
Onken et al.\citep{2004ApJ...615..645Onken} determined an empirical sample average estimate of $f =  5.5$ by assuming that the local reverberation mapped black holes are sufficiently close to their final mass due to their low mass-accretion rates. This assumption means that the $M_{\rm BH} - \sigma_{\ast}$ relationship  for AGN should coincide with that of quiescent black holes in the local Universe, allowing  for the $f$ scaling factor to be constrained. 



\subsubsection{Velocity-Delay Maps}
Given sufficiently high quality spectroscopic and photometric data obtained at high cadence, the RM technique can provide a so-called velocity-delay map, probed by the ionizing photons as they propagate through the broad-line region. This map shows how the broad line gas of a given velocity responds to continuum variations.  
By connecting the velocity to a given central distance the velocity delay maps hold important information about the three-dimensional structure and kinematics of the broad line gas projected onto our line of sight, as well as its inclination to our sight-line. One advantage of the velocity delay map is that this two-dimensional transfer function (that depends on velocity and time delay) promises to provide black hole mass measurements without the need for the $f$-factor. 

The first 1-dimensional velocity-delay maps were gleaned from the {\it  International Ultraviolet Explorer} monitoring data of NGC 5548, now one of the most well-studied Seyfert galaxies\cite{1991ApJ...371..541K, 1991ApJ...367L...5H}. In recent years the data quality and cadence have improved sufficiently to provide detailed features in the 2-dimensional velocity delay maps, showing clear signs that the broad-line region kinematics can include inflows and outflows in addition to the predominant Keplerian motion\cite{2010ApJ...720L..46B, 2010ApJ...721..715D, 2013ApJ...764...47G, 2015ApJ...806..128D}.  The most detailed and high quality velocity delay maps obtained so far are based on the 6 months campaign on NGC 5548 of AGN STORM\cite{2021ApJ...907...76H}.


\subsubsection{Radius -- Luminosity Relationship \label{sec:rlrelation}}

Based on the large body of RM work over the years, the empirical radius ($R$)--luminosity ($L$) relationship has been established. It shows that the size of the broad-line region scales\cite{2005ApJ...629...61K} with the continuum luminosity: the more powerful the central source, the larger the broad-line region. This is also seen for a single AGN: as the source luminosity changes, the size of the broad-line region changes -- known as 'breathing'\cite{2004ApJ...606..749K, 2015ApJ...801....8K, 2020ApJ...903...51W}. With increasing luminosity, the photon flux penetrates further into the broad line gas, changing the ionization structure of the gas, thereby shifting the location at which the gas is most efficiently emitting a given emission line. This consequence of photoionization physics was discussed by Baldwin et al.\cite{1995ApJ...455L.119B} in the framework of their Locally Optimized Cloud model. 

\begin{figure}
    \centering
    \includegraphics[width=0.45\textwidth]{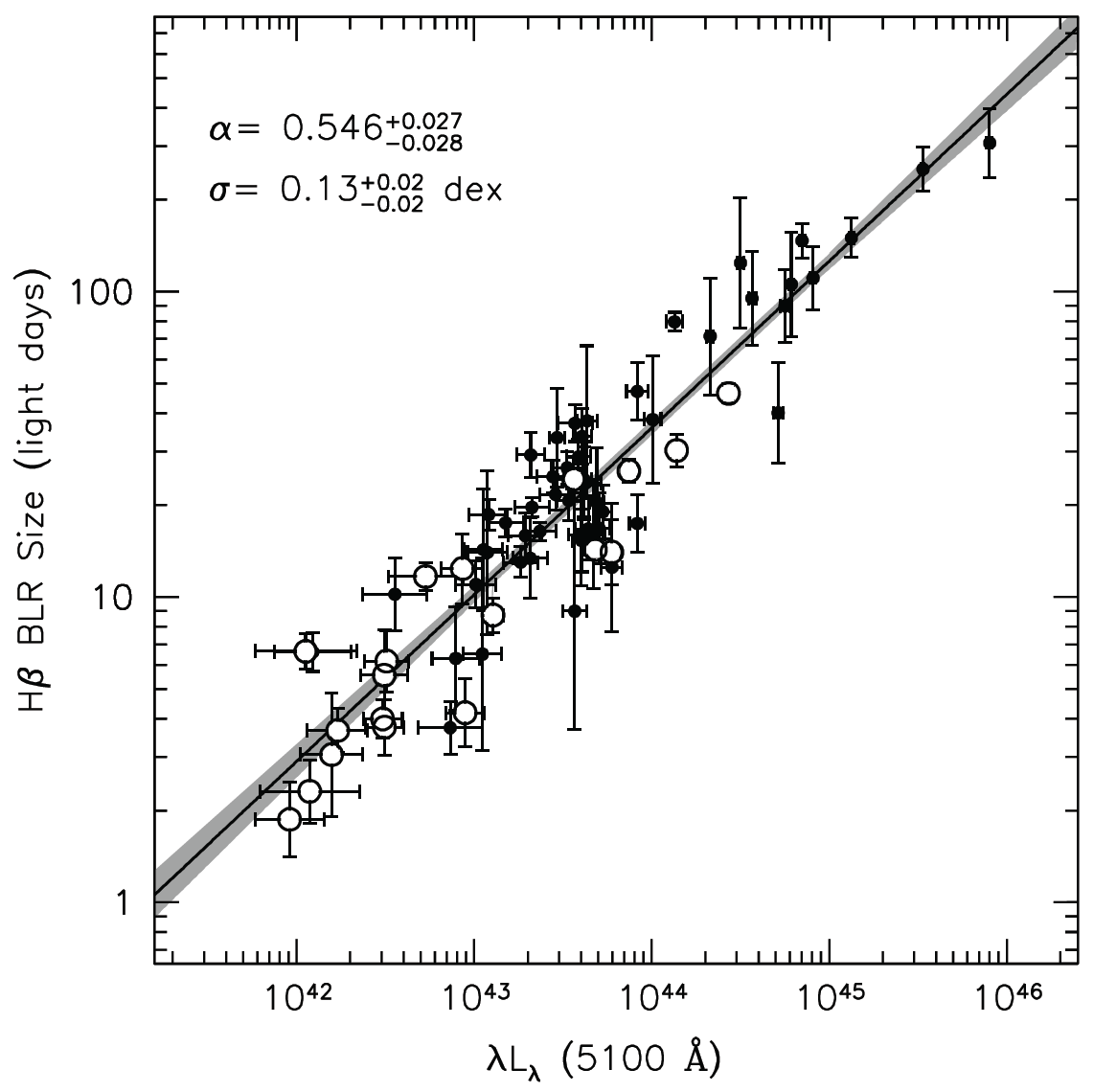}
    \hspace{0.15cm}
    \includegraphics[width=0.52\textwidth]{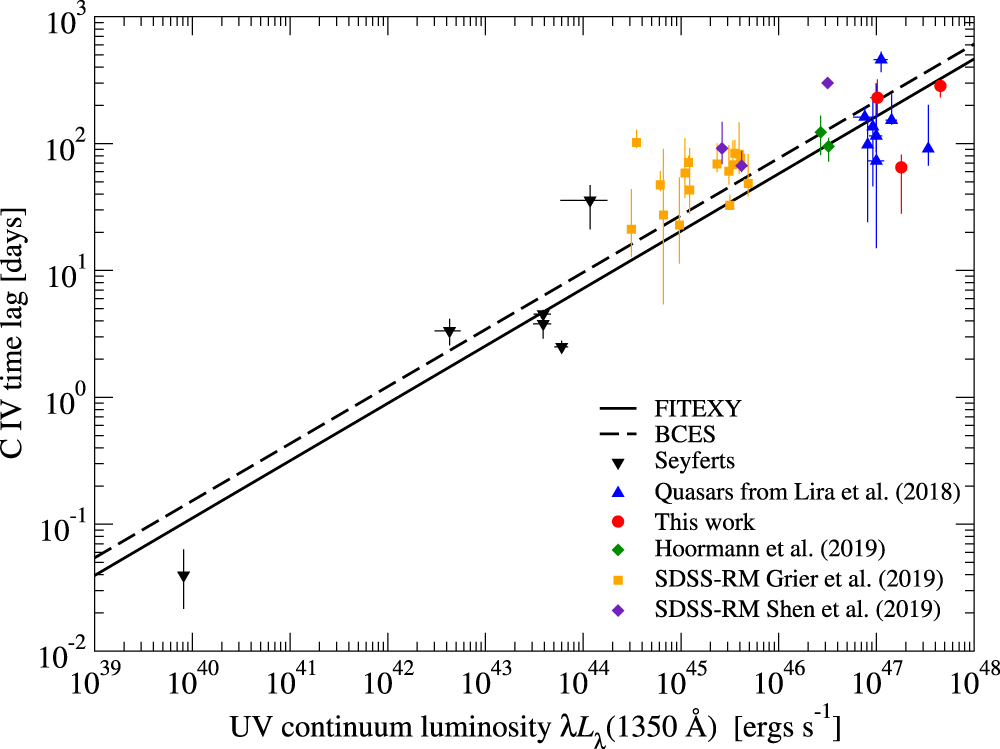}
    \caption{Radius--Luminosity relations for ({\it left panel}) the broad \hb{} emission line (Figure 11 of Bentz et al.\citet{2013ApJ...767..149B}) and ({\it right panel}) the \civ{} emission line  (Figure 9 of Kaspi  et al.\cite{2021ApJ...915..129K}). The BLR size (or time lag) is in  the AGN  restframe. In the left panel the best fit slope $\alpha$ and uncertainties are shown by the solid line and shading and listed in upper left corner. The  scatter $\sigma$ relative to  the best fit relation is also listed. For \civ{} ({\it right}) the best  fit relation is determined by two different  methods (solid and dashed lines).}
    \label{fig:rlrelation}
\end{figure}

The distance (i.e., radius) to the H$\beta$ broad emission line region correlates strongly with both the optical and the infrared continuum luminosity\citet{2009ApJ...697..160Bentz, 2013MNRAS.432..113Landt}. The optical relationship for H$\beta$ (Fig.~\ref{fig:rlrelation}, left) turns out to be quite tight once the stellar light in the host galaxy
has been corrected and improved distance measurements to the nearest AGN are adopted\cite{2013ApJ...767..149B}. Restricting the relationship to the most accurate measurements\cite{2010IAUS..267..151P}, there is a strong potential for this relationship to measure cosmic distances (\S~\ref{sec:cosdist}).


The recent increase in RM studies of distant black holes has provided better statistics to the relationship for \civ{} (Fig.~\ref{fig:rlrelation}, right), especially at high luminosities\cite{2021ApJ...915..129K}.
This relationship has a slope consistent
with 0.5 , similar to that of H$\beta$\cite{2013ApJ...767..149B}.
Although the statistics for \mgii{} lags are increasing sufficiently to test for a $R$(\mgii{})--$L$ relationship, there is currently a significant scatter\cite{2021MNRAS.507.3771Y}. The reasons  for this scatter may include the difficulties in determining the 3000\,\AA{} luminosity  and the \mgii{} line flux due to strong blending from \feii{} and  the lower responsivity of the \mgii{} line\cite{2004ApJ...606..749K}.
The recent RM surveys show that for the general AGN population the broad-line region size depends not only on black hole mass but also mass accretion rate and perhaps other properties \cite{2019ApJ...886...42D, 2020ApJ...899...73F}. In particular, distant high-luminosity quasars show \hb{} and \mgii{} delays that are shorter than expected from the local $R$(\hb)$ - L$ relationship\cite{2018ApJ...856....6D, 2019ApJ...880...46C}. 



\subsubsection{Single-epoch Scaling Relationships \label{sec:SEmass}}
While the RM technique provides the most robust mass determinations of the black holes powering AGN, it is quite resource intensive -- requiring large monitoring campaigns of individual targets. Single-epoch scaling relationships offer quick mass {\it estimates} based on measurements of an individual spectrum of the target, albeit with uncertainties\cite{2006ApJ...641..689V} as large as  a factor of several. The single-epoch mass estimates are mere approximations to the RM mass determination. While RM uses the width of the variable emission line and the measured responsivity-weighted distance to the line emitting gas (\S~\ref{sec:RM}), the single-epoch method relies on the width of the line profile in a spectrum obtained at a random luminosity level plus an estimate of $L$ from the continuum flux in said spectrum combined with the $R - L$ relation for an estimate of $R$. The method was originally developed for the \hb\ and \civ{} $\lambda$1549 emission lines\cite{1999ApJ...526..579W, 2002ApJ...571..733V, 2002MNRAS.337..109M, 2006ApJ...641..689V}. 
Many authors have since presented their own single-epoch scaling relations for the main emission lines of \ha, \hb, \mgii, \civ, and \ciii{}
\cite{2005ApJ...630..122G, 2009ApJ...699..800V, 2013ApJ...770...87P, 2017ApJ...839...93P, 2013BASI...41...61S}, each with their own advantages and disadvantages\cite{2011nlsg.confE..38V , 2013BASI...41...61S}.


The large body of work studying line width behavior with luminosity for individual AGN and across the AGN population has shown that the \mgii{} and \civ{} lines do not always behave as well as the \hb{} profile\cite{2011nlsg.confE..38V, 2013BASI...41...61S}. This has created a debate in the literature on the reliability of especially \civ{}, which also displays blueshifts relative to the systemic velocity\cite{1992ApJS...79....1T, 2002AJ....124....1R}, 
perhaps due to central outflows or disk winds\cite{2012ApJ...759...44D, 2013BASI...41...61S}. No firm consensus has yet been reached on this issue, in part as results depend on measurement techniques, data quality, and sample selection\cite{2009ApJ...704L..80D, 2011nlsg.confE..38V,  2018NatAs...2...63M, 2020ApJ...899...73F}.
However, a positive aspect of these efforts is the increased insight into the physics of the broad-line region that is ultimately obtained. 
Over the years, significant efforts have been made to decrease the scatter in these relationships and hence improve the accuracy of the mass estimates\cite{
2020ApJ...903..112D}. An important recent result shows that to avoid biased mass estimates a correction is needed because the FWHM line width depends on the source Eddington luminosity ratio\cite{2020ApJ...903..112D} --- as does the $R$--$L$ relationship\cite{2018ApJ...856....6D}  (\S~\ref{sec:rlrelation}). While a single spectrum and a single emission line is often used per AGN for mass estimates, the way forward may be to also use multiple emission line estimates (and/or multiple epochs) for each AGN to increase statistics\cite{2011nlsg.confE..38V}, as previously confirmed by the early RM studies\cite{1999ApJ...521L..95P}.






 \subsection{Polarization}
 When photons scatter off electrons or dust, they become polarized. The polarization characteristics depend on the scattering medium. Examining scattered light has allowed scientists to use the scattering medium as a mirror to 'look around the corner' and thereby reveal hidden broad emission lines in highly inclined AGN\cite{1985ApJ...297..621A}.
 It turns out that photons from the broad-line region, scattered off the dusty obscuring torus, can be used to estimate the black hole mass for AGN with source inclinations between 20$^{\circ}$ and 70$^{\circ}$, since then only one side  of the torus is visible along our line of sight, creating a predominantly  single scattering surface\cite{2015ApJ...800L..35A, 2018A&A...614A.120S}. 
 
\begin{figure}
     \centering
     \includegraphics[width=\textwidth]{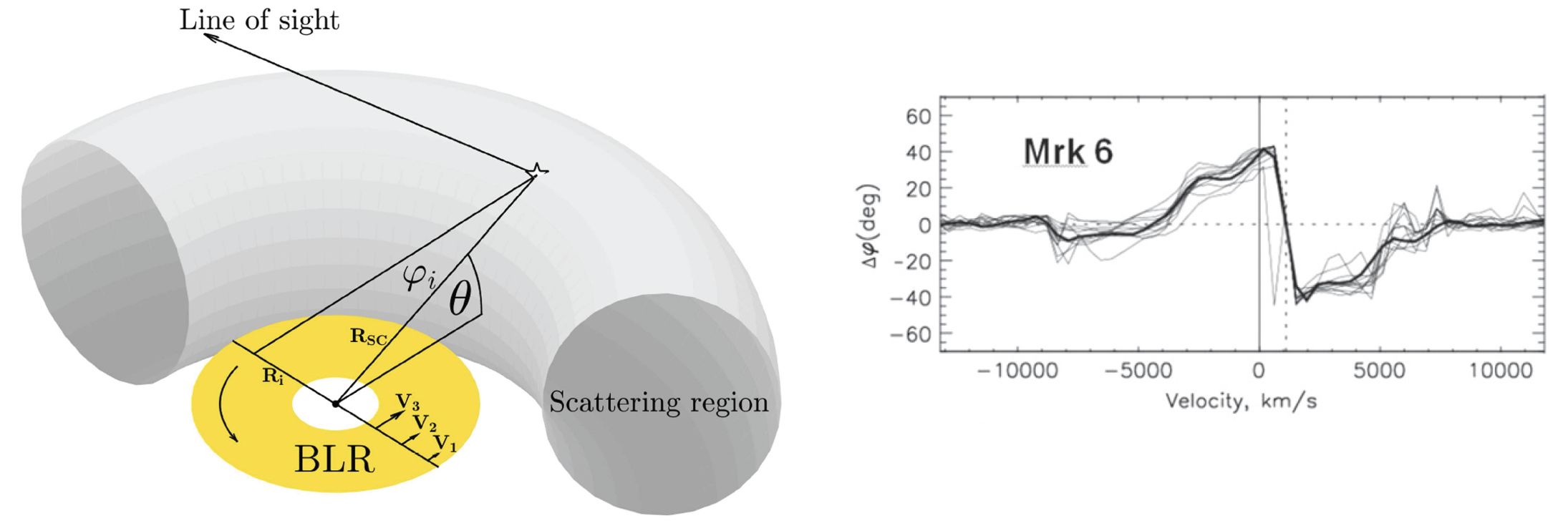}
     \caption{{\it  Left panel}: Assumed geometry of   the broad  line region (BLR) and the  scattering  region. The relevant parameters defined in the  text are marked. (Figure 1 of Savic et al.\cite{2018A&A...614A.120S}).  {\it Right panel}: Polarization angle  change as a function of velocity observed  across the \hb{} emission line region\cite{2015ApJ...800L..35A}.}
     \label{fig:polariz}
 \end{figure}

 For a Keplerian velocity field in the broad-line region, the polarization angle of the line emission (relative to the polarization angle of the continuum emission) depends on the velocity in the broad-line region. Hence, the observed polarization angle $\Delta \phi_i$ will change across the line profile (denoted by `$\phi_i$' in Fig.~\ref{fig:polariz}). The angle $\Delta \phi_i$ is related to the velocity of a gas element in the broad-line region via the relation\cite{2015ApJ...800L..35A, 2018A&A...614A.120S}
 
 \begin{equation}
     \log \left( \frac{V_i}{c}\right) = \frac{1}{2} \log \left(\frac{G M_{\rm BH} \cos^2(\theta)}{c^2 R_{\rm sc}}\right) - \frac{1}{2} \log (\tan ( \Delta \phi_i))
 \end{equation}
 where $V_i$ is the velocity of gas element $i$ in the broad-line region, $\Delta \phi_i = \phi_L - \phi_C$ is the polarization angle of photons from gas element $i$ (here $\phi_L$ and $\phi_C$ is the polarization angle of the line  and  continuum  emission, respectively), $G$ is the gravitational constant,  $\theta$ is the declination of the scattering region element of the torus, illuminated by the broad-line region, $c$ is the speed of light, and $R_{\rm sc}$ is the radial distance of the scattering element from the ionizing source, i.e. the inner radius of the torus. 
 To estimate the black hole mass, $M_{\rm BH}$, requires knowledge of $\Delta \phi_i$, $V_i$ and $R_{\rm sc}$ and $\theta$. Unless the scale height of the torus exceeds 20$^\circ$ solid angle on the sky, using the approximation $\theta \sim 0$ inflicts an uncertainty in $M_{\rm BH}$ of order $\sim$10\% or less\cite{2015ApJ...800L..35A}. While $V_i$  and $\Delta \phi_i$ are measured from the spectropolarimetric observations, $R_{\rm sc}$ can measured from reverberation mapping analysis of direct light at infrared wavelengths\cite{2014ApJ...788..159K} or from the empirical relationship between the inner torus radius and the AGN luminosity based on infrared interferometry\cite{2011A&A...527A.121K}.
  

 Similar to single-epoch scaling relationships, an advantage of this method is that only a single observation is necessary, in contrast to the reverberation mapping technique. In addition, due to the scattering process, the observed velocity of the broad-line region does not depend on the inclination of the broad-line region velocity field to our line of sight. However, this method does assume the broad-line region has a disk-like structure with a Keplerian velocity field. This is an assumption that, at present, is difficult to confirm although previous observations do argue that the CIV and Balmer line emitting broad-line regions predominantly move in a plane\cite{1986ApJ...302...56W, 2000ApJ...538L.103V}  and that the Hydrogen emitting regions display a velocity gradient perpendicular to the radio jet axis\citep{2018Natur.563..657G}.
 
 
 So far, the black hole mass of 30 AGN have been estimated by spectropolarimetric observations of the strong \ha{} emission line\cite{2019MNRAS.482.4985A} and one AGN by the polarization of the \mgii{} line\cite{2021ApJ...921L..21S}.
 Despite the unknown torus structure and broad-line region velocity field and the fact that the surface of the torus is not a sharp edge\cite{2018A&A...614A.120S},  mass estimates based on the Balmer lines are remarkably consistent with those measured from reverberation mapping\citep{2019MNRAS.482.4985A}. 
 While the polarization method is promising for obtaining independent mass estimates of distant quasars for comparison with other methods, one important limitation is the need for the broad-line region gas to move in planar Keplerian orbits and with limited radial motion\cite{2018A&A...614A.120S} ($\lesssim$\,2000 \kms{}). The good news is that any significant radial motion of gas will be evident from the polarized spectra.

 
 
 \subsection{X-ray variability}

It turns out that aspects of how the X-ray flux,  as emitted from a volume near an accreting black hole, varies holds information on the black hole mass. In particular, the amplitude of the high-frequency ($\nu >$\,10\,Hz) tail in the power density spectrum of X-ray variability -- observed for the hard spectral state --  scales inversely with black hole mass. This is observed over eight orders of magnitude in mass, from stellar mass black hole binaries to supermassive black holes\cite{2008MNRAS.383..741G}. 
This X-ray variability method
yields mass measurements consistent with reverberation-based masses albeit with larger measurement uncertainties and some scatter (of a factor $\sim$2 -- 3 \cite{2006MNRAS.370.1534N}).
Interestingly, Gierlinski et al.\cite{2008MNRAS.383..741G} speculate that the high frequency tail is an imprint of the last stable orbit near the black hole (see e.g., \S~\ref{sec:BHspin}).

 
 \subsection{Tidal Disruption Events  \label{sec:TDE}}
 
 If a star wanders too close to a black hole it will be captured by the black hole's gravitational potential and devoured by the black hole. Although rare, observations  of such events can be used to locate quiescent massive black holes lighter than $10^8$\msun{} and determine their mass. This holds exciting prospects as an avenue for finding intermediate mass black holes and low-mass supermassive black holes that are vital for constraining the early formation and evolution of black holes, yet difficult to study with stellar kinematics. This issue is addressed in Section~\ref{sec:imbhs}.
 
 If the black hole has a mass of $10^8$\msun{} or  higher, a captured main-sequence star will be swallowed whole\citep{2019ApJ...872..151Mockler}. For lower mass black holes the star will be tidally disrupted. In this case, some of the tidal debris will be dispelled from the gravitational potential at high speeds, while the remaining will temporarily form a gas streamer or an accretion disk before it eventually falls on to the black hole. The capture of the tidal debris will cause a luminous flare that can last for a few years\citep{1988Natur.333..523Rees}. The temporal evolution of the emission from the accreting debris will show a fast increase in intensity until it reaches a peak luminosity and will thereafter fade\citep{1988Natur.333..523Rees} with time, $t$, as  $t^{-5/3}$. 
 The shape of this characteristic light curve of the flare  (or `TDE afterglow') depends on, among other parameters, the black hole mass and properties of the disrupted star. In particular, the time scale of the initial rise to the peak luminosity correlates strongly with the black hole mass, allowing a relatively accurate mass determination\citep{2019ApJ...872..151Mockler}.  Of order 100 TDEs have been observed so far.\citep{2017ApJ...838..149Katie}
 
 High cadence monitoring campaigns aimed at identifying supernovae has recently allowed the discovery of a transient event in year 2020 where the start of the flare and thus the initial rise time could be accurately determined.  This  revealed the fastest initial rise recorded in a TDE to date. The data show the TDE to occur in a nearby dwarf galaxy of similar mass to the Large Magelanic Cloud with a $\sim 10^5$\msun central black hole.\citep{Angus2022}
 This is one of only a few dwarf galaxies with mass estimates independent of reverberation mapping, the single-epoch mass scaling, or stellar dynamical measurements.
\section{Measurements of Fundamental Properties of Black Holes: Spin \label{sec:BHspin}}
 
In addition to being a fundamental property of a black hole, the spin $a$ (or angular momentum $J$) also has astrophysical importance; the unitless spin parameter is defined as $a = cJ/GM_{\rm BH}^2$, where $G$ is the gravitational constant, $c$ is the light speed, $M_{\rm BH}$ is the black hole mass. An incredible amount of energy can be stored in the spin and released to drive physical processes, such as the launching and powering of either relativistic radio jets or accretion disk winds. The spin also carries information on the black hole growth history that thus has a 'fossil' record of how a black hole in the early Universe formed. When material accretes onto a black hole it deposits its angular momentum. Coherent accretion, through a single major accretion event or through a spinning disk, will then spin up or slow down the black hole, depending on its original spin state compared to the spin rate and direction of the infalling material. 
However, if  the black hole attained a significant fraction of its mass through chaotic accretion (i.e., significant gas parcels infalling from random directions and with random angular momentum directions) or chaotic merger of smaller black holes, then the spin is expected to be very low or non existent.
Because it is still an open question how the first black holes in the early Universe formed and evolved to the powerful quasars detected 
within $\sim$750 million years of Big Bang (\S~\ref{sec:firstBHs}), there is a strong interest in developing accurate and precise methods for determining the spin of massive black holes.

Because the black hole does not emit light by itself, we can only infer the mass $M_{\rm BH}$ and spin $a$ through their effect on the surroundings. For example, plasma orbiting close  to  the event horizon will move at relativistic speeds. The observed light from this plasma will not be evenly distributed around the black hole, but the approaching plasma will appear brighter. The  Event Horizon Telescope image of M87 thus confirms that the central black hole is spinning  (Fig.~\ref{fig:fig-M87BH}). Since the shape of the black hole shadow depends on both $M_{\rm BH}$  and $a$, the spin can in principle be constrained through detailed modeling of the light distribution, especially if the mass is determined independently and accurately\citep{2019ApJ...875L...5E} (see e.g., Chapter~7 for greater details). 


\begin{figure}
    \centering
    \includegraphics[width=5in]{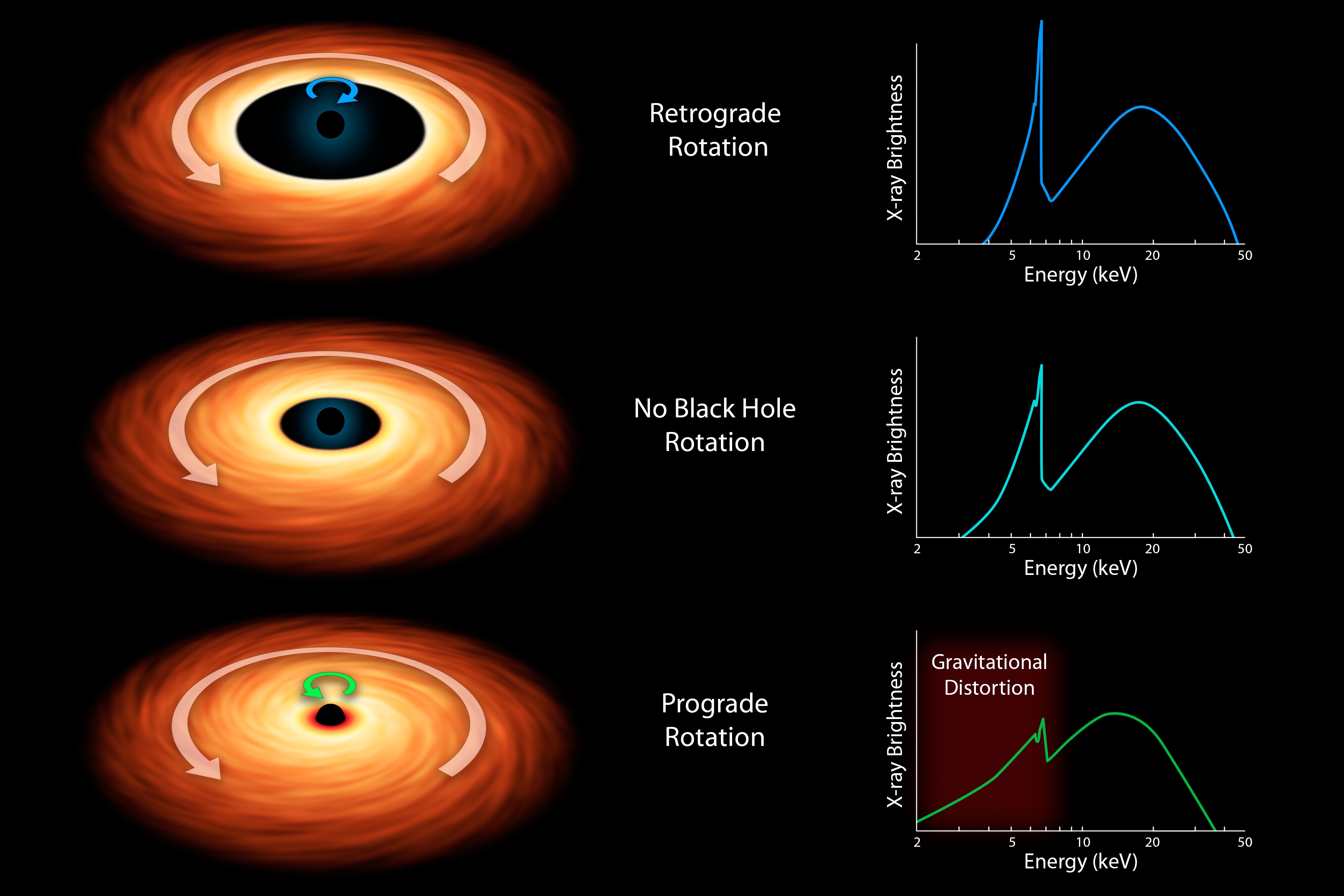}
    \caption{A black hole spinning in the same direction as its accretion disk (prograde) will pull the inner disk closer to  the event horizon.  Photons from the X-ray central corona reflected on the inner edge of the accretion disk will be subject to relativistic effects. Characteristically, broad asymmetric atomic features with a strong red tail caused by gravitational redshift will be observed (bottom panel). For retrograde spin (top panel), the innermost stable circular orbit (ISCO) will be further from the black hole than for a non-rotating case (middle panel). (Credit: NASA/JPL-Caltech)}
    \label{fig:BHspin-disk}
\end{figure}

There  are  multiple additional ways to infer or constrain the  spin  of a black hole through its effects on its environment\cite{2019NatAs...3...41R, 2021ARA&A..59..117R}. Here, we will mainly focus on the method that so far has provided actual spin estimates for massive black holes. 
This involves accreting black holes where the effects of the spin can be characterized with existing X-ray instruments.
Because the powerful gravitational field of a black hole distorts spacetime, when the black hole spins, it pulls spacetime with it in the rotation, causing 'frame-dragging'. It literally wraps spacetime closer to its central axis (Fig.~\ref{fig:BHspin-disk}). The  innermost stable circular orbit (ISCO) of accreting material will be closer to  the black hole for a spinning black hole. This inner edge of the accretion disk will reflect the high energy (X-ray) photons from  the central accretion disk corona. Atomic features, such as the iron K-$\alpha$ line at $\sim$7\,keV, act as diagnostics with prominent blueshifts and broad redshifted tails. As the black hole spin increases, the ISCO moves closer to the event horizon, the line profile broadens with a stronger redshifted tail due to the stronger gravitational redshift of the reflecting material (Fig.~\ref{fig:BHspin-disk}).

\begin{figure}
    \centering
    \includegraphics[width=3in]{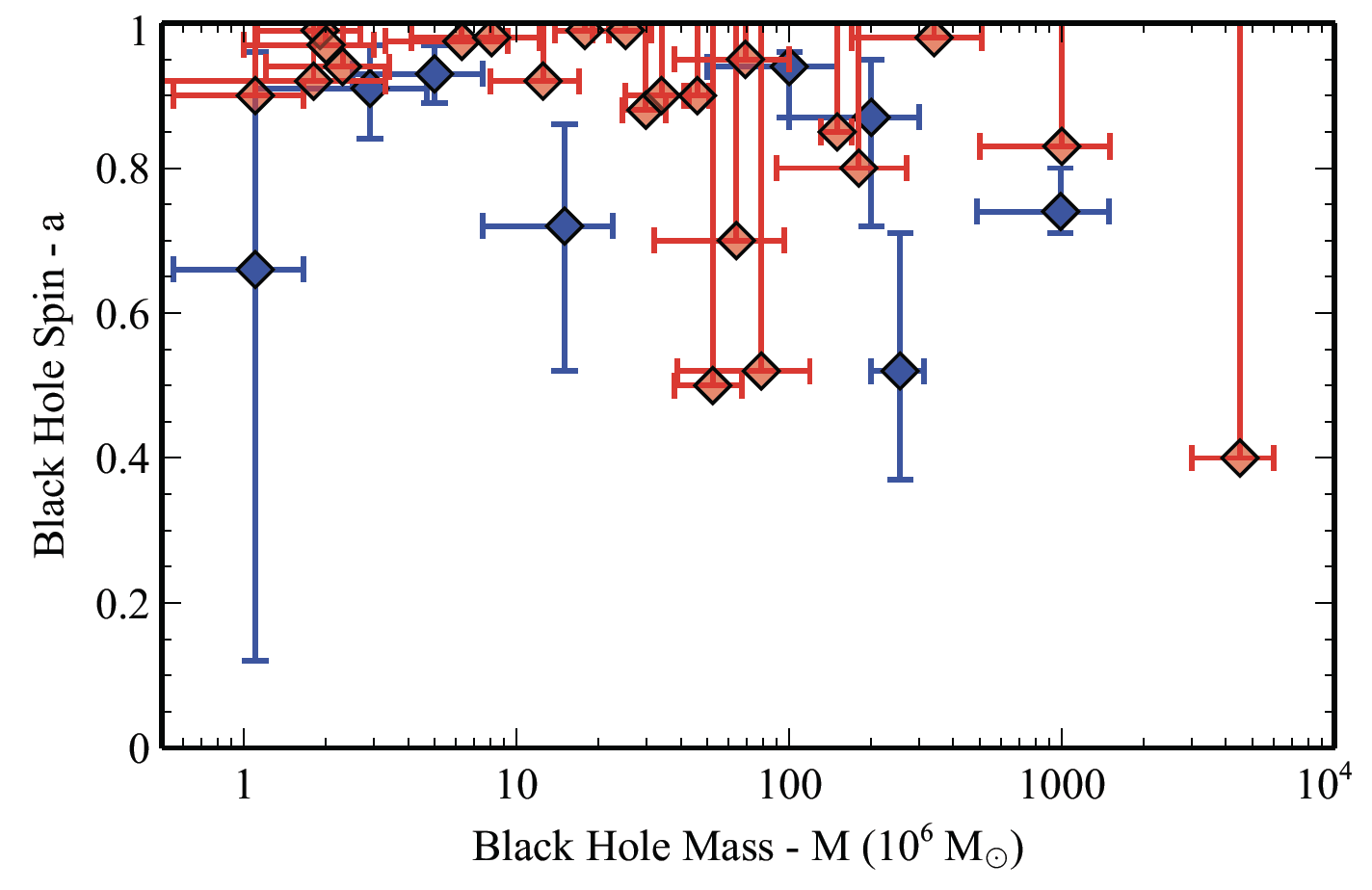}
    \caption{Spin parameter $a$ as determined from the X-ray reflection method is  shown versus black hole mass $M_{\rm BH}$ for 32 supermassive black holes with mass determinations. Red symbols denote lower limits in  $a$, blue  symbols show measurements that have  a meaningful upper bounnd (distinct  from $a$= 1).  Error bars in spin are   the 90\% confidence range while error bars in mass are 1 $\sigma$ uncertainties. (Figure 6 of Reynolds\cite{2021ARA&A..59..117R}).  }
    \label{fig:spin-mass}
\end{figure}

In a recent review, Reynolds\cite{2021ARA&A..59..117R}  notes that over 30 spin measurements have been made over the past couple of decades using the  X-ray reflection method. It might seem that black holes lighter than $\sim$30 million \msun{} tend to be fast spinning ($a > 0.9$), while the most  massive are not (Fig.~\ref{fig:spin-mass}). However, we may be seeing biases from how the objects were chosen for study (i.e.,\ selection effects). Firstly, this measurement can only be made  for accreting (i.e.,\ growing) black holes and in  this case  the higher the mass accretion rate, the brighter the AGN and the easier and more reliable is the measurement of $a$. Secondly, the most  massive black holes reside at  larger cosmic distances resulting in lower fluxes (i.e.\ lower data quality) and weaker constraints on $a$. 
Thirdly, for  a given mass accretion rate, the  accretion disk  will liberate more energy for a faster spinning black hole (the radiative efficiency depends on $a$). As a result, AGN with faster spinning black holes will naturally be over-represented in random, flux-limited samples subject to brightness limits (a common property of many astronomical object samples). Current analysis of these selection biases indicate that  the observed number distribution of $a$ is consistent with  an  equal number  of black holes with spin values between\cite{2019NatAs...3...41R} $a$=\,0.4  and $a$=\,1. 

%

The gravitational wave signal emitted when two black holes collide and merge are also affected, on the  one hand, by the spins of the two initial black holes and, on  the other hand, by the spin of the final black hole (in the 'ring-down' signal; see e.g., Chapter~8). However, extracting this information from the weak signal is challenging\cite{2021ARA&A..59..117R}. While efforts are underway to improve measurement methods, scientists will have to  await  the launch of  the Laser Interferometer Space Antenna (LISA) to have the hope of measuring the spin of colliding  supermassive black holes because ground-based gravitational wave detectors operate at a frequency insensitive to mergers of massive black holes.

%

\section{Characterizing the black hole local environment}
\label{sec:characterizing}
   
Owing to the tiny angular scales of the black hole and its immediate environment, it is only in the last few years that technological advances have allowed extremely high resolution observations that are needed to obtain definitive insight on the structure and physical properties of this environment, so important for how  the black hole feeds, grows and generate jets and/or winds that deposit enormous amounts of energy in its host galaxy and beyond. Previously, scientists have had to infer the physics through the observed time variability of the electromagnetic emission and by comparing observations with theoretical models of the  physics of the expected emission and absorption components and their properties. For these reasons, our knowledge of the  structure and detailed physics of these components,  such as  the X-ray corona, the accretion disk/flow, the broad emission line region, potential disk winds and/or the launching and collimation of radio jets, is generally  limited. In this section, we provide a brief overview of the recent advances obtained through these new technologies.
  \subsection{Milky Way / Sgr\,A* \label{sec:MWBH}}
  The proximity of the Galactic Center makes it an unparalleled site to study an individual black hole's immediate surroundings. 
     \subsubsection{Dynamical Mass Measurement from Individual Stars}
  As discussed in \S\ref{sec:mass}, mass measurements are one of the fundamental measurements we can make of black holes, and the mass of the black hole in the Galaxy can be measured like no other.  With adaptive optics and a large aperture telescope, it is possible to measure the positions and proper motions of individual stars in orbit around the black hole.  With multiple epochs, the orbits are well mapped out and appear to be ellipses with a common focus.  Such orbits describe a Keplerian motion around a point mass.
  Efforts of two independent research teams\cite{2008ApJ...689.1044G, 2010RvMP...82.3121G} using the 10\,m Keck telescope in Hawaii and the 8.4\,m Very Large Telescope in Chile over almost two decades have provided us with exquisite precision measurements that allow accurate  determination of the black hole mass\cite{2008ApJ...689.1044G, 2010RvMP...82.3121G} ($M_{\rm BH} = 4.1 \pm 0.4 \times 10^{6}\ \msun$), confirmation of general relativistic  effects such as gravitational  redshift\cite{2018A&A...615L..15G,  2019Sci...365..664D} and Schwarzchild orbital precession of stellar orbits \cite{2017PhRvL.118u1101H, 2020A&A...636L...5G}, the distance to the Galactic center\cite{2008ApJ...689.1044G} ($d = 7.7 \pm 0.4\ \units{kpc}$) plus novel insight on the small scale environment near the central mass.

  
  
     \subsubsection{RIAF/ADAF}

     \begin{figure}
      \centering
      \includegraphics[width=0.9\textwidth]{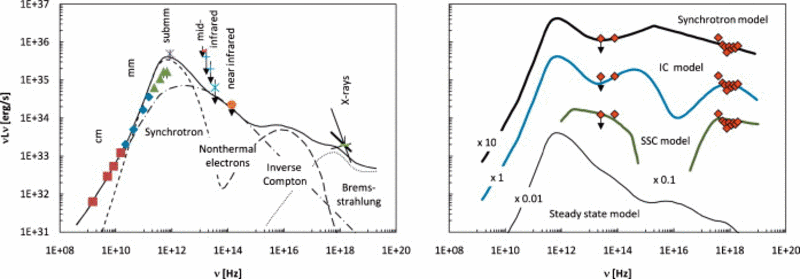}
      \caption{Spectral energy distribution of Sgr A* showing its accretion mode.\cite{2010RvMP...82.3121G}.
      }
      \label{fig:sgrastarsed}
  \end{figure}
     The proximity of Sgr A* allows us to study the accretion physics of a single source at extremely low luminosity.  The mode of accretion is thought to be one of a variety of radiatively inefficient accretion flow (RIAF) models, an early example of which is the advection dominated accretion flow (ADAF) model.  We show the spectral energy distribution (SED) of Sgr A* and representative models in Fig.\ \ref{fig:sgrastarsed}.  Observationally, Sgr A* shows a steady emission from radio to X-ray that is thought to be a contribution of synchrotron emission from thermal and non-thermal electrons plus inverse Compton scatterings by the thermal electrons and free-free emission.  In addition to the steady-state emission, Sgr A* is moderately variable and shows flaring activity in the infrared.  
     
     There is a qualitative difference in the emission seen in Sgr A* and other extremely low luminosity accreting black holes.  The cause of the difference is likely that the temperature of the gas is much higher than is seen in thin-disk or slim-disk accretion.  Many RIAF models invoke a two-temperature plasma with ions having much higher temperature than the electrons.  For a fuller discussion of accretion flows at very low luminosities, we refer the reader to reviews by Yuan \& Narayan\cite{2014ARA&A..52..529Y} and Genzel et al.\cite{2010RvMP...82.3121G} and Chapter~7 of this volume.


  \subsection{Messier 87 \& The Event Horizon Telescope }
  \label{sec:characterizing-m87eht}
\begin{figure}
      \centering
      \includegraphics[width=4.5in]{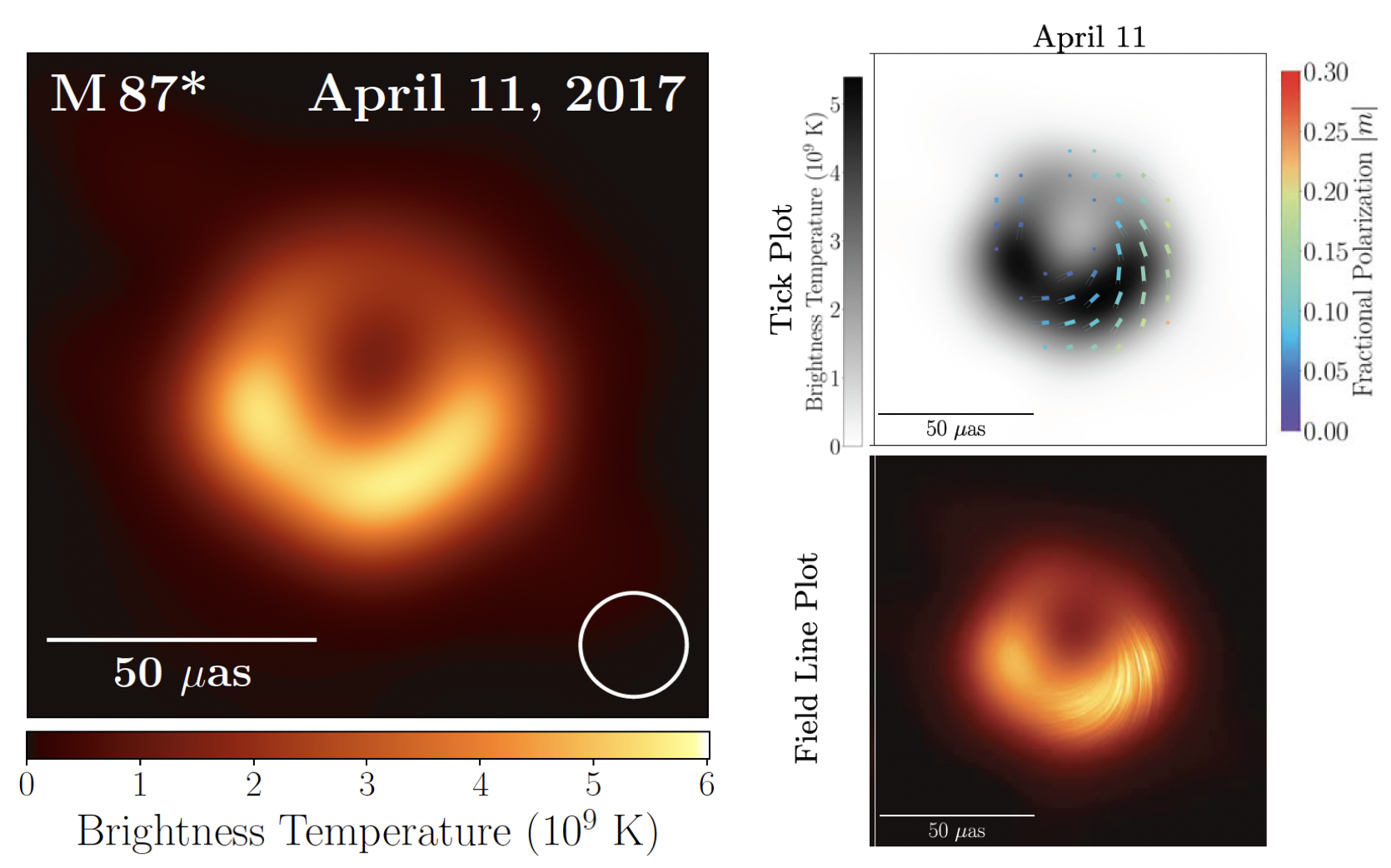}
      \caption{{\it Left panel:} The first-ever spatially-resolved image of a black hole was obtained with the Event Horizon Telescope (EHT) in 2017. This synthesized image shows radio (synchrotron) emission from plasma swirling in close proximity to the event horizon of the supermassive black hole of Messier 87, a giant elliptical galaxy in the Virgo cluster of galaxies. The central shadow is characteristic of a black hole, since it has no surface on which the light from nearby gas can be reflected\cite{1997ApJ...478L..79N}. The size of the dark disk is proportional to the black hole mass and spans  approximately 2.6 times the diameter of the black hole, as characterized by the event  horizon. The shown panel is a subset of Figure 3 of EHT Collaboration et al. \citep{2019ApJ...875L...1E}. 
      {\it Right panel:} Synthesized image of the polarized emission near the event horizon of the M87 central black hole. The polarization is not uniform across the bright ring of light, but rises to $\sim$15\% in the southwest part of the ring (i.e. lower right). The swirling pattern shows the direction of the polarization: i.e., the polarization angle changes with radial distance. This right panel is a sub-panel of Figure 7 of EHT Collaboration et al.\cite{2021ApJ...910L..13E}. 
      }
      \label{fig:fig-M87BH}
  \end{figure}

  While scientists have long suspected the existence of black holes (\S\ref{sec:intro}), 
  the first definitive and direct evidence of the existence of black holes came with the first direct detection in September 2015 of gravitational waves from the merger of two compact objects (Abbott et al.  2016) with a waveform that indisputably confirmed that the collision occurred between two black holes with masses of 36\msun{} and 29\msun{} (See also Chapter~8). 
  The first direct evidence of the existence of {\it supermassive} black holes came with the first spatially resolved image ever obtained of a compact object. The image was synthesized from data obtained with the Event Horizon Telescope (EHT), an international collaboration that uses a global network of sub-millimeter wavelength telescopes\footnote{https://eventhorizontelescope.org/} that can work together as a single telescope with the size of Earth, using the technique of interferometry. In April 2017, EHT observed the central compact object in Messier 87 (or M87), a giant elliptical galaxy 53.5 million light years away in the Virgo galaxy cluster, at 1.3 millimeter wavelengths at the exquisite angular resolution\citep{2019ApJ...875L...1E, 2019ApJ...875L...2E} of 25 $\mu$arcsec. 
  The image -- presented to the community at an international press conference on April 10, 2019 -- shows a clear, central dark shadow surrounded by a glow from plasma accreting onto the black hole (Fig.~\ref{fig:fig-M87BH}). 
  A high precision measurement of the black hole angular gravitational radius of $GM/Dc^2 = 3.8 \pm 0.4 \,\mu$as and mass of $(6.5 \pm 0.7) \times 10^9$ \msun{} was obtained from the angular size of the shadow\citep{2019ApJ...875L...6E} ($D = 16.8 ^{+0.8}_{-0.7}$\,Mpc is the distance to M87). 
  This confirms that M87 possesses one of the most massive black holes in the Universe known to us. Such large black holes are also seen in the distant powerful quasars detected at cosmic distances (that is, the black holes powering the quasars are active at earlier epochs in the history of our Universe\cite{2020ARA&A..58...27I} also discussed in Chapter 9). With the clear detection of a radio jet emanating from the center of M87\cite{2018ApJ...855..128W}, there is little doubt that M87 must have been a powerful quasar earlier in its lifetime.
  
  This independent, high-precision black hole mass measurement from EHT is important as it helps to settle the inconsistency between earlier mass measurements of M87 based on two different dynamical methods, namely based on stars\cite{2011ApJ...729..119G} and ionized gas\cite{2013ApJ...770...86W}. The EHT measurement lends support to the notion that stellar kinematics provides a more robust mass determination. However, time and further analysis will reveal if also the ionized gas dynamical method can be improved to provide consistent results. 
  

  The non-uniform illumination of the glow around the black hole shadow (Fig.~\ref{fig:fig-M87BH}) is caused by rotation of the black hole\citep{2019ApJ...875L...1E, 2019ApJ...875L...5E}, 
  which drags spacetime with it at relativistic speeds, causing the plasma approaching us along our line of sight to appear brighter (\S~\ref{sec:BHspin}).  While the axis of rotation is not uniquely determined from the EHT data, it is consistent with a direction parallel to the radio jet observed at parsec scales\cite{2018ApJ...855..128W, 2021ApJ...911L..11E}.  The exact mechanism for jet launching has not yet been settled, but the data are consistent\citep{2019ApJ...875L...5E} 
  with the extraction of spin energy from the black hole by strong magnetic fields through the Blandford-Znajek mechanism\cite{1977MNRAS.179..433B}.

  In March 2021 the EHT collaboration published the first image of polarized light from the plasma near the event horizon of the black hole, based on further analysis of the 2017 data\cite{2021ApJ...910L..12E, 2021ApJ...910L..13E}. 
  The polarized light carries important information on the strength and direction of the magnetic fields near the black hole event horizon, suspected crucial for the launching of the relativistic radio jet.  While analysis of the signal is not trivial, it does show a strong (1--30 Gauss) poloidal (i.e., radial and vertical) magnetic field component, subjected to depolarization (i.e., a weakening of the polarization degree) by electrons in and around the synchrotron-emitting plasma. General relativistic magneto-hydro-dynamical modeling shows that the accreting material is strongly magnetized (a 'magnetically arrested accretion flow') and that the magnetic field is dynamically very important. The mass accretion rate is inferred to be a feeble $\sim$\,$10^{-4}$\msun/year (i.e., 1 \msun{} per 10.000 years).

  At present, the shadow of supermassive black holes can only be imaged in spatially resolved observations for two galaxies, namely M87 and our Milky Way galaxy\footnote{EHT observed the Milky Way black hole, Sgr A*, in 2017 along with M87.  The image was presented to the public in a press release on May 12th, 2022. The image reveals a shadow in the center of  a  bright ring of emission.  The image confirms that the central dark object in Sgr A* is indeed a black hole and that the physics of supermassive black holes is similar on  the scales of this shadow despite the  factor 1600 difference in mass between the M87 and Sgr A* black holes.  See Chapter 7 for more.}. There are other prospects for similar observations such as using lensing in binary SMBHs\cite{2022PhRvL.128s1101D}.
  The EHT collaboration is working to improve the resolution by operating more antennas at higher frequencies and by adding further telescopes to the network.
  While the ability to resolve the black hole shadow in distant galaxies depends strongly on the distance and the mass of the black hole, these upgrades will increase the number of nearby AGN for which we can hope to study the properties of the central radio plasma at scales slightly larger than the event horizon, but still sufficiently close to the black hole to provide new insight on the physics of the central engine.


  

  \subsection{Results from the GRAVITY instruments \label{sec:AGN-gravity}}
  The infrared interferometry instrument GRAVITY at the European Southern Observatory's Very Large Telescope   in Chile was one of the instruments used to map the motion of stars near the Milky Way black hole (\S~\ref{sec:MWBH}). This instrument has also provided ground-breaking insight on the structure of the black hole central engine powered AGN and quasars.
  
      \subsubsection{The size and structure of the hot dust component}
\begin{figure}
      \centering
      \includegraphics[width=5in]{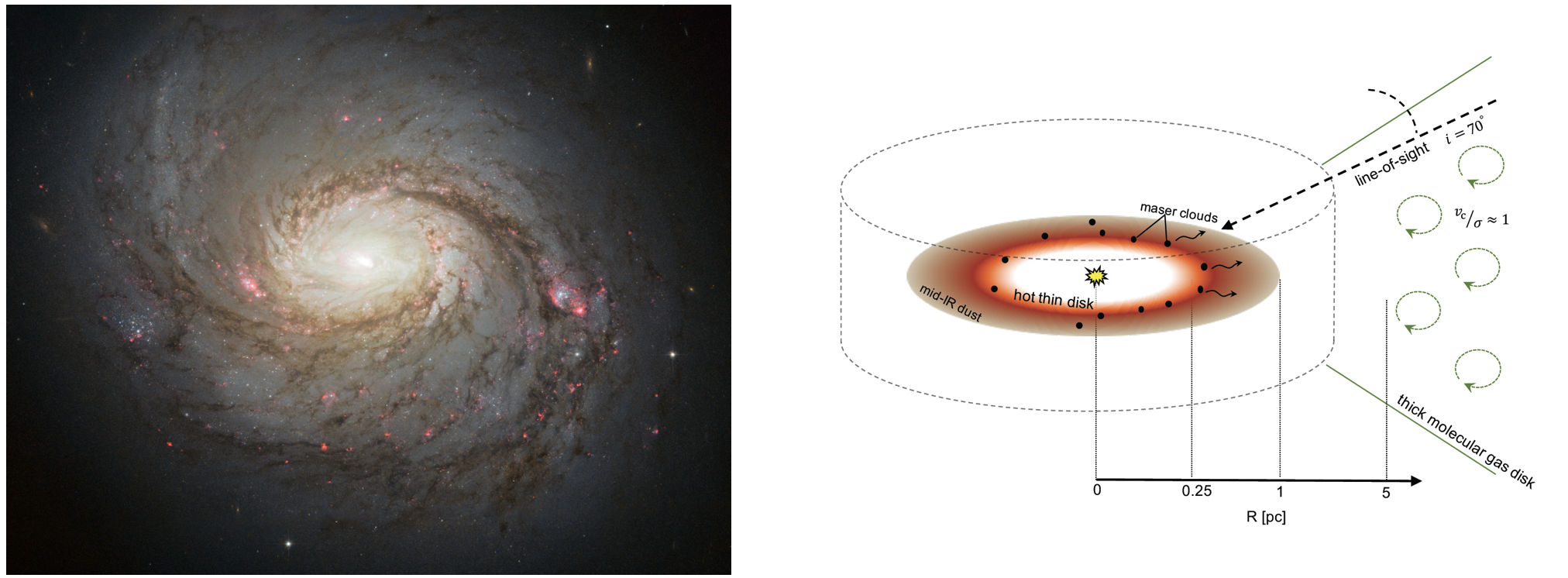}
      \caption{{\it Left panel:} Composite image, obtained by {\it the Hubble Space Telescope}, of nearby galaxy NGC 1068 that harbors a hidden supermassive black hole in  its center. Blue colors represent bright young stars that ionize gas in their vicinity, which then glows by emitting Balmer line emission (red hues). Dark bands are dust and dense gas that absorb the blue stellar light.
      Credit: NASA, ESA \& A. van der Hoeven. {\it Right panel:} Sketch illustrating the central structures observed by GRAVITY. The $K$-band emission from hot dust traces the inner regions of a thin disk of hot dust and gas near the dust sublimation radius ($\sim0.25\,pc)$. The hot dust is co-spatial with the central water masers that reside in the thin disk, stretching out to 1 parsec (Figure 7 of Pfuhl et al.\cite{2020A&A...634A...1G}). The purported thick dusty torus resides at larger radial distances from  the black hole.}
      \label{fig:HST_ngc1068}
  \end{figure}

  The first results from observing AGN with GRAVITY were presented on the nearby, archetypical Seyfert type 2 galaxy NGC\,1068, a well studied AGN and starburst galaxy located circa 45 million light years away (Fig.~\ref{fig:HST_ngc1068}). The observations, at an unprecedented 0.2 parsec angular resolution at a wavelength of 2 $\mu m$, show the dust-sublimating region (i.e., the hot dust near the sublimation radius) as a thin ring-like structure\cite{2020A&A...634A...1G}, residing at the inner rim of the thin disk occupied by water masers. This result was quite surprising because it had long been expected that the hot dust, observed at wavelengths longward of the $1\mu m$ inflection in the spectral energy distribution (Fig.~\ref{fig:elvisSED}), take origin in the inner face of a geometrically thick torus of dense gas and dust (e.g., Fig.~\ref{fig:polariz}). This torus blocks our direct view of the central regions in this highly inclined AGN. Instead, a thin disk of dense gas and dust exists inside this torus. This disk is not to be confused with the central accretion disk residing much closer to the black hole, emitting strongly in the UV-optical regime. 

  
  A sample of eight additional nearby AGN were observed with GRAVITY during 2017--2019\cite{2020A&A...635A..92G}, partially resolving the hot dust emitting region. The measured size of the hot dust region, in the range 0.3 -- 0.8 $\mu$arcsec, is seen to increase with the AGN bolometric luminosity over four orders of magnitude. This confirms that the dust sublimation radius increases with increasing AGN power, as previously established based on measured time delays\cite{1989ApJ...337..236C,2014ApJ...788..159K, 2014ApJ...784L...4H, 2014cosp...40E1215H}. 

     \subsubsection{Size and velocity structure of the broad emission line region}
     With  a size of less than 0.1 milli-arcsecond for even the most nearby AGN, the broad emission line region is too small to be imaged directly with individual telescopes; its size is expected to be smaller than the dust-sublimation radius. Until recently, scientists were primarily limited to studying the physics of this region through variability of the continuum and line emission (\S~\ref{sec:RM}) or via analysis of spectroscopic data. Given these constraints, the nature of this region, its geometry and kinematics, and its relationship to the rest  of the central engine are still unknown. For example, it is still debated whether the broad-line region is (a) the atmosphere of the accretion disk or of the inner face of the obscuring  torus, (b) an integral part of the accretion flow; (c) part  of an outflowing disk wind; or (d) if it is a separate entity all together. 
     However, observations with GRAVITY are now allowing scientists to peer deeper into the heart of AGN, to map the dominating velocity structure of the broad-line region, providing the first direct measurements of spatially resolved velocity gradients across the broad-line region\cite{2018Natur.563..657G}. The GRAVITY instrument combines the light from the four 8 meter VLT telescopes to obtain interferometric amplitudes  and phases. The amplitudes measure the angular extent of the broad-line region while the phase determines the position on the sky.


     \begin{figure}
         \centering
         \includegraphics[width=3in]{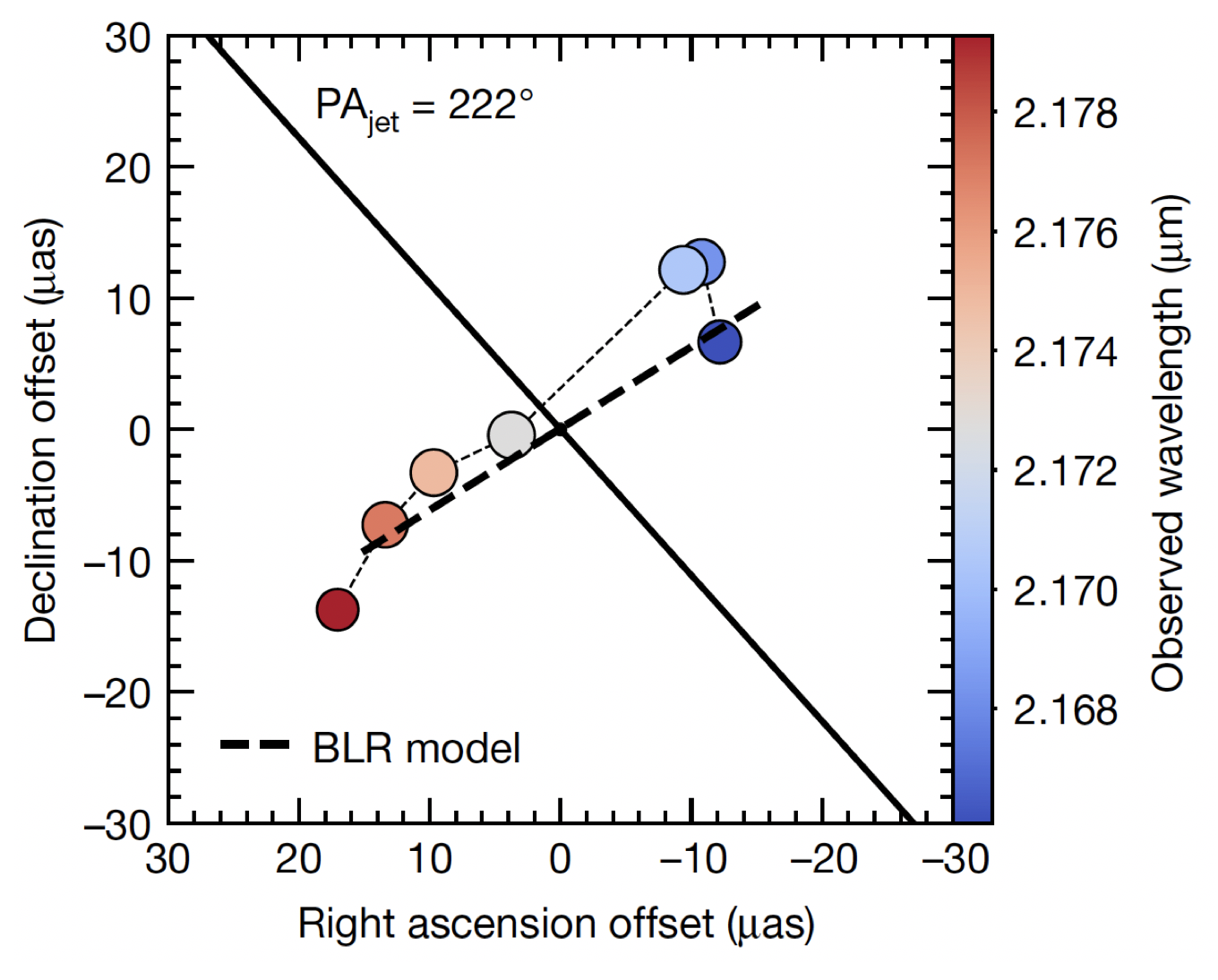}
         \caption{Spatially resolved velocity gradient of the broad-line region of the radio-loud quasar, 3C\,273, as observed by GRAVITY in  the Hydrogen Paschen alpha emission line. The velocity gradient shows that the broad-line region is dominated by ordered motion around the central black hole in a plane essentially perpendicular to  the radio jet axis. (Figure 1 of Sturm et al.\citep{2018Natur.563..657GSturm})}
         \label{fig:3c273vel}
     \end{figure}
     
     The first GRAVITY observations of a powerful AGN with broad emission lines were made of 3C\,273, one of the most powerful radio-loud quasars. As the first quasar to be discovered~\citep{1963Natur.197.1037H, 1963Natur.197.1040S}, 3C273 is in our local extragalactic neighborhood (at a redshift of 0.158 or a co-moving distance of 660 million parsec, corresponding to a distance of 2 billion light years). 
     The phase shifts of the Hydrogen Paschen alpha (Pa-$\alpha$) emission line specifically show the broad line gas in ordered rotation around an axis parallel to the radio jet (Fig.~\ref{fig:3c273vel}). This confirms the conclusions reached in previous studies that find larger velocity dispersion in the Hydrogen Balmer beta (H\,$\beta$) and \civ{} $\lambda$ 1549 broad emission lines for  radio sources that are more  inclined relative to our line of sight\citep{1986ApJ...302...56Wills,2000ApJ...538L.103Vestergaard}.
     The GRAVITY observations also clearly establish that the broad-line region is physically separated from the obscuring torus. The mean size of  the Pa-$\alpha$ emitting region is measured to be 0.12$\pm$0.03 parsec with an inner radius of 0.03 $\pm$0.01 parsec, a factor of 3 smaller than the size of the continuum dust radius measured with the (earlier) VLT interferometry instrument AMBER\citep{2018Natur.563..657GSturm}. 

     The more recent GRAVITY observations of two other AGN, IRAS 09149-6206 and NGC\,3783, paint a  similar general picture of a disk-like broad-line region in orbit around the black hole\citep{2021A&A...648A.117GNGC3783} 
     that is much more compact than the hot dust emitting region.  
     Dexter et al.\citep{2020A&A...635A..92GDexter} also find the broad-line region sizes of five other AGN, observed with GRAVITY, significantly smaller than the hot dust emitting region. 
     It will be interesting to see if future GRAVITY observations can shed more light on the relationship between the hot dust emitting component in the central engine and the broad-line region.



\section{Co-evolution of Black Holes and Host Galaxies}
\label{sec:coevolution}
Tight correlations between the properties of black holes and the galaxies that host them strongly indicate that the evolution of galaxies and their black holes are strongly entwined.  The most compelling set of evidence in favor of co-evolution is the set of scaling relations between the black hole mass and the host galaxy properties.  The mass of black holes ($M_{\rm BH}$) can be measured (see \S\ref{sec:mass}) and when it is plotted against some galaxy properties, there is a strong positive correlation.  These correlations are most important with galaxy bulge luminosity ($M_{\rm BH}$--$L_{\mathrm{bulge}}$ relation), galaxy bulge mass ($M_{\rm BH}$--$M_{\mathrm{bulge}}$ relation), and galaxy bulge velocity dispersion ($M_{\rm BH}$--$\sigma_{\ast}$ relation).  

The relations between black hole mass and host galaxy property are all well described by power laws of the form $M_{\rm BH} = A X^\beta$, where $X$ is one of the galaxy properties, $A$ is the intercept, and $\beta$ is the power-law index.  The best-fit  values of $A$ and $\beta$ for these relations has been updated many times over the years both with increased numbers of black hole mass measurements and more informed isolation of the sample.  The current understanding is that the black hole scaling relations hold best for ellipticals and classical bulges (not pseudo-bulges) while omitting galaxies that are `mergers in progress,' which have not settled into a final elliptical/classical bulge\cite{2013ARA&A..51..511K}.  With such a sample, the $K$-band $M_{\rm BH}$--$L$ relation is 
\begin{equation}
    M_{\rm BH,9} = (0.554^{+0.067}_{-0.059}) L_{K,\mathrm{b},11}^{1.22 \pm 0.08},
\end{equation}
where $M_{\rm BH,9}$ is the black hole mass in units of $10^9\ \msun$ and $L_{K,\mathrm{b},11}$ is the bulge $K$-band luminosity\citep{2013ARA&A..51..511K} in units of $10^{11}\ L\subsun$.  The $M_{\rm BH}$--$\sigma$ relation is 

\begin{equation}
M_{\rm BH,9} = (0.310^{+0.037}_{-0.033}) \sigma_{200}^{4.38 \pm 0.29},
\end{equation}
where $\sigma_{200}$ is the bulge stellar velocity dispersion\citep{2013ARA&A..51..511K} in units of $200\ \kms$.  Both of these relations have intrinsic scatter measured to be about 0.3 dex.


   \begin{figure}
       \centering
       \includegraphics[width=0.49\textwidth]{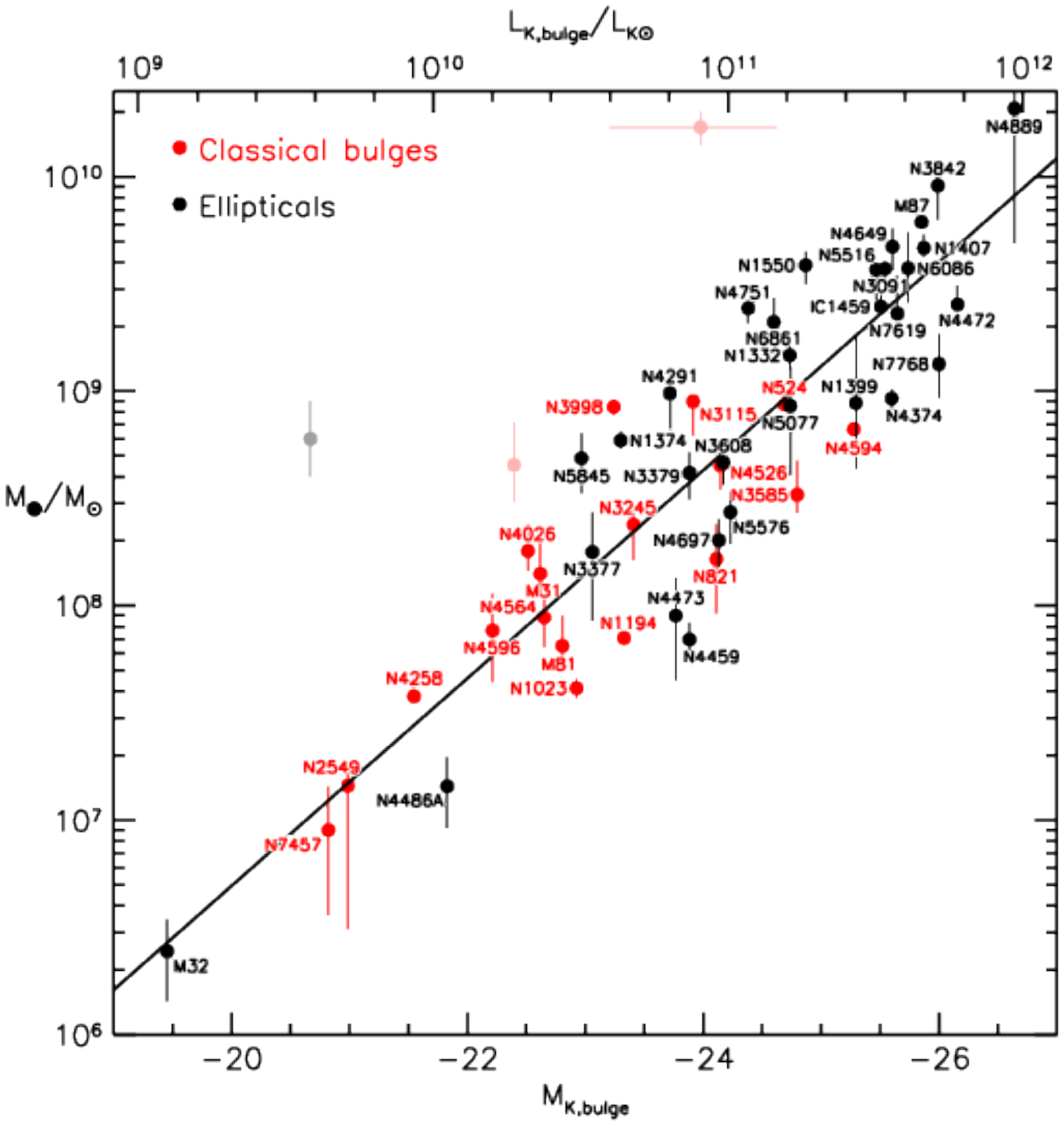}
       \includegraphics[width=0.49\textwidth]{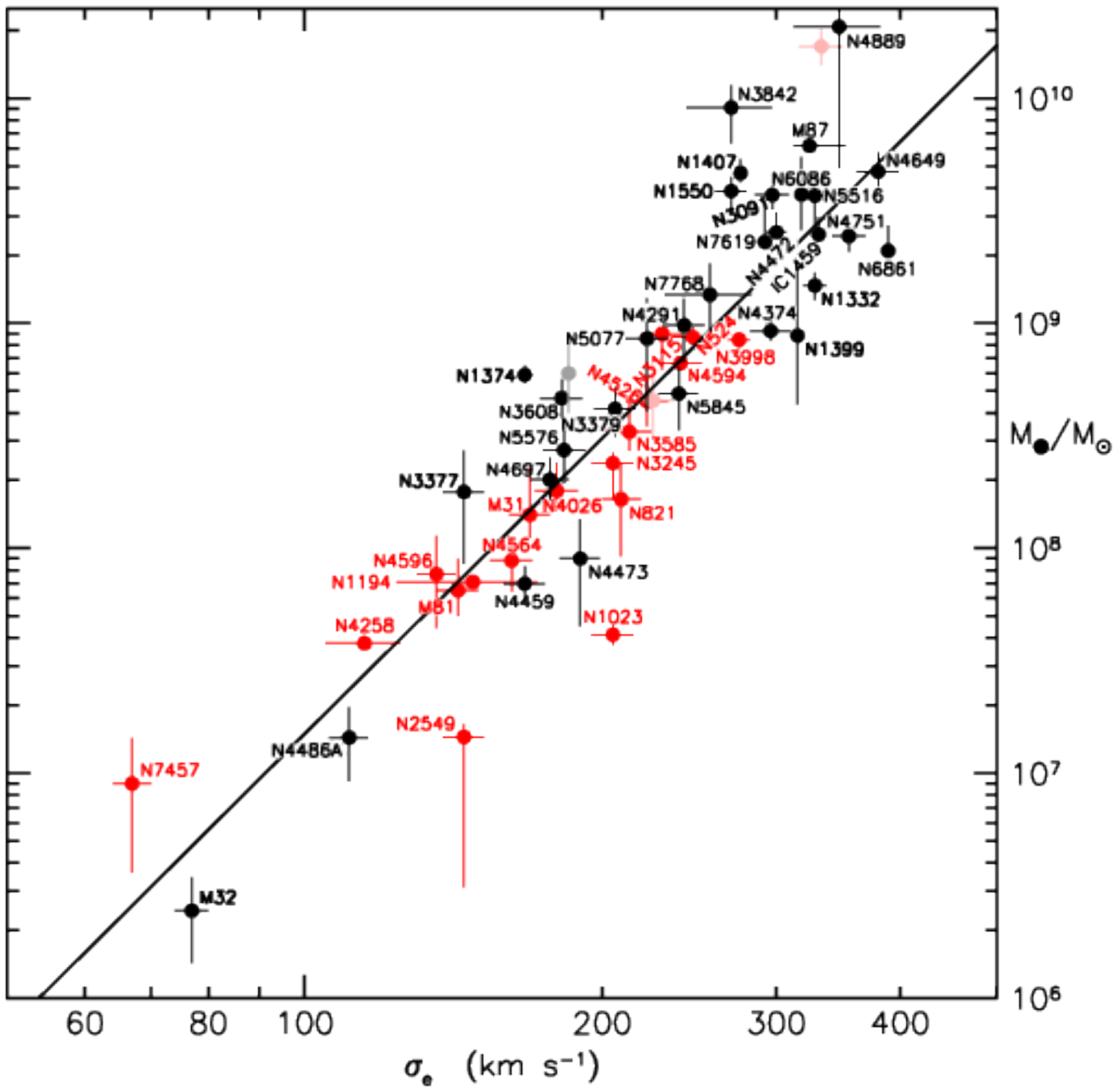}
       \vskip -1.6cm
       \caption{$M_{\rm BH}$--$L_K$ and $M_{\rm BH}$--$\sigma$ relations for dynamically measured black holes. The relation is shown only for classical bulges and elliptical galaxies as pseudobulges do not show the same low-scatter relation. (From Kormendy \& Ho 2013)}
       \label{fig:mlmsigma}
   \end{figure}

That there is a positive correlation between $M_{\rm BH}$ and any of these is expected: it makes sense that the largest black holes would be in the largest galaxies.  What is surprising is the relatively small intrinsic scatter in these relations despite the vast differences in size of black holes and size of galaxy bulges.  A $10^{8}\ \msun$ black hole has Schwarzschild radius approximately 2 Astronomical Units (AU), and a typical host galaxy of such a black hole would be an elliptical with half-light radius of order $\sim1$ kpc.  So even though the black hole  size is $\sim10^{8}$ times smaller than the host galaxy, it is possible to predict the black hole mass to within a factor of 2--4 by measuring the host galaxy properties.  This leads us to conclude that there is something about the evolution of the two that is tightly coupled.

While the evidence for coevolution in the form of the $M_{\rm BH}$--$\sigma$ relation was originally defined by quiescent black holes in our local Universe, there is evidence that this relationship also holds for actively accreting black holes, as one would expect if they are the same beasts.  For the few nearby AGN for which reliable bulge stellar velocity dispersion, $\sigma_{\ast}$, measurements can be made (given the bright nuclear source), the black hole masses determined by reverberation mapping (\S~\ref{sec:RM}) scale with  $\sigma_{\ast}$ with a similar slope and scatter ($\sim$0.3\,dex) as observed for quiescent galaxies\citep{2001ApJ...555L..79Ferrarese,2004ApJ...615..645Onken}. However, the 'AGN-only relation' has a slight zero-point shift toward lower black hole mass, consistent with the fact that the black holes are still growing. For convenience, reverberation-based masses are at present bootstrapped to the $M_{\rm BH}$--$\sigma_{\ast}$ relation for quiescent black hole  (\S~\ref{sec:f-factor}). 


Most theoretical ideas about the underlying cause of black hole coevolution use some form of self-regulated feedback.  As gas in a galaxy halo cools and falls into the stellar component of the galaxy, much of it will form stars, and some of it will fall all the way to the center onto the black hole.  This gas thus adds to the stellar mass and light as well as the black hole mass.  Both the formation of stars and black hole accretion can lead to processes that prevent further falling of galactic halo gas onto the galaxy.  Star formation is accompanied by supernovae, which can heat up and/or expel the gas from the galaxy bulge.  Black hole accretion is accompanied by powerful quasars that can drive strong gas outflows or by long-lived jets that continuously deposit energy and momentum into the interstellar medium (see next section for details).  Thus black hole accretion can also heat up and/or expel gas from the galaxy bulge.  If either process (star formation or black hole accretion) is sufficiently efficient, then it will cut off the supply of cold gas that is forming stars and feeding the black hole -- ending the growth of both the stellar component(s) and the black hole.  This general picture is likely to be correct in some way, but the details are poorly understood.  For example, it is thought that jet-driven feedback is the dominant process in the largest galaxies and that star-formation-driven feedback is dominant in the smallest galaxies.  It is possible, however, that even the smallest galaxies require some form of AGN feedback to fully explain the galaxy luminosity function.  The final conclusion is the same regardless of the driver of the feedback: the source of mass for the black hole and the stellar bulge is the same (namely, halo gas), and the termination of the mass accretion of black hole and stellar bulge is the same.  Thus both black hole and galaxy stellar content have coevolved to their $z = 0$ state.


Direct observational evidence of the processes responsible for the co-evolution is lacking. 
However, some  insight into the underlying cause of the scaling relations may be gained by looking at the outliers in the relations.  One important category of outliers is pseudobulges.  Classical bulges are created relatively quickly out of mergers between disk galaxies and result in kinematically hot and geometrically round distribution of stars at the center of the galaxy.  Pseudobulges, however, appear superficially to resemble classical bulges, but were formed over a much longer time through a variety of non-merger secular processes and remain very disky in terms of their kinematics and surface brightness profiles\citep{2004ARA&A..42..603K}.  The black holes in pseudobulges tend to fall below (sometimes far below) the black hole scaling relations when compared to classical bulges and elliptical galaxies.  Because classical bulges and elliptical galaxies are the result of mergers, it seems reasonable to look for an underlying connection through the physics of mergers, which are especially efficient at bringing gas from the outer regions of a galaxy to the center.

It is worth noting that our evidence of coevolution comes almost entirely from $z=0$ sources even though the coevolution must have happened over a long timescale.  Dynamical black hole mass measurements (\S~\ref{sec:mass}) can only currently be  made  for the nearest galaxies, and reverberation mapping at present is by and large bootstrapped to the scaling relations (\S~\ref{sec:f-factor}).  On top of this, many low-luminosity AGN are  
disproportionately found in pseudobulges, which do not follow the same relation. Measurements of the velocity dispersion of CO molecular gas, as a proxy attempt to gauge the bulge velocity dispersion, at early epochs (redshifts 5 --  6) show low velocities, suggesting significantly overmassive black holes\citep{2019ApJ...880....2WangF}. The strong caveat here is that the dynamics of these young systems are not understood. In fact, the gas may not be virialized but still be chaotically infalling  as part of the galaxy  assembly process\citep{2019ApJ...887...40WangR}. So thus far we have less observational certainty as to what the black hole scaling relations look like at $z > 0$.  With the launch of the {\it James Webb Space Telescope} and the eventual construction of 30-m class ground-based telescopes with adaptive optics (\S~\ref{sec:future}), it will be possible to make direct dynamical measurements of $M_{\rm BH} > 10^9\ \msun$ black holes out to a redshift of $z \approx 1.5$.  With such measurements it will be possible to learn whether black holes or the galaxies grow first.

\section{Black hole fueling mechanisms at low redshift \label{sec:fueling}}
What makes a black hole shine as an AGN  -- and thus be detectable at cosmic distances -- is the gas that it feeds on. As noted earlier (\S~\ref{sec:intro}), there is a direct link between the mass accretion rate $\dot{M}$ and the luminosity $L$ of the AGN: $L = \eta \dot{M} c^2$, where $\eta \approx 0.1$ is the energy conversion efficiency and $c$ is the light speed.   Black holes in local Seyfert galaxies ($L \lesssim 10^{44}$\,\ergs) typically accrete $10^{-3} - 10^{-2}$ \msun/year while $\sim$10 \msun{} per year is needed to power luminous quasars\citep{1997iagn.book.....Peterson} ($L \geq 10^{46}$\,\ergs). Because high spatial resolution is required to study the fueling mechanisms, we predominantly have insight from  AGN at low redshifts.


\begin{figure}
    \centering
    \includegraphics[width=9cm]{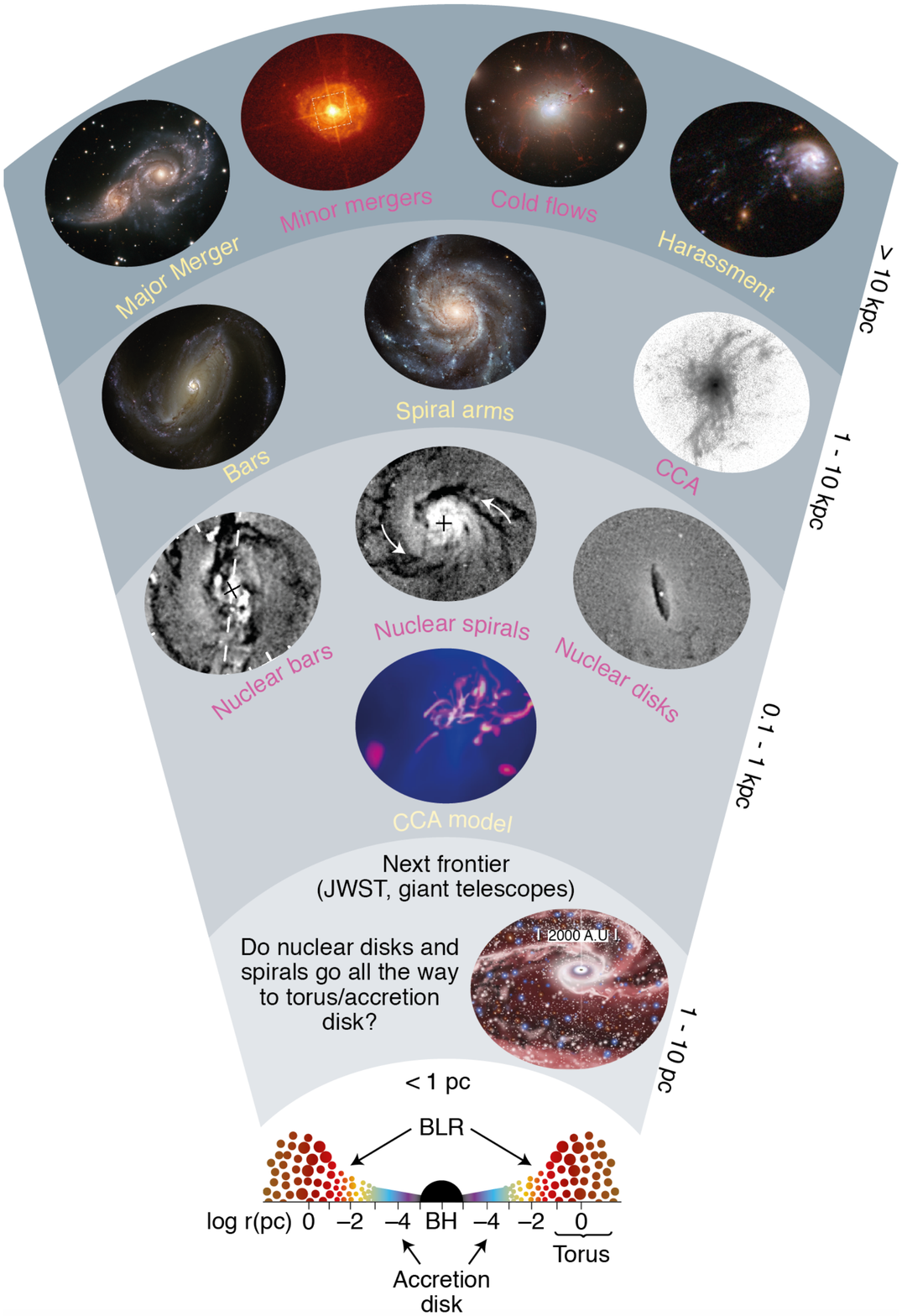}
    \caption{Summary of the expected mechanisms that cause gas inflows at different spatial scales, measured from the black hole.   Galaxy interactions are responsible for galaxies gaining external gas for group galaxies, while cold gas streamers (chaotic cold accretion, CCA) occur in larger groups and galaxy clusters.  Internal (secular) processes, like spiral gas structures, stellar bars and nuclear disks likely provide the necessary negative torques on smaller scales that allows the gas to sink to the very center, and feed the central supermassive black hole. Adapted from Figure 8 of Storchi-Bergmann \& Schnorr-M\"uller~\cite{2019NatAs...3...48S}}
    \label{fig:BH-fuelling}
\end{figure}

%

Different mechanisms can drive gas from large spatial scales (in the host galaxy or beyond) down to the immediate black hole vicinity. The specific mechanism in action depends largely on the environment of the host galaxy, whether it resides isolated in the field, in a group of galaxies or in a dense environment, like a galaxy cluster.  A black hole in an isolated galaxy can only  feed on gas already in the galaxy  (or near the galactic halo), while a host  galaxy in a denser environment can obtain additional gas through cold gas streamers in the hot intergalactic medium in clusters (a.k.a. chaotic cold accretion) or through galaxy interactions or galaxy mergers\citep{2019NatAs...3...48S}. 
While clusters contain the densest environment with large amounts of gas and a large number of galaxies (up to 1500-2000), most of the gas is hot, radiating in X-rays\citet{2007ARA&A..45..117Mcnamara}. A galaxy that falls into a cluster, like the Virgo Cluster where M87 resides (\S~\ref{sec:characterizing-m87eht}), is more likely to lose its cold gas through ram pressure stripping and heating by the intergalactic medium\citep{2000Sci...288.1617Quilis}, than to be able to steal gas from another cluster galaxy\citep{2007gitu.book.....SparkeGallagher}. Also, the galaxy velocities are much too high ($\sim$1000\,\kms) to invoke gravitational gas stripping. These high velocities are induced by the gravitating mass of the cluster, which is typically $10^{14} - 10^{15}$ \msun{},
a factor 100--10,000 more massive than galaxy groups.
It is more likely that a galaxy will accrete new cold gas in a sparser group environment (less than 100 galaxies) through galaxy interactions due to the lower galaxy velocities (about 300--400\,\kms{}) in the shallower gravitational potential compared to that of a cluster of galaxies. The lower velocities result in a longer duration  interaction during fly-bys, allowing gravity to work for longer, thereby invoking the physical process called dynamical friction (an every-day analog is air-resistance against bikers and parachutes). Simple calculations of the exchange of momentum and energy show that the kinetic energy removed when a galaxy is decelerated is transferred to the stars and gas in the other galaxy. This disturbance of the energy balance of the gas and stars can drive the gas -- in-situ or new -- to the center of the galaxy\citep{2014ApJ...792..101Davies}. 


The most efficient mechanism for driving cold gas residing in the host galaxy from large galactic scales to the central black hole remains a mystery. Evaluation is challenged by the different dynamical time scales associated with gas on different spatial scales seen in relation to the currently observed black hole activity level. 
It takes tens of millions of years or more for the gas to travel from galactic scales, so gas that powers an active black hole right now has little (or no) relation to the amount of gas present at the largest spatial scales. 
Also, if in energetically stable orbits, the gas will not move inward (or outward). So, how much gas is available at large galactic scales is no indication for the current or future black hole activity in a particular galaxy. 
Indeed, the amount of gas on large scales is observed to be similar for galaxies with quiescent and with active black holes.
It turns out that AGNs tend to have relatively more molecular gas within the central $\sim$200 parsecs\citep{2013ApJ...768..107Hicks}. 
It makes sense that the shorter dynamical time scales of gas close to  the center is more relevant for the activity we see now.

Extensive research on this topic has found that the main mechanisms that drive gas all the way to the galaxy center likely include cold gas streamers, spiral arms and stellar bars on large scales, and on smaller scales nuclear disks, nuclear gas spirals and nuclear bars that can exert negative torques on the orbiting gas and thereby remove angular momentum (i.e., rotational energy) from the gas, making it sink to the center due to the gravitational force of the material within its orbit. Figure~\ref{fig:BH-fuelling} provides an overview of the different physical mechanisms thought to be at work at different spatial scales. A comprehensive overview of theoretical and observational studies detailing these results is provided by the review of Storchi-Bergmann \& Schnorr-M\"uller\citep{2019NatAs...3...48S}.

Observations of nearby galaxies show that not just cold ($<$100\,K) and warm ($\sim$2000\,K) molecular gas but also hot ($\sim$10,000\,K) ionised gas flows toward the center at rates of 0.01 -- 10\,\msun/yr, 
$10^{-5}$ -- $10^{-3}$\,$\msun{}$/yr, and 0.01 -- 3\,\msun/yr, respectively.  
These mass inflow rates should be taken only as {\it guideline estimates} because these measurements are difficult and based on studying residuals after subtracting assumed models of ordered rotation and outflows that dominate the kinematics.  
In fact, the central gas kinematics is quite complex. For example, the warm molecular gas displays complicated central gas kinematics with a mixture of inflows, outflows and ordered orbital motion. This gas is observed within the central $\sim$100-200 parsecs in several nearby galaxies\citep{2013ApJ...768..107Hicks, 2014ApJ...792..101Davies}  and possibly extends all the way to the central torus. 
Also, since these mass inflow rates are based on a small, select number of galaxies where these measurements are possible, they may not be representative of the AGN population especially beyond our local Universe. Even with the significant uncertainties\citep{2019NatAs...3...48S}  (factors of 5--10) 
these numbers do show that much more gas is deposited in the center  than  is required to fuel the AGN activity ($\sim 10^{-3}$--$10^{-2}$\,\msun{}/yr). This means that  there is plenty of gas  to fuel the black hole activity as well as circumnuclear star formation, observed in many AGN\citep{2022MNRAS.509.4653Dahmer}. However, not all the centrally deposited gas may reach the black hole.
The powerful radiation from the central accretion disk will exert radiation pressure on the inflowing gas. Depending on the gas density and temperature of the inflowing gas, parts of this gas may be ejected from  the nucleus in powerful outflows that may reach the host  galaxy again. These outflows may deposit sufficient kinetic energy and momentum in the interstellar medium of the galaxy to either stimulate  or quench star formation, a process known as 'black hole feedback', addressed in the next section. 


\section{Observational Evidence for Feedback \label{sec:feedback}}
The energy generated by accreting black holes can affect the gaseous environment in the host galaxy and between galaxies. Shocks induced by outflows and jets of molecular gas or radio plasma can make molecular gas clouds collapse and fragments to form stars.   A good example of such 'positive feedback' is the jet-triggered star formation observed in the  radio galaxy 4C\,41.17\citep{1990ApJ...363...21Chambers, 2020A&A...639L..13Nesvadba}. However, momentum and heat energy from outflows and shocks plus radiation from the accretion process can also have a profound adverse impact on the black hole environment where the cold gas reservoir is heated, dispersed or ejected from the galaxy  and/or the  galaxic nucleus, leaving little or no fuel for further star formation and black hole growth. This (positive and negative) feedback is thought to be the underlying physics of the self-regulated black hole growth and star formation\citet{2015ARA&A..53..115KingPounds} manifested in the observed $M_{\rm BH} - \sigma$ relation (\S~\ref{sec:coevolution}).
Further details of this vast research topic can be obtained from the review articles on observational evidence of AGN feedback from outflows of ionized gas\citep{2012ARA&A..50..455Fabian} and from cold (atomic and molecular) gas\citep{2020A&ARv..28....2V}. 



\begin{figure}
    \centering
    \includegraphics[width=0.463\textwidth]{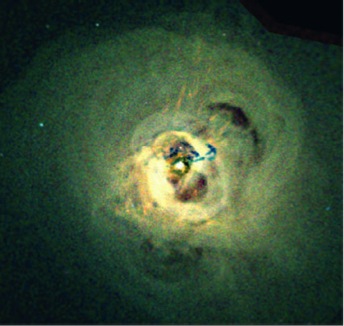}
    \includegraphics[width=0.44\textwidth]{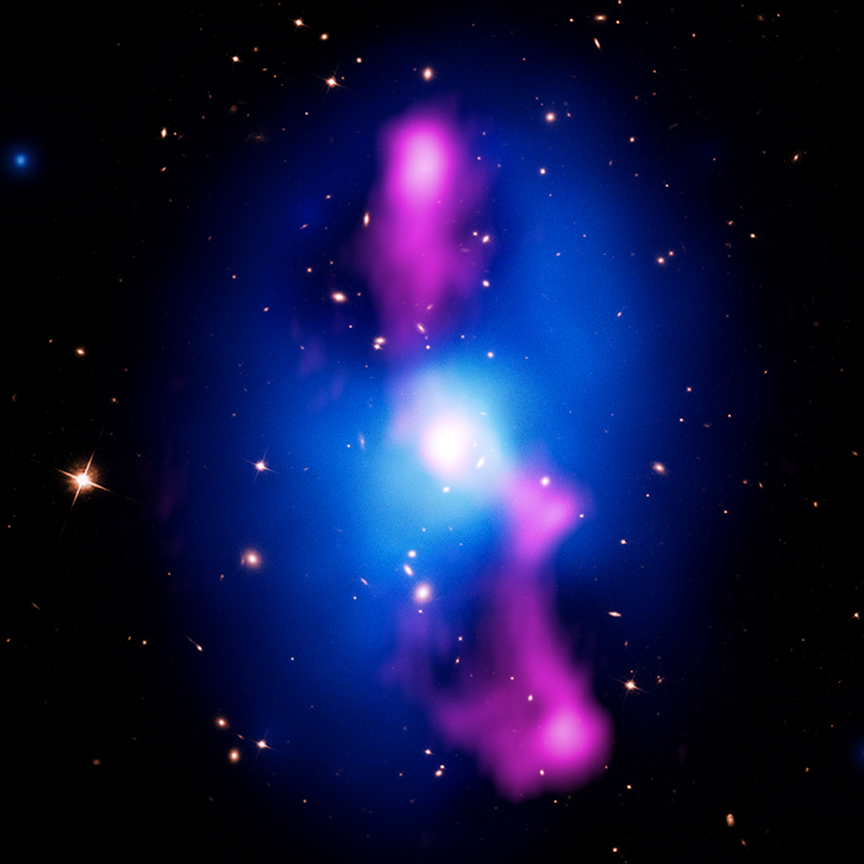}
    \caption{{\it Left: Chandra} X-ray image of the Perseus Cluster with cavities or `bubbles' (dark arcs) in the X-ray emission (green and yellow). (Figure 6 of Fabian\citet{2012ARA&A..50..455Fabian} and Figure 1 of McNamara \& Nulsen\citet{2007ARA&A..45..117Mcnamara}). {\it Right:} Composite image of MS0735.6$+$7421 cluster showing optical emission from  stars (white;  {\it Hubble Space Telescope}), X-ray emission  (blue; {\it Chandra X-ray Telescope}) and radio  emission (red;  the Very Large Array). Figure~2 of McNamara \& Nulsen\citet{2007ARA&A..45..117Mcnamara}.}
    \label{fig:clusterbubbles}
\end{figure}

Two modes of feedback 
are often considered: radio (or kinetic) mode and quasar (or radiative) mode. In the radio mode, the mechanical energy of radio jets in low-rate accreting AGN heats the surrounding medium. This is common in giant elliptical galaxies where  the jet and central engine serve to keep the interstellar and intergalactic medium hot, thereby limiting or preventing star formation. Clear evidence of this feedback is observed in massive ($10^{14} - 10^{15}$\msun{}) X-ray cooling clusters of galaxies. Cavities (or bubbles) in the hot intracluster medium are created by the jets from the central radio source\citet{2007ARA&A..45..117Mcnamara,2012ARA&A..50..455Fabian} 
(Figure~\ref{fig:clusterbubbles}). Heat dissipation from the central source and the $P$d$V$ work of the jets is sufficient to account for the energy deficit\citet{2007ARA&A..45..117Mcnamara}  
thought to exist in the 1980s - 1990s,  an issue coined 'the cooling-flow problem'. The vast amount of research on this topic can be accessed through the reviews (and references  thereto) covering observations of  clusters\citet{2007ARA&A..45..117Mcnamara, 2012ARA&A..50..455Fabian, 2012ARA&A..50..491Putman}, and theoretical considerations of the underlying physics\citet{2015ARA&A..53..115KingPounds}.

Quasar mode feedback is a powerful process thought to exist at early epochs (redshifts of 2--3) in the evolutionary history of galaxies and AGN and mostly involves pushing around cold gas, vacating it from the center or from the host galaxy through powerful winds of ionized gas\citep{2008ApJS..175..356Hopkins} 
or radio plasma\citep{2013Sci...341.1082Morganti}. Quasars certainly should have the power\citep{2012ARA&A..50..455Fabian, 2015ARA&A..53..115KingPounds} to alter and significantly shape the evolution of the host galaxy, but wide-spread direct evidence of this quasar mode feedback in action is harder to come by\citep{2012ARA&A..50..455Fabian}. While the required ultra-fast winds have indeed been detected in X-rays\citep{2015Natur.519..436Tombesi} 
and at UV wavelengths, the longevity of the winds and the mechanism by which the winds energetically couple to the interstellar medium are still very uncertain.  The UV evidence is in  the  form of broad absorption troughs in the blue wings of UV high-ionization and  low-ionization lines (observed in $\sim$20\% of quasars), covering a wide range in line-of-sight velocities up to  tens of thousands of \kms{}. The troughs are due to radiative acceleration of dense material --  seen in absorption along our  line  of sight to the continuum source -- from the vicinity of the black hole\citep{1995Natur.376..576Arav}, perhaps as a wind from the accretion disk\citep{1995ApJ...451..498Murray}. There is evidence that a large fraction of these broad absorption outflows are physically located at distances of 100\,parsecs or more from the  black hole\citep{2018ApJ...857...60Arav}, supporting the notion that the disk winds  that are driving these outflows can impact the host galaxy environment (e.g.,  through a `snow plough effect').
Evidence for powerful ionized gas outflows with significant kinetic  energy or momentum, as detected in emission at optical and infrared wavelengths, are starting to accumulate\citep{2016MNRAS.459.3144Zakamska, 2020ApJ...894...28DaviesRebecca,  2021MNRAS.504.4445VaynerZakamska} for AGN at redshifts 2 -- 3.
And recent observations confirm that the UV broad absorption lines and the ionized gas outflows, as probed by optical broad and blueshifted \oiii{} line emission, in fact trace the same outflows for a sample of extremely red\citep{2020MNRAS.495..305XuZakamska} quasars. These AGN are thought to be young quasars, reddened by large amounts of dusty gas along our line  of sight, perhaps due to outflows\citep{2017MNRAS.464.3431Hamann}.








\begin{figure}[h]
    \centering
    \includegraphics[width=0.75\textwidth]{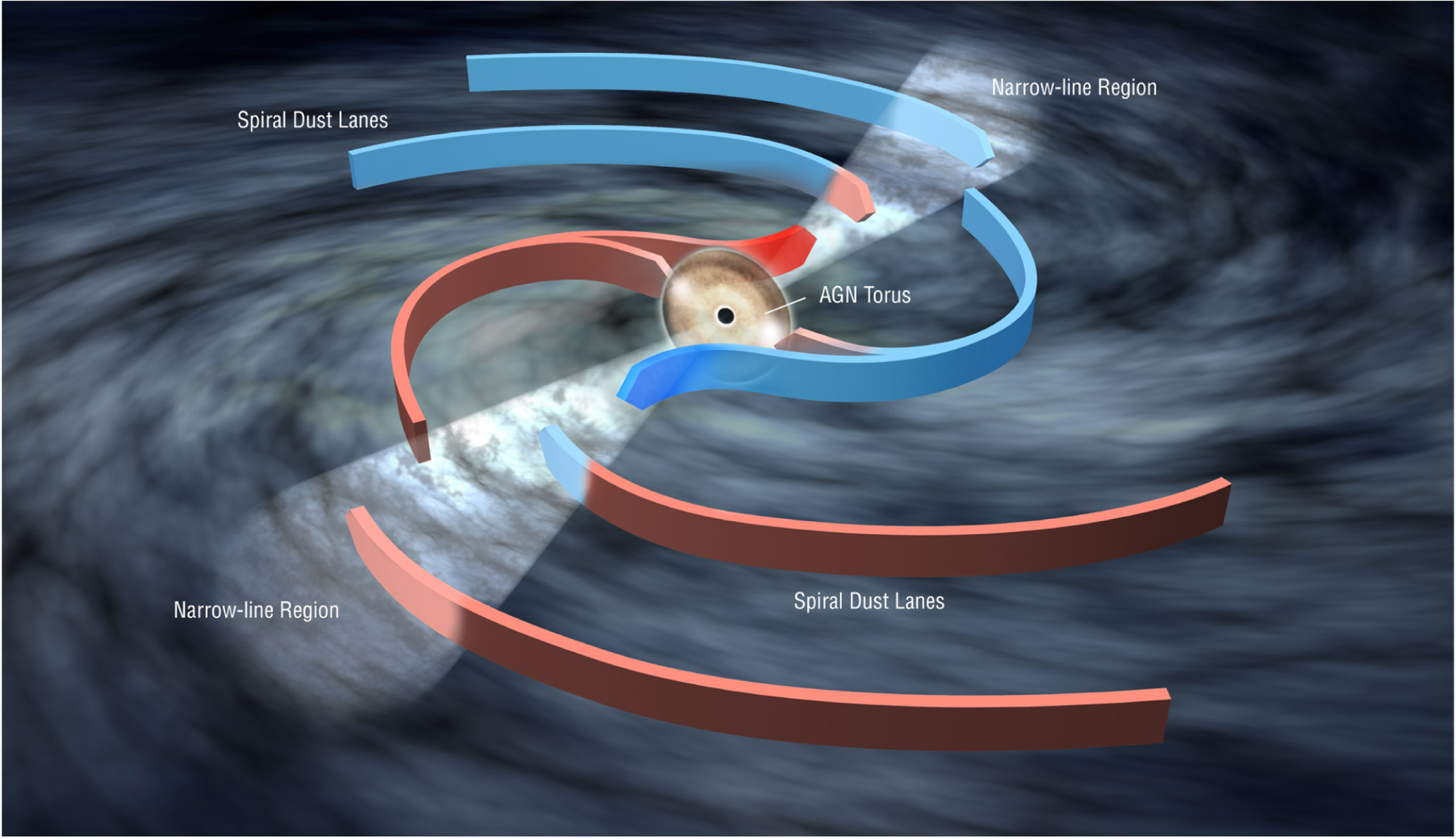}
    \caption{Cartoon explaining the observations of  Mrk\,573: When the AGN turns on a bi-cone of radiation illuminates the atomic  and molecular gas orbiting in the spiral arms. At small radial  distances from the nucleus ($<$\,750\,parsecs)  the illuminated gas in the spiral arm is accelerated radially outward, creating an outflow.  At larger distances yet the gas is not accelerated, simply ionized. (Figure~16 of Fischer  et al.\citep{2017ApJ...834...30Fischer}) Red bands move away from Earth, blue bands move toward Earth.}
    \label{fig:fischer}
\end{figure}

At lower redshifts and in the nearby universe, powerful quasars are more sparse. The AGN population is dominated by the less luminous Seyfert galaxies and LINERs. While blueshifted UV absorption lines are detected in Seyferts, the velocity widths are quite low (a few thousand \kms{}) compared to the mildly relativistic velocities observed in broad absorption line quasars. This has raised the question to what extent black hole feedback impacts the host galaxy for these AGN, especially beyond the circumnuclear region ($>$\,0.5 kpc).  
Indeed, it seems that especially low luminosity AGN ($L_{BOL} <  10^{43}$\ergs{}) are not sufficiently powerful to accelerate the in-situ gas in the galaxy much  beyond the central $\sim$200\,pc. The threshold luminosity\citep{2017MNRAS.467.2612W} required for the outflows to overcome the host galaxy gravitational potential lies somewhere between  $10^{43} - 10^{45}$\,\ergs{}. 

In recent years, spatially resolved observations with modern integral field units have provided further insight on the topic\citep{2018MNRAS.480.3993BaronNetzer}. 
Evidence of outflowing gas is in particular seen  in the narrow  line region, a  low-density region of low-velocity ($\lesssim$1000 \kms{}) ionized gas extending from $\sim$1 parsec to distances of 10 -- 1000\,parsecs. 
The observed mass outflow rates of this gas are comparable to the mass inflow rates or higher, suggesting that nuclear outflows plow into and accelerate gas located at larger galactic distances\citep{2011ApJ...739...69MullerSanchez, 2015ApJ...799...83Crenshaw}. In the case of Mrk\,573 (Figure~\ref{fig:fischer}) nuclear emission radially accelerates in-situ gas that orbits with the stellar component at central distances $<$750\,parsecs, but not the more distant galactic gas\citep{2017ApJ...834...30Fischer}.
This in-situ gas acceleration is confirmed in other nearby AGN\citep{2021ApJ...910..139Revalski} 
and may well be common. Interestingly, for these nearby AGN, it is the X-ray wind that has the kinetic power to  drive this acceleration, not the \oiii{}  wind\citep{2021MNRAS.505.3054Trindade}.
With the typically  short  reach ($<$1\,kpc) of the ionized outflow,  it is, however, unclear if these winds are commonly capable of clearing the  galactic  bulge of cold gas\citep{2015ApJ...799...83Crenshaw, 2017ApJ...834...30Fischer}. For about half the nearby AGN\citep{2011ApJ...739...69MullerSanchez} 
the observed kinetic  power of  the ionized  outflows is significantly less than the expected 0.5\% of the  bolometric luminosity required\citep{2005Natur.433..604DiMatteo}. 










\section{Highest redshift black holes \label{sec:firstBHs}}

At  present $\sim$200 quasars and 6 quasars are known to reside at epochs  when the Universe was a mere 940 Myr and 770 Myr old, respectively (i.e., redshift $z$\,$\gtrsim$\,6 and $z \gtrsim$\,7), based on deep surveys undertaken over the past $\sim$20 years\citep{2021ApJ...923..262Yang}.
The mass of their central black holes are estimated to be $10^9$ \msun{} or higher based on restframe UV-optical spectroscopy or by assuming Eddington-limited accretion. 
Observations reveal limited evolution in the spectral energy distribution of quasars out to the highest redshifts\citep{2006NewAR..50..665Fan}, indicating fast evolution of the most massive black holes at early times.
These quasars must be the very tip of the iceberg of the population of massive black holes at those epochs\citep{2020ARA&A..58...27Inayoshi}, since less luminous (and less massive) systems would not be detectable.
The mere presence of such massive black holes so early in the history of the Universe places stringent constraints on the theory of supermassive black hole formation\citep{2012Sci...337..544Volonteri} as well as the duty cycle of black hole growth by accretion, since only the fraction of black holes that are actively accreting (i.e., luminous) will be detectable (Chapter~9). 



  
   The question of how the black holes, powering the highest redshift quasars,  came to be so massive in such a short time since the Big  Bang has puzzled scientists for the past couple of decades. Insight can be obtained by examining their impact on  the gaseous environment beyond their host galaxies\citep{2020ARA&A..58...27Inayoshi, 2021ApJ...917...38Eilers}. 
   The powerful radiation from the quasar will ionize the Hydrogen atomic gas in a region in the vicinity of the quasar, allowing Lyman-$\alpha$ photons to escape the otherwise predominantly neutral intergalactic medium at the earliest epochs (redshift $\gtrsim 6$). This is evident in the observed spectrum of high redshift quasars.
   The emission from the quasar is absorbed by intervening atomic Hydrogen along our line of sight. This results in a large number of absorption lines appearing at a range of wavelengths (called the `Lyman-$\alpha$ forest'), shortward of the Lyman-$\alpha$ transition of the quasar. At some wavelength blueward of the quasar Lyman-$\alpha$ emission line, the intervening absorption lines  suddenly disappear on account of the Hydrogen ionization by the quasar. This is known as the proximity effect. 
   
   The quasar lifetime can be estimated from the size of this ionized `bubble': a black hole that has grown by accretion for an extended time, producing ionizing photons in the process, should have a larger ionized bubble than a much younger black hole.   The first measurements show a small number of the black holes at redshifts beyond 6 with unusually small bubbles, suggesting that they have not lived long\citep{2021ApJ...917...38Eilers} ($\lesssim  10^{4-5}$\,years).
   Since the bubble size also depends on other parameters, such  as the density of the extra-galactic gas and the black hole duty cycle (the fractional time that the black hole is `on'), scientists have compared observations with simulations to interpret the results. Such work suggests that these small ionized bubbles are not due to a low duty cycle\cite{2020MNRAS.493.1330DaviesFred}.  If correct, this will place additional constraints on when the first massive black  holes  formed, leaving little time to grow these giants. Observational evidence that the first supermassive black holes are born large ($M_{\rm BH} \sim 10^3$  \msun{}) has been looked  for,  but  has  not yet been confirmed\citep{2017MNRAS.469..448Bowler}.   Theoretical considerations on this topic is covered by  Chapter~9. 
   

   Since quasars are very powerful objects, they are obvious candidates  for  the  source of  ionizing radiation during the epoch of reionization. 
   However,  it  is now  clear  that the number density of quasars at the end of the reionization epoch (redshifts of 6 -- 9) is much too low to  be the predominant ionizing  source; the much more numerous population of star forming galaxies  is  the likely source\citep{2006ARA&A..44..415Fan}.
   
\section{Black Hole Growth Across Cosmic Time \label{sec:cosmic-growth}}
Understanding black hole growth on cosmological timescales informs our understanding of the connections between black holes and their host galaxies, the initial seed masses of black holes, the role of black holes in feedback processes, and understanding cosmological structure growth.  One way to study the growth of black holes is to assume that black hole mass scaling relations with host galaxy properties continue at high redshift.  This assumption can be useful, e.g., for predicting gravitational wave rates or for estimating the distribution of Eddington ratios.  An alternative is to use the observed radiative output of accreting black holes to study the population as a whole as a function of redshift.

    \subsection{Quasar number density}
    The simplest exercise in understanding black hole growth is to count the number of quasars in a given volume.  It was recognized as early as the 1960s that the number density at $z=1$ is at least two orders of magnitude larger than in the local Universe\cite{1968ApJ...151..393S}, assuming an unchanging distribution of quasar luminosities.  Deeper surveys soon indicated that the luminosity function was evolving\cite{1990ARA&A..28..437H}.
    
    \subsection{The So{\l}tan argument}
    \label{soltan}
    As we indicated in \S\ref{sec:intro}, we now know that all massive, quiescent galaxies contain relics of potential previous quasars.  This connection was first suggested by Lynden-Bell\cite{1969Natur.223..690L}, but the connection between the luminosity function of quasars in the distant universe and mass density of black holes in the local universe was not made until the pioneering work of So{\l}tan\cite{1982MNRAS.200..115S}.  Assuming a fixed radiative efficiency, $\eta$, for a black hole accreting at a rate $\dot{M}$ the luminosity of a quasar is $L = \eta \dot{M} c^2$.  By looking at the luminosity distribution of quasars as a function of redshift, an integral argument on the total mass accreted by black holes can be made.  This is now known as the So{\l}tan argument, and one of the resulting key insights is that nearly all of the $z=0$ black hole mass density was accreted during these luminous  quasar phases over cosmic time onto the black holes.  Thus studying quasar luminosity functions over cosmic time will tell us about black hole growth over cosmic time.  In nearly every waveband, the number density of quasars increases rapidly from $z=0$ to $z=1$--$2.5$ and then more gradually declines to higher redshift\cite{2015A&ARv..23....1B, 2019ApJ...884...30WangF}.  
    \subsection{Quasar luminosity functions}
    Quasar luminosity functions, describing the number density of quasars as a function of redshift and luminosity, are measured in radio, infrared/optical, and X-ray bands.  Each has its benefits.  Radio quasar luminosity function directly probe quasars' potential impact in terms of radio-mode feedback.  Infrared/optical surveys provide the best combination of breadth and depth of coverage, and infrared is proving to be especially reliable in isolating quasars from host galaxy light\cite{2006NewAR..50..665Fan}.  X-rays are almost always observed when selecting quasars by radio or infrared/optical and therefore offer an unambiguous way to find quasars\cite{2015A&ARv..23....1B}.  X-ray also is less susceptible to absorption, compared to optical; and it does not suffer from confusion with star formation like radio\cite{2005ARA&A..43..827B}.  A common theme seen in all bands of quasar luminosity functions is that as redshift increases, the peak of the luminosity function shifts to higher luminosity\cite{2009MNRAS.396.1537Labita}.  Thus the typical quasar at low redshift is less luminous than at high redshift and, for fixed radiative efficiency, this implies that the typical quasar is less massive at low redshift.  This `quasar downsizing' implies that the most massive black holes grow quickly early on and then mostly shut down\cite{2009MNRAS.396.1537Labita}.


    \subsection{Black hole mass  functions}
    
Empirical black hole mass functions -- the observed  space density of black holes as a function of redshift and  mass -- are also vital tools for characterizing the cosmic black hole growth and evolution.
Mass scaling relationships (\S~\ref{sec:SEmass}) have allowed the determination of the mass function for actively accreting black holes for  the  redshift  and mass ranges over which AGN and quasars are known.
This mass function has been characterized using a suite of black hole samples covering a range of distances from the local to the more distant Universe\citep{2012AdAst2012E...7KellyMerloni}.
While the mass functions more directly characterize the black hole demographics compared to luminosity functions, they are limited to the black holes that grow by accreting material at high rates. Another challenge is that the lower mass boundary is a priori not well defined because the black holes are not selected based on their mass,  but based on the luminosity of the AGN, since all AGN are discovered and selected from flux-limited surveys.  

Nonetheless, the observed mass functions support the notion of a maximum black hole mass\citep{2004ApJ...601..676V} of $\sim 10^{10}$\msun{}, by displaying a sharp decline in number density above this mass\citep{2012AdAst2012E...7KellyMerloni}. The mass functions also clearly reveal the aforementioned `cosmic downsizing' trend: the massive black holes are only actively accreting at earlier epochs while at lower redshifts essentially only low-mass black holes are active\citep{2009ApJ...699..800V},  explaining why the lower luminosity AGN, such as Seyfert  galaxies, are the dominant type of AGN in the nearby Universe. While the most massive black holes live fast and furious as AGN at early epochs,  
the low-mass black holes may need more time to grow sufficiently massive to power an AGN owing to their lower mass accretion rate.

    \subsection{Obscured growth}
    One potential difficulty in mapping out cosmic growth of quasars is when there is strong obscuration of the quasar light from some medium probably close to the black hole.  The number of quasars that are significantly obscured is large, potentially a majority.  Thus accounting for them is necessary to fully determine their cosmic growth.  The main difficulty is that the fraction of obscured quasars is not constant with luminosity or redshift.  Obscured quasars are also in host galaxies with larger amounts of gas and dust. Our missing the obscured quasars thus results in a systematic selection effect.  Galaxies with larger amount of gas and dust also tend to be star-forming and/or merging.  Therefore, characterizing the full connection between galaxy evolution and quasar activity requires taking into account the obscured population.  For a more detailed look into the different techniques for identifying obscured quasars see the review by Hickox \& Alexander\cite{2018ARA&A..56..625H}.

\section{Intermediate-Mass Black Holes}
\label{sec:imbhs}
Intermediate-mass black holes (IMBHs) are those with mass between stellar-mass black holes ($M_{\rm BH} < 10^2\ \msun$) and supermassive black holes ($M_{\rm BH} > 10^{6}\ \msun$). The boundaries of the definition vary by scientist, but are usually set so as not to include black holes that could have formed from a single massive star's evolution and those commonly found in centers of galaxies.  
In the context of supermassive black holes, IMBHs are astrophysically relevant as precursor stages to current supermassive black holes\cite{2017IJMPD..2630021M}. 
As stated in \S\ref{soltan}, the So{\l}tan\cite{1982MNRAS.200..115S} argument has shown that most of the mass in black holes at $z = 0$ was accreted onto the black holes over cosmic time.
Thus most black holes started less massive than they are as observed in the current epoch.  The consideration of Eddington-limited growth of black holes and the observations of $10^9\ \msun$ black holes at $z\sim7$ also requires IMBHs as their starting, or `seed' mass or else there is insufficient time to grow to the observed masses\cite{2012Sci...337..544Volonteri}.  Despite the astrophysical expectation that IMBHs exist,  observational evidence of their existence has proven difficult to come by.


Efforts to find IMBHs largely follow observational techniques of stellar-mass and supermassive black holes: dynamical evidence and evidence of accretion.
Dynamical evidence comes from both stellar and gas kinematics as well as reverberation mapping.   Greene, Strader \& Ho\citet{2020ARA&A..58..257GreeneStraderHo} compiled a handful of IMBHs of mass $M_{\rm BH} <10^6\ \msun$ at the center of galaxies.  Two of them (NGC 205 and NGC 4395) have reported masses $M_{\rm BH} < 10^4\ \msun$, but neither is fully certain as there are other mass estimates consistent with larger masses \citep{2020ApJ...892...68N, 2019NatAs...3..755W}.  There are also about a dozen galaxies with stellar dynamical measurements of a central black hole that place an upper limit such that any black hole would have to be an IMBH, but the observations are also consistent with a lack of a black hole \citep{2020ARA&A..58..257GreeneStraderHo}.
As described in \S\ref{sec:mass}, dynamical mass measurements at a specified distance are more difficult for smaller black holes  because of their smaller sphere of influence.

\begin{figure}
     \centering
     \includegraphics[width=0.49\textwidth]{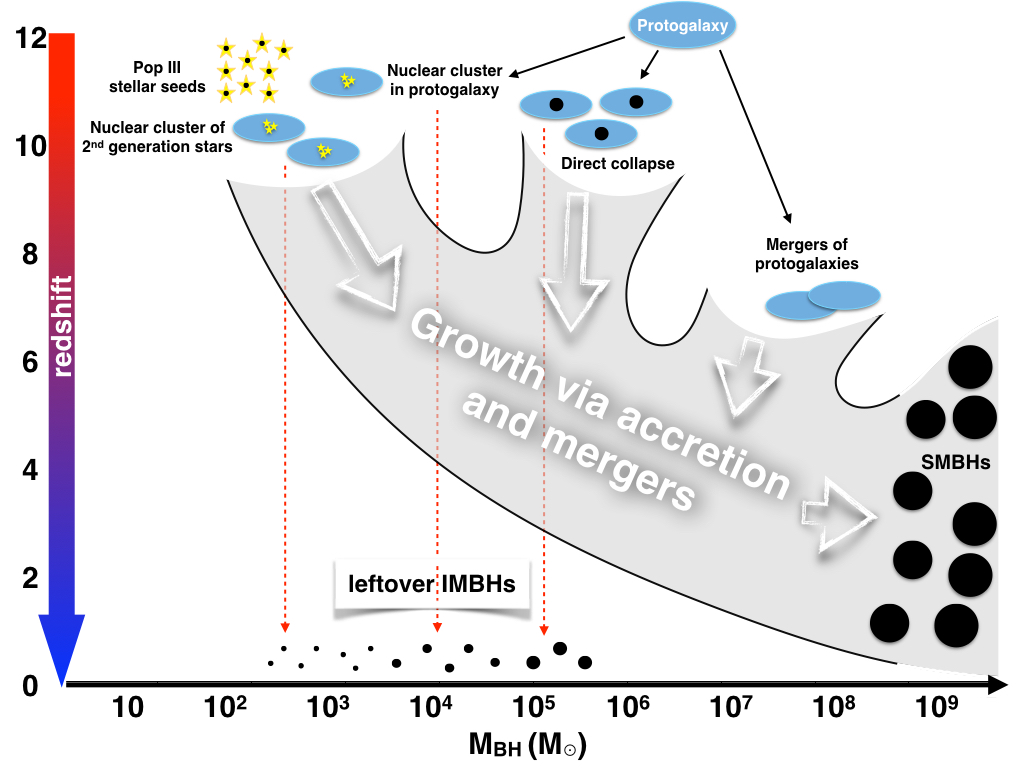}
     \includegraphics[width=0.49\textwidth]{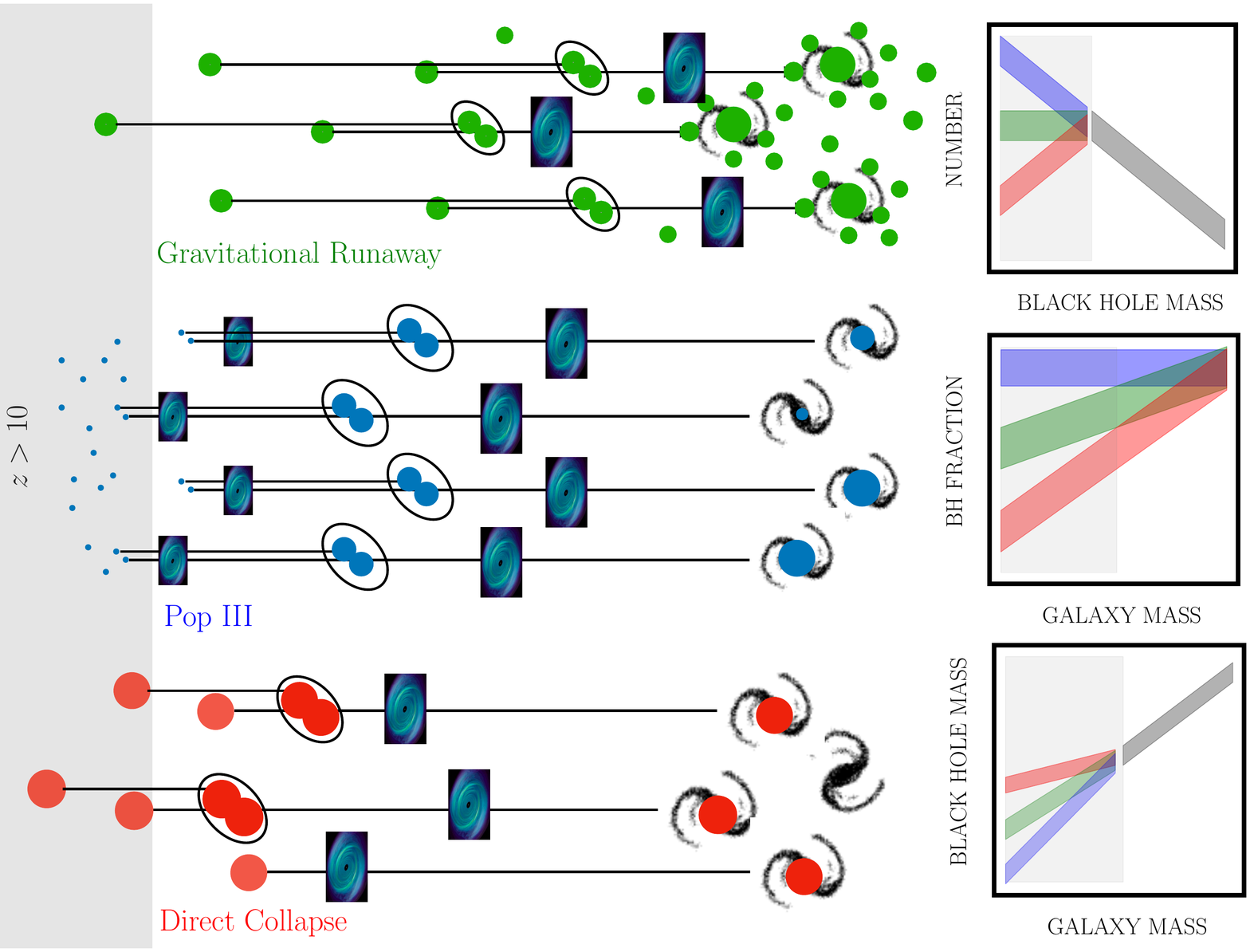}
     \caption{Schematic diagrams showing possible formation channels of black holes that lead to intermediate-mass black holes. {\it  Left panel}: Mass evolution of black holes over redshift. From Mezcua\cite{2017IJMPD..2630021M}. \textit{Right panel}: Schematic showing how the different black hole seeding mechanisms lead to different mass functions at different epochs.  These different mass functions are, in principle, observationally distinguishable at intermediate masses. From Greene et al.\cite{2020ARA&A..58..257GreeneStraderHo}}
     \label{fig:imbh}
 \end{figure}

Over one hundred candidate IMBHs have been identified from single-epoch observations of AGN line widths  (\S~\ref{sec:SEmass}) indicating masses below $10^6\ \msun$ \citep{2003ApJ...588L..13F, 2004ApJ...607...90B, 2004ApJ...610..722G, 2013ApJ...775..116R, 2014AJ....148..136M}.  A key challenge is that the spectral signatures of low-mass AGNs can be mimicked by other types of sources.  Application of the BPT diagrams (see diagrams of Section~\ref{sec:bpt}) used
to determine whether the emission lines are ionized by an accreting black hole or star formation (\S~\ref{sec:bpt}) are problematic when used in the smaller galaxies that host these candidate IMBHs. The lower metallicities of the host dwarf galaxies may lead to more high-mass X-ray binaries and more O stars with radiation fields  sufficient (especially when clustered) to explain the line ratios seen in the candidates \citep{2019A&A...622L..10S}.  An additional issue is that stellar winds and supernovae can produce line widths of $\sim300\ \units{km\ s^{-1}}$. Multiwavelength follow-up is needed to confirm the AGN nature of the candidates, and most of them do not show positive evidence of being an AGN.  Finally, there is concern that the H$\alpha$ line usually used to estimate the black hole mass in these systems may come from  non-uniform, non-virialized gas, invalidating the mass estimate\cite{2020ARA&A..58..257GreeneStraderHo}.

The chance detection of a star being tidally disrupted by a light supermassive black hole ($M_{\rm BH} < 10^8$\msun) is an alternative way to identify IMBHs. As discussed in section~\ref{sec:TDE}, the characteristics of the light curve hold information on the mass of not only the disrupted star but also of the disrupting black hole. While the number of such black holes discovered this way is currently quite low\cite{2020ARA&A..58..257GreeneStraderHo},  the next generation high-cadence photometric surveys (e.g., the Rubin
Observatory’s Legacy Survey of Space and Time) hold great potential for identifying a large number of IMBHs. This approach is demonstrated by the recent discovery of a candidate IMBH of $\sim 10^5$\msun{} in a dwarf galaxy\citep{Angus2022}.

Evidence of IMBH candidates has also been found outside of galaxy nuclei in stellar clusters and as ultraluminous X-ray sources (ULXs).  Several Galactic and one extragalactic globular cluster have been modelled via the stellar dynamical techniques and resulted in mass estimates as large as $10^4\ \msun$\cite{2020ARA&A..58..257GreeneStraderHo}.  Many globular clusters have been modeled by multiple teams, but none has multiple mass estimates that are significantly above zero.  
ULXs are off-nuclear X-ray sources with inferred luminosities above $10^{39}\ \units{erg\ s^{-1}}$, which corresponds to the Eddington limit of a $10\ \msun$ black hole \citep{2017ARA&A..55..303K}.  Above $L_X = 10^{40}\ \units{erg\ s^{-1}}$, a natural explanation is an IMBH.  Detailed observations of ULXs has shown that many are likely accreting neutron stars with anisotropic emission, invalidating the luminosity inference.  HLX-1 is one ULX that has survived to stand as best explained by an accreting IMBH \citep{2012Sci...337..554W}.

Despite the lack of confirmed IMBHs observed electromagnetically, the search continues as it is critical for understanding a number of open questions.  The seed mass distribution of supermassive black holes is unknown. 
The small black holes at the centers of the smallest galaxies are potentially closer to their seed masses on account of having undergone fewer mergers and accreted less material.  Measurement of the full mass distribution of black holes would enable inferences about the distribution of seed masses, and thus allow discrimination between competing supermassive black hole formation models\cite{2020ARA&A..58..257GreeneStraderHo}.

\section{Pairs of Black Holes}
\label{sec:pairs}
Since every massive galaxy hosts a black hole and massive galaxies merge, there is the obvious expectation that more than one black hole can exist in a galaxy at the same time. The first serious consideration of this idea was made theoretically by Begelman et al.\cite{1980Natur.287..307B} and observationally by Faber et al.\cite{1989ApJS...69..763F}  
Theoretically, the idea is relatively straightforward, but observations of pairs of black holes are relatively rare.

\subsection{Theoretical Overview}
\label{sec:pairs:theory}
The theoretical picture has not qualitatively evolved much in the last 40 years.  When two black-hole-hosting galaxies merge with each other, the galaxy merger does what it will to the dark matter, stars, and gas of the galaxies\cite{2008ApJ...676...33D}.  The black holes go through several  phases with the black hole separation monotonically decreasing with time:

\begin{enumerate}
    \item When the galaxies are interacting but still obviously separate galaxies at separations $a \gtrsim40\ \units{kpc}$, the black holes are almost completely unaffected by the galaxy-scale interaction. 

\item  Once tidal fields are strong enough to induce significant torque on the gas ($a \sim 20\ \units{kpc}$), cold gas can flow to the center of the galaxies and induce substantial accretion onto the black holes. 

\item  After the second pericenter passage, the galaxies are usually no longer individual galaxies but a merging system.  At this point, there are two black holes in a single galaxy with separation $a \sim 10\ \units{kpc}$, depending on the mass of the galaxies. 

\item As the merged galaxy settles, dynamical friction of stars acting on the black holes causes the black holes' orbits within the galaxy potential to lose energy.  If the mass ratio  ($q_{\mathrm{gal}} = M_2 / M_1 \le 1$) of the two galaxies is far from unity, then the black hole in the more massive galaxy (and presumably the more massive black hole) may not be far from the center of the recently merged galaxy. 

\item  If conditions are not unfavorable, dynamical friction can remove enough energy such that the two black holes become a gravitationally bound binary with separation of order $a \sim G M_{\mathrm{tot}} \sigma_*^{2}$, where $M_{\mathrm{tot}}$ is the total mass of the binary and $\sigma_{*}$ is the local stellar velocity dispersion. Typical values of $M_{\mathrm{tot}} = 2 \times 10^{8}\ \msun$ and $\sigma_* = 200\ \kms$ result in $a \approx 20\ \units{pc}$. 

\item Gravitational interactions with stars on orbits within $\sim10a$ will tend to take energy away from the binary, causing its orbit to tighten or `harden.'  An early key theoretical concern (catchily named the `final parsec problem') was whether there exist sufficient stars on orbits to harden the binary to the next phase.  Most modern numerical simulations agree that some combination of gas dynamical drag and non-axisymmetry is sufficient to overcome this problem \cite{2021arXiv210903262B, 2019NewAR..8601525D}. 

\item  When the binary's separation is small enough that gravitational wave emission is significant, then its evolution decouples from the stars and its orbit decays at a rate (under the quadrupole approximation)
\begin{equation}
\langle{\frac{da}{dt}}\rangle = -\frac{64}{5}\frac{G^3 m_1 m_2 (m_1 + m_2)}{c^5 a^3 (1 - e^2)^{7/2}}\left(1 + \frac{73}{24}e^2 + \frac{37}{96}e^4 \right),
\end{equation}
where $a$ is the semimajor axis, $e$ is the eccentricity, and $m_1$ and $m_2$ are the masses\cite{1964PhRv..136.1224P}.
Gravitational wave emission will continue until the two black holes merge (see Chapter~8). 

\item During the final stages of merger, gravitational waves  may be anisotropically emitted because of mass asymmetry and/or spin asymmetry.  This anisotropy and conservation of the momentum carried by the gravitational waves leads to a recoil of up to several thousands of $\kms$.  Though most recoils will be of much smaller magnitude, several thousand $\kms$ is sufficient to eject the coalesced black hole from the galaxy.

\end{enumerate}

At any point in the above sequence of events until coalescence, we can refer to the two black holes as a \emph{black hole pair}.  When not gravitationally bound, they are referred to as \emph{dual black holes}; and only when gravitationally bound are they \emph{binary black holes}.  If the black holes are active, then we refer to them as dual or binary AGNs.  The timescales for all of the phases after the galaxy has merged are highly uncertain.  Original calculations of dynamical friction timescales may have overestimated the efficiency and thus underestimated the amount of time it takes for the black holes to become bound.

Observationally, inferring the presence of black hole pairs  becomes more difficult the smaller the separation.  The techniques for finding them are organized here by decreasing physical separation of the black holes.

\subsection{Spatially resolving pairs}
\label{sec:pairs:spatial}
The most straightforward way to identify pairs is by spatially resolving them.  Inferring the presence of quiescent black holes is notoriously difficult (\S \ref{sec:mass}); so, black hole pairs are almost always found as AGN pairs.  Dual AGN can be identified in the same types of ways that single AGN are identified: by SED analysis, line ratios indicating ionization by an AGN, substantial X-ray emission, or by AGN radio emission.  The presence of an X-ray point source emission is not sufficient to infer the presence of an AGN because X-ray binaries can also appear as X-ray point sources.  However, point-source X-ray emission above about $10^{40}\ \units{erg\ s^{-1}}$ almost always is a result of AGN activity; so it is frequently used for final vindication of AGN status of otherwise ambiguous optical or infrared signatures.  To this end, much effort has gone into using \emph{Chandra} as it has the best spatial resolution of any X-ray telescope. Also,  advanced statistical techniques are used to infer the presence of pairs with separations smaller than the native pixel resolution.  

The separations probed by these techniques are limited by angular resolution and have thus far found dual AGN with separations as small as $\sim100\ \units{pc}$ in optical and X-ray.
At radio wavelengths, VLBI has the highest angular resolution achieved, and it thus has the potential to resolve widely separated binary black holes.  There are several binary candidates revealed from VLBI observations, but only one (0402+379\cite{2004ApJ...602..123M}) has a secondary that is obviously not part of an elongated jet structure.  It also has proper motion consistent with a binary orbit at speeds expected for its projected 7 pc separation \cite{2006ApJ...646...49R}.

\subsection{Varstrometry}
Varstrometry is a technique that uses high-precision astrometry to infer the presence of more than one source in an unresolved system by looking for variability in the astrometric center.  If the components of the system have uncorrelated,  intrinsic variability, then the astrometric center of the system will vary over time.  The angular variability in the  astrometric center depends on the flux ratio and intrinsic variabilities of the component sources, but if the components are comparable to each other, the  astrometric variability amplitude is expected to be
\begin{equation}
    \sigma_{\mathrm{astro}} \approx \frac{\theta}{2}\frac{\sqrt{\langle{\Delta f}\rangle}}{\bar{f}},
\end{equation}
where $\theta$ is the angular separation of the sources, $\bar{f}$ is the average total flux of the sources, and $\sqrt{\langle{\Delta f}\rangle}$ is the rms flux variability.  Typical fractional AGN variability is 10\%, and GAIA's astrometric precision of 10 mas correspond to angular separations of 0.2\arcsec, and has been used to find approximately 20 dual AGN at redshifts $z > 2$.\cite{2020ApJ...888...73H, 2022ApJ...925..162C}

\subsection{Spectro-astrometric techniques}
A third technique combines spatially distinguishing sources with separation in wavelength in two-dimensional spectra.  Dual AGN will have some relative velocity to each other if one or both is infalling or recoiling but at speeds that are comparable to or smaller than the width of lines used to identify the AGN.  The angular separations are similarly small compared to the angular resolution of the instrumentation, but when combined, the two AGN appear as sources separated in the two-dimensional spectra\cite{2012ApJ...753...42C}.  AGN emission lines, however, can be complicated.  Narrow lines in particular can have complex distributions in space and/or velocity. Thus dual AGN candidates identified with spectro-astrometric techniques usually need to be followed up with X-ray or radio imaging to confirm the presence of an AGN\cite{2015ApJ...806..219C}.

\subsection{Spectroscopic techniques}
At separations unresolvable by imaging, the high orbital velocity of binary AGN can be identified spectroscopically.  The AGN must be identified by their broad emission line regions, which consist of the gas assumed to be tightly bound to the black holes.  The width of the broad lines, however, is necessarily broader than the relative velocity of the black holes and thus makes identification of two broad-line-emitting AGN very difficult.  If only the secondary AGN has significant broad-line emission, as has been predicted by a number of theoretical works\cite{2009MNRAS.393.1423C, 2011MNRAS.415.3033R, 2014ApJ...783..134F}, then it would spectroscopically appear as a broad-line region offset in velocity relative to the narrow emission lines, which will be at the systemic velocity of the host galaxy.  Nearly 100 offset broad line AGN have been identified in SDSS data, but confirmation of these as binary AGN is difficult because of the complex nature of broad emission line region.  One alternative hypothesis is  a single-arm spiral instability in the broad emission line region.  Continued monitoring of the offset broad line AGN could eventually reveal orbital motions, but with  expected periods of up to a few centuries for separations $a \sim 0.1\ \units{pc}$, other tests are needed.\cite{2017MNRAS.468.1683R}

\subsection{Light curve analysis}
At smaller separations still, time-domain astronomy may be able to identify variable emission resulting from binary AGN.  Variable accretion rates arising from torques generated by the binary black holes plus Doppler boosting of emitting gas will result in variable continuum emission.  A large number of candidates have been identified with this technique\cite{2015MNRAS.453.1562G, 2016MNRAS.463.2145C, 2022arXiv220611497C}, but separating intrinsic AGN variability from binary AGN variability is difficult.\cite{2016MNRAS.463.2145C}

\subsection{Gravitational waves}
Gravitational waves are more fully covered in Chapter~9, but they are an obvious way to detect binary black holes.  They will be strong sources in the LISA band at milihertz frequencies as individual sources and in pulsar timing array bands at nanohertz frequencies as a stochastic signal or as individual sources.

\subsection{Recoiled and recoiling AGN}
Finally, a recoiled or recoiling black hole would be indirect evidence for a binary black hole.  Positive evidence for a galaxy lacking a black hole is very difficult, but positive evidence for an AGN off-center from the host galaxy is possible.  Several candidates have been identified\cite{2021arXiv210903262B}.

\section{Cosmic Distance Measurements \label{sec:cosdist}}
Scientists have long investigated ways in which quasars could serve as cosmic distance indicators given their high luminosity and ubiquitous presence especially at large cosmic  distances. The challenge is to identify a property that can be used as a standard candle or standard ruler.

One of the first studies is that of J. Baldwin\cite{1977ApJ...214..679B} who examined if the quasar continuum luminosity correlates sufficiently strongly with other quasar properties that they could be used as a standard candle.  J. Baldwin discovered that the strength of the CIV $\lambda$1549, in particular, decreases with increasing continuum luminosity, an effect now known as the Baldwin effect. However, the  scatter in the relationship is too large for this relationship to be a useful standard candle. Contributing to the scatter are light travel time effects between the continuum source and the broad-line region for individual objects\citet{1992AJ....103.1084PoggePeterson}.

Since then, proposals have been made over the years to use different properties of quasars to measure cosmic distances. Some examples include emission from the black hole accretion disk\citep{1999MNRAS.302L..24C}, 
the broad emission line widths\cite{1999MNRAS.308.1150R}, superluminal motion in radio jets\cite{2000ApJ...535..575H}, the angular diameter of the broad-line region\cite{2002ApJ...581L..67E} and by identifying quasars emitting at the Eddington limit\cite{2014MNRAS.442.1211M}. However, these early suggestions were not taken further. 
More recently, our improved understanding of quasars and active galactic nuclei have enabled further progress and a few promising techniques have been proposed.



In 2010 the relationship between the size of the broad-line region, measured by reverberation mapping, and the optical luminosity (\S~\ref{sec:rlrelation}) had become sufficiently tight\cite{2009ApJ...697..160Bentz, 2013ApJ...767..149B, 2010IAUS..267..151P}  that Watson et al.\cite{2011ApJ...740L..49W} proposed to use the relationship in a reversed manner. The Radius -- Luminosity relation is commonly used to infer the size of the broad-line region for an estimate of the  mass of the black hole in distant AGN where reverberation mapping is unsuitable or simply impractical (\S~\ref{sec:SEmass}) .
Instead of using the continuum luminosity from the AGN to infer the size of the broad-line region, Watson et  al.\  proposed to use reverberation  mapping results to infer the source luminosity. By comparing with the observed continuum flux received from the source  the cosmic (luminosity) distance can thus be inferred. 
By using the restframe ultraviolet continuum luminosity instead of the optical luminosity the uncertainty in the distance modulus can be decreased\cite{2015ApJ...801....8K}  (from $\Delta \mu = 0.33$ magnitudes to $\Delta \mu = 0.26$ magnitudes) due to AGN varying more with larger amplitudes in the ultraviolet regime.


Scientific groups are currently undertaking monitoring programs, in collaboration with  surveys targeting supernovae searches. The goal  is to perform reverberation mapping  of distant AGN and quasars (\S~\ref{sec:RM}) with a hope to also be able  to constrain cosmic distances in addition to measuring black hole masses and the size of the broad-line region in  these galaxies.   While it appears that Baryonic Acoustic Oscillation measurements  are currently outperforming the reverberation mapping method, the latter method is currently  where the  Supernovae method was in the 1980s, leaving room for further improvements.

As it turns out, the physical structure of black hole powered central engine is suitable as a cosmic distance gauge in other ways.  H\"onig et al.\citep{2014Natur.515..528Honig} show that the size of the hot-dust emitting region of AGN can be used as a standard ruler if the physical size $R$ is determined from time delays and the angular dimension $\theta$ is measured from infrared interferometry. This geometric technique has been dubbed the dust-parallax method. The method measures the angular diameter distance, $D_A$. The principle of this method is easily understood in a non-expanding, local volume of space. Then the geometric distance is simply $D = R/\theta$. This method was used to measure a $D_A$ of 19\,Mpc to the  local galaxy\citep{2014Natur.515..528Honig} NGC\,4151. 
Recently, a high-precision distance of 15.8 $\pm$ 0.4 Mpc was measured to NGC\,4151 using long-period Cepheid variable stars\citet{2020ApJ...902...26Yuan}.  The poor agreement of the dust-parallax distance to this robust measurement suggests that the systematics of the dust parallax technique is insufficiently understood.

Similarly, by combining measurements of the broad-line region angular size from observations with the GRAVITY instrument (\S~\ref{sec:AGN-gravity}) with the physical size measured via time delays by reverberation mapping, scientists have now measured the angular diameter distance $D_A$ to two other local AGN, 3C\,273\citep{2020NatAs...4..517W} 
and NGC\,3783\citep{2021A&A...654A..85G}. 
This advance can prove quite important since some of the nearby AGN are sufficiently nearby that their redshifts do not accurately reflect their cosmic distances. This is because the gravitational influence of nearby mass structures, such as galaxy clusters, give these AGN a significant perculiar velocity. Accurate distances are crucial for the determination of luminosities and black hole masses\cite{2019ApJ...885..161B} and the feedback power of the black hole\cite{2019NatAs...3...48S}.





Most recently, Lusso et al.\cite{2020A&A...642A.150L} show that the luminosity properties of distant quasars are intrinsically stable enough to act as a reliable standard candle. The authors show that the relationship between the X-ray and UV luminosities is sufficiently tight over four orders of magnitude in luminosity to be a cosmic distance measure when using only optically selected quasars with homogeneous spectral energy distributions that show no signs of reddening or obscuration and are free of flux-limit biases. This method is showing promise as it yields distance estimates matching those based on Supernovae measurements, \mbox{currently observed to a redshift of 1.5.}

\section{Future Outlook \label{sec:future}}

Technological advances have in the past several years pushed black hole research into an entirely new era, allowing the detection of gravitational waves from colliding black holes and the first high-resolution imaging of the shadows of two supermassive black holes using sub-millimeter interferometry that spatially resolves structure on scales near the  size of the black hole event horizon. This is possibly just a taste of the new breakthroughs and discoveries to come when the ground-based gravitational wave detectors are expanded and upgraded\footnote{For example,  for the Laser Interferometry Gravitational-wave Observatory: https://www.ligo.caltech.edu/page/ligo-lab-statement-long-term-future-observing-plans; links  to other gravitational-wave observatories (Virgo, Kagra, GEO600) are also provided on this website.}. The Event Horizon Telescope will also be upgraded\footnote{https://eventhorizontelescope.org/technology} to be more sensitive (larger collecting area and wider frequency band width) and to operate at a higher frequency. This allows  sharper images and provides access to other supermassive black holes than those in centers of the Milky Way and M87 galaxies. While the EHT may not image the  shadow  of a significant number of  supermassive black holes, its exquisite angular resolution means that EHT observations can be used to probe other aspects of  black hole and AGN physics than the plasma at and near the event  horizon. This is quite promising. This is the first time that scientists will have access to the spatial scales where most of the physics of the AGN central engine take place. However, the limitation is that this new window of discovery space is at sub-millimeter and millimeter wavelengths\citep{2019ApJ...875L...1E},  
which are not necessarily the best probes of the physics of the winds and outflows.
Granted, gravitational waves from coalescing supermassive black holes cannot be detected until the planned {\it Laser Interferometry Space Antenna}\footnote{https://lisa.nasa.gov/} ({\it LISA}) is launched in  $\sim$2037. {\it LISA} will operate at much lower frequencies (0.1 mHz and 1 Hz) than ground-based detectors  (10 Hz to 1 kHz), so {\it LISA} will be sensitive to the wider orbits of binary supermassive black holes. With {\it LISA}  it will in principle be possible to characterize the frequency of occurrence of colliding supermassive black holes and thus characterize their growth history by mergers, thereby providing important constraints on the mass distribution of the first black hole seeds. By combining electromagnetic observations with gravitational wave observations (see Chapter~9) scientists will be able to enhance the astrophysical return of both types of investigations. 

ESO's Very Large Telescope infrared interferometry instrument, GRAVITY, will also undergo an upgrade\footnote{https://www.mpe.mpg.de/ir/gravityplus} that will make it much more sensitive, reaching a $K$-band magnitude of 22, compared to the current limit\footnote{https://www.eso.org/sci/facilities/paranal/instruments/gravity/inst.html} of 17\,mag. This is expected  to enable mass determinations of supermassive black holes out to a redshift of about 2 and to measure the central mass of nearby candidate intermediate mass black holes with stellar kinematics. The upgraded instrument, GRAVITY+, is expected to come  online  in $\sim$2025.

 Hopes are high among scientists that observations\footnote{https://www.stsci.edu/jwst/science-execution/approved-programs} with the newly launched {\it James Webb Space Telescope} ({\it JWST}), sensitive in the near- and mid-infrared wavelength regime, will provide significant, further insight into not only the properties of the first supermassive black holes in the Universe and their host galaxies but also into their collective effect on the environment at those early times. 
 {\it JWST} may also provide the first accurate characterization of the \msigma{} relationship for galaxies residing at distances beyond the local Universe.
  
The Vera C. Rubin Observatory\footnote{Formerly known as the Large Synoptic Survey Telescope (LSST), https://www.lsst.org/ } will be a game changer for the discovery of variable celestial sources and transients, including tidal disruption events and AGN.  With first light in 2023, this 8.4\,m optical telescope will image the entire sky every few days. The planned high cadence photometry will be helpful for (a) characterizing the light curves of new tidal disruption events and the time variability of new and known bona-fide AGNs; (b) demographic studies of low- and intermediate mass black holes; (c) preparing follow-up spectroscopic studies of the accretion physics and stellar dynamical mass measurements with other large aperture telescopes; and (d) periodic binary AGN with short periods.

By the middle of the next decade (mid 2030's) very large ground-based telescopes will have had their first light: namely, ESO's Extremely Large Telescope\footnote{https://elt.eso.org/} (ELT, 39 meter), the Giant Magellan Telescope\footnote{GMT will consist of seven 8.4 meter segmented mirrors with equivalent resolving power of a 24.5 meter single-mirror telescope with collecting area of a 22.0 meter mirror; https://www.gmto.org/}  (GMT, $\sim$23 meter), the Thirty Meter Telescope\footnote{https://www.tmt.org/} (TMT, 30 meter). Their large telescope apertures will provide exquisite angular resolution and an amazingly large collecting area suitable for (a) detailed studies of the physics and growth of distant supermassive black holes and their role in galaxy evolution; and (b) for identifying and weighing intermediate-mass black holes. In particular, integral field spectrographs with adaptive optics such as HARMONI at ELT and GMTIFS at GMT, promise significant advances that allow determination of stellar and gas dynamics on small spatial scales that are needed to weigh intermediate-mass and supermassive black holes and to characterize the physics responsible for black hole fueling, growth and feedback onto the host galaxy.

Looking further into the future, NASA is planning for its next flagship mission the `Large UV-Optical Infrared Surveyor'\footnote{https://www.luvoirtelescope.org/} ({\it LUVOIR}) potentially to be launched no earlier than in the late 2030's with either an 8-m or a 15-m mirror. In short, the angular resolution and sensitivity will be like the {\it Hubble Space Telescope} on steroids. While one of the primary aims is to study extra-solar planets, its capabilities will also be suitable for weighing intermediate mass black holes with stellar and gas dynamics in the nearby Universe and supermassive black holes beyond that.

Significant advances are also expected from new X-ray mission concepts.
One of the primary goals of {\it ATHENA} (the {\it Advanced Telescope for High-ENergy Astrophysics}\footnote{https://www.the-athena-x-ray-observatory.eu/}) X-ray Mission, recently selected for construction by the European Space Agency with a tentative launch date in the 2030s, is to provide a complete census and characterization of how black holes have grown over cosmic time and of the physical processes responsible for this growth. 
{\it Lynx}\footnote{https://www.lynxobservatory.com/} is an ongoing concept study of a very ambitious,  transformative X-ray mission aimed to achieve high angular resolution, high sensitivity, high survey speed and large field of view. One of its main science goals is to study in detail how material accretes onto the first black holes formed in the Universe and to observe in real time the collision of black holes, localized by gravitational wave detectors weeks before the actual collision. 
If {\it Lynx} is selected, one may hope that its active mission will overlap with that of {\it LUVOIR} for optimal science return of each of these two flagship missions.

While individual modern telescopes with advanced optics and detector technology in each of their specialized wavelength regime will bring promising advances,  the combination of ongoing and future wide-area surveys at X-ray  (e.g. e-Rosita\footnote{https://www.mpe.mpg.de/eROSITA}, {\it ATHENA}), optical-infrared (e.g., {\it Euclid}\footnote{https://www.esa.int/Science\_Exploration/Space\_Science/Euclid\_overview}), infrared (e.g., {\it WFIRST-Roman}\footnote{https://roman.gsfc.nasa.gov/}) and radio (e.g., Square Kilometer Array, SKA\footnote{https://www.skatelescope.org}) wavelengths can provide powerful observational constraints on the primordial black hole population, let alone the physical conditions of the early Universe\citep{2022ApJ...926..205Cappelluti}.

In summary, the immediate and more distant future  is looking particularly bright for shedding light on the darkest, most powerful, and possibly the most enigmatic objects that we currently know to exist in our Universe.

\bibliographystyle{ws-rv-van}
\bibliography{bhbibs}

\end{document}